\documentstyle[preprint,12pt,prd,aps,epsf,epsfig,feynmf,boxedminipage,pifont,amssymb]{revtex}
\draft
\def\bfgamma{\mbox{\boldmath $\gamma$}}
\def\bfnabla{\mbox{\boldmath $\nabla$}}

\def\bfsigma{\mbox{\boldmath $\sigma$}}
\def\bfxi{\mbox{\boldmath $\xi$}}

\def\bfPi{\mbox{\boldmath $\Pi$}}
\def\lQ{\Lambda_{\rm QCD}}
\newcommand{\one}{1\!\!{\rm l}}
\newcommand{\onec}{1\!\!{\rm l}_c}
\newcommand{\ones}{1\!\!{\rm l}_s}
\newcommand{\nn}{\nonumber}
\newcommand{\be}{\begin{equation}}
\newcommand{\ee}{\end{equation}}
\newcommand{\bea}{\begin{eqnarray}}
\newcommand{\eea}{\end{eqnarray}}

\def\als{\alpha_{\rm s}}
\def\siml{{\ \lower-1.2pt\vbox{\hbox{\rlap{$<$}\lower6pt\vbox{\hbox{$\sim$}}}}\ }} 
\newcommand{\MS}{\overline{\rm MS}}

\def\lla{\langle\!\langle}
\def\rra{\rangle\!\rangle}
\newcommand{\Appendix}[1]%
    {%
     \section{#1}%
      }

\begin{document}
\tighten
\preprint{\small \tt 
CERN-TH/2002-179 ~ 
IFUM-719-FT  ~
UB-ECM-PF 02/15
\vspace{2cm}}
\title{\bf Inclusive Decays of Heavy Quarkonium to Light Particles} 
\author {Nora Brambilla$^1$, Dolors Eiras$^2$, Antonio Pineda$^2$, Joan
  Soto$^2$ and Antonio Vairo$^3$}
\address{$^1$ INFN and Dipartimento di Fisica dell'Universit\`a di Milano \\
  via Celoria 16, 20133 Milan, Italy}
\address{$^2$ Dept. d'Estructura i Constituents de la Mat\`eria and IFAE,
  U. Barcelona \\ Diagonal 647, E-08028 Barcelona, Catalonia, Spain}
\address{$^3$ Theory Division, CERN, 1211 Geneva 23, Switzerland}
\maketitle\begin{abstract}
\noindent
We derive the imaginary part of the potential NRQCD Hamiltonian up to
order $1/m^4$, when the typical momentum transfer between the heavy
quarks is of the order of $\lQ$ or greater, and the binding energy $E$
much smaller than $\lQ$. We use this result to calculate the
inclusive decay widths into light hadrons, photons and lepton pairs,
up to ${\mathcal O}(mv^3\times (\lQ^2/m^2,E/m))$ and ${\mathcal
O}(mv^5)$ times a short-distance coefficient, for $S$- and $P$-wave
heavy quarkonium states, respectively. We achieve a large reduction in
the number of unknown non-perturbative parameters and, therefore, we
obtain new model-independent QCD predictions.  All the NRQCD matrix
elements relevant to that order are expressed in terms of the wave
functions at the origin and six universal non-perturbative
parameters. The wave-function dependence factorizes and drops out in
the ratio of hadronic and electromagnetic decay widths. The universal
non-perturbative parameters are expressed in terms of gluonic field-strength 
correlators, which may be fixed by experimental data or,
alternatively, by lattice simulations. Our expressions are expected to
hold for most of the charmonium and bottomonium states below
threshold. The calculations and methodology are explained in detail
so that the evaluation of higher order NRQCD matrix elements in this
framework should be straightforward. An example is provided.
\end{abstract}

\vfill\eject

\tighten

\section{Introduction}

Heavy Quarkonium is characterized by the small relative velocity
$v$ of the heavy quarks in their centre-of-mass frame. This small
parameter produces a hierarchy of widely separated scales once
multiplied by the mass $m$ of the heavy particle: $m$ (hard), $mv$
(soft), $mv^2$ (ultrasoft), $\dots$. In general, we have $E\sim mv^2
\ll p \sim mv \ll m$, where $E$ is the binding energy and $p$ the
relative three-momentum.

The use of NRQCD \cite{nrqcd} allowed a factorization of the physics
due to the scale $m$ from the one due to smaller scales. Moreover, it
allowed the description of heavy quarkonium inclusive decays into
light fermions, photons and leptons, in terms of matrix elements of
local 4-quark operators, in a systematic way. These 4-quark operators
are of two types: colour-singlet and colour-octet operators.  The
matrix elements of the colour-singlet operators can be related in a
rigorous way with quantum-field theory defined quarkonium wave
functions \cite{nrqcd}.  Intuitively, these wave functions should be
related with the wave functions that appear in a Schr\"odinger-like
formulation of the bound-state system, namely two heavy quarks
interacting through a potential. On the other hand, the colour-octet
ones were thought to have no parallel in such formulation. In either
case, even though there had been a lot of relevant work in obtaining
the QCD potential in terms of Wilson loops \cite{potentials}, it was
not known how to obtain the systematic connection between NRQCD and a
Schr\"odinger-like formulation in the non-perturbative case, or even
whether it existed and, if so, under which circumstances.  Even in the
perturbative case, for which expressions for the potential existed at
lower orders in the past \cite{potpert}, a clean and simple derivation
of such Schr\"odinger-like formulation incorporating perturbative
ultrasoft gluons was not clear once higher order calculations in
$\als$ were required.

The observation that NRQCD still contains dynamical scales that are
not relevant to the kinematical situation of the lower-lying states in
heavy quarkonium (those with energy scales larger than the ultrasoft
scale) \cite{Mont} (see also \cite{nucl}), paved the way towards the
resolution of the questions above. Indeed, it was realized that
further simplifications occur if we integrate them out, and the
resulting effective field theory was called pNRQCD \cite{Mont}.  The
degrees of freedom of pNRQCD depend on the interplay between the
characteristic scales of the given non-relativistic system, namely
$E$, $p$ and the momentum transfer $k$, and the characteristic scale
of non-perturbative physics in QCD, which will be denoted by
$\lQ$. Therefore, how a Schr\"odinger-like formulation develops, and
thus how the NRQCD 4-fermion matrix elements will show up within this
framework, depends on the specific kinematic situation considered.

When the typical momentum transfer $k$ is much larger than $\lQ$,
$k\sim p \gg E \gtrsim \lQ$, the pNRQCD Lagrangian \cite{Mont,long}
contains not only the singlet field, which is also present in the
Schr\"odinger-like formulation, but also the octet field, ultrasoft
gluons and light quarks.  The matching from NRQCD to pNRQCD
(integration of the soft scale) can be done in perturbation theory. In
nature, this situation is relevant to the $\Upsilon (1 S)$ and the
$t$-$\bar t$ production near threshold.  If in addition $E \gg \lQ$,
we are entirely in the weak-coupling regime ($E\sim m\als^2$, $p\sim
k\sim m\als$) where non-perturbative effects can be parameterized by
local condensates \cite{VLP}. In this regime pNRQCD has been used to
obtain the complete set of logarithmic corrections to the QCD static
potential at three loops \cite{short}, the complete set of logarithmic
corrections to the very heavy quarkonium spectrum at ${\cal
O}(m\als^5)$ \cite{logs} (see also \cite{KP}), the resummation of logs
at the same order \cite{RGstatic,RG} and, very recently, the (almost)
complete spectrum of very heavy quarkonium at ${\mathcal O}(m\als^5)$
\cite{KPSS}.  We can still use the same pNRQCD Lagrangian for systems
with $E\gtrsim \lQ$.  Then, however, some of the calculations in
pNRQCD cannot be carried out perturbatively and the non-perturbative
effects can no longer be parameterized by local condensates (see
\cite{VLP,long}).

When the typical momentum transfer $k \gtrsim \lQ$ and the binding
energy is small, namely $E \ll \lQ$, the degrees of freedom of pNRQCD
are the singlet field and pseudo-Goldstone bosons (pions), if hybrids
and other degrees of freedom associated with heavy--light meson pair
threshold production develop a mass gap of ${\mathcal O}(\lQ)$, as it
is assumed in Refs. \cite{M1,M2,long} and in what follows.  If we
ignore Goldstone bosons, which play a negligible role in the present
analysis, we recover the celebrated Schr\"odinger-like picture of
quark and antiquark interacting through a potential. Therefore, the
pNRQCD Lagrangian reads \cite{M1,long}:
\begin{eqnarray}
& & {\mathcal{L}}_{\rm pNRQCD}= 
{\rm Tr} \,\bigg\{ {\rm S}^\dagger \left( i\partial_0 - h  \right) {\rm S} \bigg \} \, ,
\label{lpnrqcd}
\end{eqnarray}
where $h$ is the pNRQCD Hamiltonian, to be determined by matching to
NRQCD.  In general, one should be able to obtain the binding energies
and the total decay widths from the real and imaginary parts of the
complex poles of the propagator.  At the accuracy we are aiming at in
this paper the total decay width of the singlet heavy-quarkonium state
may be defined as:
\be
\Gamma= - 2\, {\rm Im} \, \langle n,l,s,j| h  |n,l,s,j \rangle,
\label{imag}
\ee
where $|n,l,s,j \rangle$ are the eigenstates of the real part of the Hamiltonian $h$.

In this paper we will be concerned with this situation and will
consider in full detail not only the calculation in the general case
(A) $\lQ \lesssim k$ (Section \ref{qmmatching}), but also the
particular situation (B) $\lQ \ll k$ (Section \ref{twostepmatching}):

A) $\lQ$ is smaller than or of the order of $k$. In this case, the
(non-perturbative) matching to pNRQCD has to be done in a single step.
This case has been developed in a systematic way in
Refs. \cite{M1,M2}.  As a consequence the complete set of potentials
up to order $1/m^2$ could be finally calculated \cite{M1,M2},
including a $1/m$ potential, which had been missed so far and
completing (and in some cases correcting) the previous expressions
obtained in the literature \cite{potentials} for the $1/m^2$
potential.  Most of the charmonium and bottomonium states below
threshold are expected to be in this situation.

B) $\lQ$ is much smaller than the typical momentum transfer $k$. In
this case, the degrees of freedom with energy larger than or similar
to $k$ can still be integrated out perturbatively. This leads to an
intermediate EFT that contains, besides the singlet, also octet fields
and ``ultrasoft'' gluons (meaning gluons with energies $\lesssim\lQ$
here) as dynamical degrees of freedom \cite{Mont,long}; it has the
same Lagrangian as pNRQCD in the weak coupling regime.  We will call
this EFT pNRQCD$^\prime$.\footnote{Note the change of name with
respect to section 5 of \cite{long}.}  The octet and ``ultrasoft''
gluon fields are eventually integrated out by the (non-perturbative)
matching to pNRQCD \cite{long}.

In either case, it remained to be seen how the matrix elements of the
4-fermion operators are encoded in this formulation. This was
especially needed for the octet ones since, as mentioned before, it
was thought that they could not be accommodated in a
Schr\"odinger-like formulation.  However, in \cite{pw}, we have shown
that, by using pNRQCD, it is indeed possible to relate the matrix
elements of the colour-octet operator with the wave function at the
origin and additional bound-state independent non-perturbative
parameters. This was done for the specific case of $P$-wave quarkonium
decays.  Here, we will apply the same method to express all the NRQCD
matrix elements relevant to inclusive $S$-wave quarkonium decays into
light hadrons, photons and lepton pairs at ${\mathcal
O}(c(\als(m))mv^3\times (\lQ^2/m^2,E/m))$ ($c(\als(m))$ being a
function of $\als(m)$ computable within perturbation theory). This
reduces the number of unknown parameters for the total decay widths of
charmonium and bottomonium states below threshold by roughly a factor
of 2, which allows us, in turn, to formulate several new
model-independent predictions.  Particularly important is the fact
that our formalism allows the physics due to the solution of the
Schr\"odinger equation, which appears entirely in the wave-function,
to be disentangled not only from the short-distance physics at scales
of ${\mathcal O}(m)$, but also from the gluonic excitations with an
energy of ${\mathcal O}(\lQ)$. As a consequence, the wave-function
dependence drops out in the ratio of hadronic and electromagnetic
decay widths.  For this class of observables the reduction in the
number of non-perturbative parameters in going from NRQCD to pNRQCD is
even more dramatic, since only the (six) non-perturbative universal
parameters appearing at this order in pNRQCD are needed.

\medskip

Finally, we would like to mention the dynamical situation when the
binding energy is positive and of the same order of magnitude as the
momentum transfer $k$, namely when $E \gtrsim \lQ \sim k$.  In this
case degrees of freedom with energy $\sim \lQ$ cannot be integrated
out.  States close to and beyond heavy--light meson pair threshold are
expected to be in this situation. The results of this paper do not
apply, in principle, to this case.

\medskip

The paper is arranged as follows. Section \ref{secnrqcd} reviews some
aspects of NRQCD that are relevant to the rest of the paper. Section
\ref{qmmatching} provides a detailed description of the computation of
the ``spectrum'' of NRQCD, in particular the ground state, in the
$1/m$ expansion in the general case. It is meant for the reader
interested in learning the techniques involved in this type of
computations. The description of pNRQCD, its power counting and the
relation between the computation of Section \ref{qmmatching} and the
Hamiltonian in pNRQCD are given in Section \ref{pNRQCD}. Section
\ref{twostepmatching} provides a detailed description of the matching
between pNRQCD and NRQCD in the particular case $E \ll \lQ \ll k$.
This section may help the reader who is not willing to go through the
general case in Section \ref{qmmatching}, but still wants to get a
flavour of the kind of calculations we are performing. Section
\ref{results} summarizes our results.  The reader who is only
interested in our final results and wants to skip any computational
detail may jump directly to this section.  Section \ref{mind} displays
some model-independent predictions that follow from our results. We
finally draw our conclusions in Section \ref{conclusions}. A number of
appendices complement the main body of the paper. Appendix \ref{appA}
recalls the 4-fermion NRQCD operators at ${\mathcal O}(1/m^4)$.
Appendix \ref{appdir} gives the general formula relating an arbitrary
NRQCD matrix element with the computation in pNRQCD. Appendix
\ref{ARG} gives the leading log renormalization group running of the
imaginary parts of the 4-fermion NRQCD operators matching
coefficients. Appendix \ref{Appreg} shows how to deal with ill-defined
products of distributions within dimensional regularization. Appendix
\ref{Appunit} shows how unitary transformations can relate different
forms of the pNRQCD Hamiltonian.

\medskip

\section{NRQCD}
\label{secnrqcd}
NRQCD is obtained from QCD by integrating out the heavy-quark mass scale $m$ \cite{nrqcd}. 
The NRQCD Lagrangian can be written as follows:
\be 
\label{lNRQCD}
{\cal L}_{\rm NRQCD} = {\cal L}_{\rm g} + {\cal L}_{\rm light} 
+ {\cal L}_{\rm 2-f} + {\cal L}_{\rm 4-f},
\ee
where ${\cal L}_{\rm g}$ involves only gluon fields, 
${\cal L}_{\rm light}$ involves light-quark and gluon fields, and 
${\cal L}_{2n-{\rm f}}$ are the terms in the Lagrangian with $2n$ 
heavy-quark fields.

The NRQCD Lagrangian can be organized as a series expansion in
$\als(m)$ and in the inverse of the heavy-quark mass $1/m$. Powers of
$\als(m)$ are encoded into the Wilson coefficients of NRQCD.

In this paper, we aim at a description of heavy quarkonium inclusive
decays into light hadrons and electromagnetic decays, whose appearance is due
to the imaginary terms of the NRQCD Lagrangian. It is convenient,
then, to split the Lagrangian into the Hermitian (real) and the  
anti-Hermitian (imaginary) part:
\be
{\cal L}_{\rm NRQCD} = {\rm Re}\,{\cal L}_{\rm NRQCD} 
+ i\, {\rm Im}\,{\cal L}_{\rm NRQCD}, 
\ee
where
\be
{\rm Re}\,{\cal L}_{\rm NRQCD}= {\cal L}_{\rm g} + {\cal L}_{\rm light} 
+ {\cal L}_{\rm 2-f} + {\rm Re}\,{\cal L}_{\rm 4-f},
\label{relag}
\ee
and
\be
{\rm Im}\,{\cal L}_{\rm NRQCD} ={\rm Im}\,{\cal L}_{\rm 4-f}.
\ee
The operators responsible for heavy-quarkonium decays are the NRQCD 
4-fermion operators whose matching coefficients carry an imaginary part.
For our purposes, it is sufficient to consider 
either dimension 6 or dimension 8 4-fermion operators:
\be
{\rm Im}\,{\cal L}_{\rm NRQCD} ={\rm Im}\,{\cal L}_{\rm 4-f} = {\rm
Im}\,{\cal L}_{\rm 4-f}^{d=6} + {\rm Im}\,{\cal L}_{\rm 4-f}^{{\rm EM}\,d=6}
+ {\rm Im}\,{\cal L}_{\rm 4-f}^{d=8} + {\rm Im}\,{\cal L}_{\rm
4-f}^{{\rm EM}\,d=8}. 
\ee
With the superscript EM, we indicate operators responsible for the
electromagnetic decays. More explicitly, we have  
\bea
{\rm Im}\,{\cal L}_{\rm 4-f}^{d=6}
&=& {{\rm Im}\,f_1({}^1S_0) \over m^2} \, O_1({}^1S_0) 
+ {{\rm Im}\,f_1({}^3S_1) \over m^2} \, O_1({}^3S_1) 
+ {{\rm Im}\,f_8({}^1S_0) \over m^2} \, O_8({}^1S_0) 
\nn
\\
&&
+ {{\rm Im}\,f_8({}^3S_1) \over m^2} \, O_8({}^3S_1), \\
{\rm Im}\,{\cal L}_{\rm 4-f}^{{\rm EM}\,d=6}
&=& {{\rm Im}\,f_{\rm EM}({}^1S_0) \over m^2} \,O_{\rm EM}({}^1S_0) 
+ {{\rm Im}\,f_{\rm EM}({}^3S_1) \over m^2} \,O_{\rm EM}({}^3S_1), \\
{\rm Im}\,{\cal L}_{\rm 4-f}^{d=8}
&=& {{\rm Im}\,f_1({}^1P_1)   \over m^4}  O_1({}^1P_1)
+ {{\rm Im}\,f_1({}^3P_{0}) \over m^4}  O_1({}^3P_{0})
+ {{\rm Im}\,f_1({}^3P_{1}) \over m^4}  O_1({}^3P_{1}) 
 \nonumber \\
&& 
+ {{\rm Im}\,f_1({}^3P_{2}) \over m^4}  O_1({}^3P_{2})
+ {{\rm Im}\,g_1({}^1S_0)   \over m^4}  {\cal P}_1({}^1S_0)
+ {{\rm Im}\,g_1({}^3S_1)   \over m^4}  {\cal P}_1({}^3S_1) 
 \nonumber \\
&& 
+ {{\rm Im}\,g_1({}^3S_1,{}^3D_{1}) \over m^4}  {\cal
P}_1({}^3S_1,{}^3D_{1})
+ [ O_1 \rightarrow O_8, {\cal P}_1 \rightarrow {\cal P}_8, f_1 \rightarrow f_8, g_1 
\rightarrow g_8], \\
{\rm Im}\,{\cal L}_{\rm 4-f}^{{\rm EM}\,d=8}
&=& {{\rm Im}\,f_{\rm EM}({}^1P_1)   \over m^4}  O_{\rm EM}({}^1P_1)
+ {{\rm Im}\,f_{\rm EM}({}^3P_{0}) \over m^4}  O_{\rm EM}({}^3P_{0})
+ {{\rm Im}\,f_{\rm EM}({}^3P_{1}) \over m^4}  O_{\rm EM}({}^3P_{1}) \nonumber \\
&& 
+ {{\rm Im}\,f_{\rm EM}({}^3P_{2}) \over m^4}  O_{\rm EM}({}^3P_{2})
+ {{\rm Im}\,g_{\rm EM}({}^1S_0)   \over m^4}  {\cal P}_{\rm EM}({}^1S_0)
+ {{\rm Im}\,g_{\rm EM}({}^3S_1)   \over m^4}  {\cal P}_{\rm EM}({}^3S_1) \nonumber \\
&& 
+ {{\rm Im}\,g_{\rm EM}({}^3S_1,{}^3D_{1}) \over m^4} {\cal P}_{\rm
EM}({}^3S_1,{}^3D_{1}). 
\eea
The definitions of the hadronic operators can be found in \cite{nrqcd}.
For ease of reference, we recall them in Appendix \ref{appA}, where 
we also give the definitions of the electromagnetic operators.

The distinction between hadronic and electromagnetic operators is
somewhat artificial. In general the 4-fermion operators listed in Eqs.
(\ref{O1singS})-(\ref{PtripoctSD}) are all the dimension 6 and 8
operators needed to describe decays into light hadrons and/or hard
electromagnetic particles. The information needed in order to describe
decays into hard electromagnetic particles is encoded into the
electromagnetic contributions to the matching coefficients.  The
electromagnetic operators defined in \cite{nrqcd} arise from singling
out the operators accompanying the matching coefficients whose
imaginary parts correspond to pure electromagnetic decays and
inserting into them the QCD vacuum ($\vert {\rm vac} \rangle \langle
{\rm vac}\vert$). This insertion guarantees that when calculating with
these operators in NRQCD, no contamination from soft strong
interactions will occur. Hence, the electromagnetic operators encode
all the relevant information needed in order to calculate the
quarkonium total decay width to electromagnetic particles
only. However, one might also be interested in the decays to hard
electromagnetic particles and soft light hadrons.  In this case, the
complementary to the above projector, namely $1-\vert {\rm vac}
\rangle \langle {\rm vac}\vert$ should be considered. In this paper,
however, we will restrict to the processes, and therefore to the
operators, originally considered in \cite{nrqcd}.

\medskip

The Hermitian piece of the NRQCD Lagrangian can also be written in a $1/m$ expansion:
\be
{\rm Re} \, {\cal L}= {\cal L}^{(0)}+{1 \over m}{\cal L}^{(1)}+{1 \over
m^2}{\rm Re}\,{\cal L}^{(2)}+\cdots
\,.
\ee 
At order $1/m$ the different pieces of Eq. (\ref{relag}) read
\bea
&&{\cal L}_{\rm 2-f}= \psi^\dagger \Biggl\{ i D_0 + \, {{\bf D}^2\over 2 m}
+ c_F\, g {{\bf \bfsigma \cdot B} \over 2 m}
\Biggr\} \psi 
+ \chi^\dagger \Biggl\{ i D_0 - \, {{\bf D}^2\over 2 m} 
- c_F\, g {{\bf \bfsigma \cdot B} \over 2 m}
\Biggr\} \chi 
\,,
\\ \nn
&&
{\cal L}_{\rm g} = 
- {1\over 4} G^a_{\mu \nu} G^{a\,\mu \nu} 
,
\\ 
&&{\cal L}_{\rm light}  = \sum_{j=1}^{n_f} \bar{q}_j i {D \!\!\!\!/} q_j,
\nn
\\
&& {\rm Re}\,{\cal L}_{\rm 4-f} =0,
\nn
\eea
where $\psi$ is the Pauli spinor field that annihilates the fermion and $\chi$
is the Pauli spinor field that creates the antifermion,
$i D^0=i\partial_0 -gA^0$, $i{\bf D}=i\bfnabla+g{\bf A}$,
 ${\bf B}^i = -\epsilon^{ijk}G^{jk}/2$; for
later use, we also define ${\bf E}^i = G_{0i}$ and 
$[{\bf D \cdot, E}]={\bf D \cdot E} - {\bf E \cdot D}$.
The chromomagnetic matching coefficient $c_F$ is known at
next-to-leading order and its value can be found in \cite{ABN}.
Concerning the explicit expression of the ${\cal O}(1/m^2)$ Lagrangian, see 
Ref. \cite{M2} for the operators without light quarks and Ref. \cite{ManBauer} 
for the operators including light fermions.

\section{The NRQCD ``spectrum'' in the $1/m$ expansion}
\label{qmmatching}
We assume to be in the situation $\lQ \siml mv$ in which the
matching to pNRQCD cannot be performed within a perturbative
expansion in $\als$. Nevertheless, it can be done by assuming an
expansion in $1/m$, within the Hamiltonian 
formalism of \cite{M1,M2}, to which we refer for further details.
We may divide the procedure into three steps:
\begin{itemize}
\item[1)]{The spectrum of the NRQCD Hamiltonian, made of quarkonium and gluonic 
excitations between heavy quarks, is evaluated order by order in $1/m$ starting 
from the static configuration. This will be done in Secs. \ref{sec3a}--\ref{sec3e}.} 
\item[2)]{The quantum-mechanical matrix elements are expressed in terms of
gluonic field correlators. This will be done in Sec. \ref{corr}.}
\item[3)]{The excitations of order $mv^2$ are identified as the degrees of freedom 
of pNRQCD. The matching to pNRQCD is performed by integrating out the
excitations of order $\lQ$ and $mv$. This will be done and discussed in Sec. 
\ref{pNRQCD}.}
\end{itemize}

\subsection{The NRQCD Hamiltonian}
\label{sec3a}
The NRQCD Hamiltonian without light fermions has been
worked out up to ${\cal O}(1/m)$ in Ref. \cite{M1} and up to ${\cal O}(1/m^2)$ 
in Ref. \cite{M2}, to which we refer for the explicit expressions. 
In the following we will consider the inclusion of light fermions. 

The inclusion of light fermions produces new terms in the Hamiltonian
of pure gluodynamics. In the static limit, we have:
\be
H^{(0)}=H^{(0)}(n_f=0)-\sum_{j=1}^{n_f} \int d^3{\bf x}\, \bar{q}_j \, i {\bf D }\cdot {\bfgamma} \, q_j \, .
\ee
The next corrections in the Hamiltonian, due to light fermions, 
appear at ${\cal O}(1/m^2)$ and have been  
considered in Ref. \cite{ManBauer}. We will
not need their explicit expressions in this paper. 
We will only need the expressions
of the Hermitian part of the NRQCD Hamiltonian up to order $1/m$: 
\be
{\rm Re}\, H = 
{H^{(1)} \over m}= - {1 \over 2m} \int d^3{\bf x} \psi^\dagger \left( {\bf D}^2 
+ g \, c_F \, \bfsigma \cdot {\bf B}\right) \psi
+ {1 \over 2m} \int d^3{\bf x} \chi^\dagger \left({\bf D}^2 
+ g \, c_F \, \bfsigma \cdot {\bf B} \right) \chi,
\label{H1nrqcd}
\ee
and of the imaginary part of the NRQCD Hamiltonian up to order $1/m^4$:
\be
{\rm Im}\, H= {{\rm Im}\, H^{(2)}\over m^2} + 
{{\rm Im}\, H^{(4)}\over m^4},
\label{H24}
\ee
where ${\rm Im}\,H^{(2)} = {\rm Im}\,H_{4-f}^{(2)}$, 
${\rm Im}\,H^{(4)} = {\rm Im}\,H_{4-f}^{(4)}$ and 
\bea
{H_{4-f}^{(2)}\over m^2} &=&  - \int d^3{\bf x}\,
\left(
{\cal L}_{4-f}^{d=6} +{\cal L}_{4-f}^{{\rm EM}\,d=6}
\right)
, \\
{H_{4-f}^{(4)}\over m^4} &=&  - \int d^3{\bf x}\,
\left(
{\cal L}_{4-f}^{d=8} +{\cal L}_{4-f}^{{\rm EM}\,d=8}
\right)
 .
\eea
The Gauss law, constraining the physical states $\vert {\rm phys} \rangle$, 
reads:
\be
{\bf D}\cdot {\bfPi}^a \vert {\rm phys} \rangle = 
g (\psi^\dagger T^a \psi + \chi^\dagger T^a \chi+\bar{q}\gamma^0T^aq)
\vert {\rm phys} \rangle, 
\label{gausslaw}
\ee
where $\bfPi^a$ is the canonical momentum conjugated to ${\bf A}^a$. 
In Ref. \cite{Hatfield}, general details about Hamiltonian
quantization can be found and in Refs. \cite{M1,M2} details specific to our case.

\subsection{The NRQCD spectrum at ${\cal O}(1/m^3)$}
\label{sec3b}
Let us call $H=H^{(0)}+H_I$ the NRQCD Hamiltonian, $H^{(0)}$ being its static part and 
\be
H_I= {H^{(1)}\over m} + {H^{(2)}\over m^2}  + \cdots \; .
\ee
We call $|\underbar{n}; {\bf x}_1 ,{\bf x}_2\rangle^{(0)} $ the eigenstates of 
$H_0$, $E_n^{(0)}$ the corresponding eigenvalues,  
$|\underbar{n}; {\bf x}_1 ,{\bf x}_2\rangle $ the
eigenstates of $H$, and $E_n$ the corresponding eigenvalues within a strict expansion
in $1/m$. This means that they satisfy the analogue of the Schr\"odinger equation:
\be 
H |\underbar{n}; {\bf x}_1 ,{\bf x}_2\rangle = \int d^3x_1^\prime d^3x_2^\prime 
|\underbar{n}; {\bf x}_1^\prime ,{\bf x}_2^\prime \rangle 
E_n({\bf x}_1^\prime,{\bf x}_2^\prime; {\bf p}_1^\prime, {\bf p}_2^\prime)
\delta^{(3)}({\bf x}_1^\prime-{\bf x}_1)\delta^{(3)}({\bf x}_2^\prime-{\bf
  x}_2).
\label{bornschroe}
\ee
With $n$ we indicate a generic set of conserved quantum numbers. 
Note that the heavy-quark positions ${\bf x}_1$ and ${\bf x}_2$ 
are conserved quantities only with respect to the zeroth order Hamiltonian $H_0$. 
The states $|\underbar{n}; {\bf x}_1 ,{\bf x}_2\rangle $ are normalized
according to
\be  
\langle \underbar{m}; {\bf x}_1 ,{\bf x}_2|\underbar{n}; {\bf y}_1 ,{\bf y}_2\rangle = 
\delta_{nm} \delta^{(3)} ({\bf x}_1 -{\bf y}_1)\delta^{(3)} ({\bf x}_2 -{\bf
  y}_2),
\label{normstate}
\ee
and we define 
\be
N^{1/2}_n({\bf y}_1,{\bf y}_2;{\bf p}_1,{\bf p}_2) \delta^{(3)}({\bf y}_1 -
{\bf x}_1) \delta^{(3)}({\bf y}_2 - {\bf x}_2) = 
{}^{(0)} \langle \underbar{n}; {\bf y}_1 ,{\bf y}_2|\underbar{n}; {\bf
x}_1 ,{\bf x}_2\rangle.
\label{prono}
\ee
The above three equations (\ref{bornschroe})--(\ref{prono}) may be used in
order to determine the three unknown quantities 
$|\underbar{n}; {\bf x}_1 ,{\bf x}_2\rangle $, 
$E_n$ and  $N_n({\bf y}_1,{\bf y}_2;{\bf p}_1,{\bf p}_2)$  
recursively using quantum-mechanical perturbation theory around the static solution. 
For this purpose a convenient way to rewrite Eqs. (\ref{bornschroe})--(\ref{prono}) 
is ($E_n(x;p) \equiv E_n({\bf x}_1,{\bf x}_2;{\bf p}_1,{\bf p}_2)$ and 
$E_n^{(0)}(x) \equiv E_n({\bf x}_1,{\bf x}_2)$):\footnote{
A slightly different set of equations can be found in Ref. \cite{M2}.} 
\bea 
& & 
N_n({\bf y}_1,{\bf y}_2;{\bf p}_1,{\bf p}_2) \delta^{(3)}({\bf y}_1 -
{\bf x}_1) \delta^{(3)}({\bf y}_2 - {\bf x}_2) = 
\delta^{(3)}({\bf y}_1 - {\bf x}_1) \delta^{(3)}({\bf y}_2 - {\bf x}_2)
\nn\\
& & 
\quad 
- \sum_{m\neq n} \int\!\! d^3z_1\int\!\! d^3z_2 {
\langle \underbar{n}; {\bf y}_1 ,{\bf y}_2|\underbar{m}; {\bf z}_1 ,{\bf z}_2\rangle^{(0)} 
\over E_m^{(0)}(z)-E_n^{(0)}(x)}
\nn\\
& &  
\quad
\times \bigg\{  \int\!\! d^3x_1'\int\!\! d^3x_2'  
\,^{(0)}\langle \underbar{m}; {\bf z}_1 ,{\bf z}_2|\underbar{n}; {\bf x}_1'
,{\bf x}_2'\rangle
(E_n(x';p')-E_n^{(0)}(x')) 
\delta^{(3)}({\bf x}_1' - {\bf x}_1) \delta^{(3)}({\bf x}_2' - {\bf x}_2) 
\nn\\
& & 
\qquad\qquad\qquad 
- ^{(0)}\langle \underbar{m}; {\bf z}_1 ,{\bf z}_2|H_I
|\underbar{n}; {\bf x}_1,{\bf x}_2\rangle \bigg\},  
\label{norm1}
\\
\nn\\
& & 
|\underbar{n}; {\bf x}_1,{\bf x}_2\rangle = 
 \int\!\! d^3z_1\int\!\! d^3z_2 |\underbar{n}; {\bf z}_1,{\bf z}_2\rangle^{(0)} 
N_n^{1/2}({\bf z}_1,{\bf z}_2;{\bf p}_1,{\bf p}_2) 
\delta^{(3)}({\bf z}_1 - {\bf x}_1) \delta^{(3)}({\bf z}_2 - {\bf x}_2) 
\nn\\
& & 
\quad 
+ \sum_{m\neq n} \int\!\! d^3z_1\int\!\! d^3z_2 {  
|\underbar{m}; {\bf z}_1,{\bf z}_2\rangle^{(0)}  \over E_m^{(0)}(z)-E_n^{(0)}(x)}
\nn\\
& & 
\quad 
\times \bigg\{  \int\!\! d^3x_1'\int\!\! d^3x_2' 
\,^{(0)}\langle \underbar{m}; {\bf z}_1 ,{\bf z}_2|\underbar{n}; {\bf
x}_1',{\bf  x}_2'\rangle
(E_n(x';p')-E_n^{(0)}(x')) \delta^{(3)}({\bf x}_1' - {\bf x}_1)
\delta^{(3)}({\bf x}_2' - {\bf x}_2)
\nn\\
& & 
\qquad\qquad\qquad 
- ^{(0)}\langle \underbar{m}; {\bf z}_1 ,{\bf z}_2|H_I
|\underbar{n}; {\bf x}_1,{\bf x}_2\rangle \bigg\},
\label{state}\\
\nn\\
& & 
\int\!\!d^3x_1'\int\!\! d^3x_2' N_n^{1/2}({\bf y}_1,{\bf y}_2;{\bf p}_1,{\bf p}_2)
\delta^{(3)}({\bf y}_1 - {\bf x}_1') \delta^{(3)}({\bf y}_2 - {\bf x}_2')
E_n(x';p') \delta^{(3)}({\bf x}_1' - {\bf x}_1) \delta^{(3)}({\bf x}_2' - {\bf x}_2)
\nn\\
& & 
=  E_n^{(0)}(y)  N_n^{1/2}({\bf y}_1,{\bf y}_2;{\bf p}_1,{\bf p}_2)
\delta^{(3)}({\bf y}_1 - {\bf x}_1) \delta^{(3)}({\bf y}_2 - {\bf x}_2)
+ {}^{(0)}\langle \underbar{n}; {\bf y}_1 ,{\bf y}_2|H_I
|\underbar{n}; {\bf x}_1,{\bf x}_2\rangle .
\label{norm}
\eea
By means of the above equations it is formally possible to obtain, 
within the framework of a $1/m$ expansion, the 
``energies'' and the ``states'' of any excitation of the NRQCD Hamiltonian.

\medskip

Up to ${\cal O}(H_I^3)$, the energy of a generic state labelled $n$ is
given by: 
\bea 
& & E_n(y;p)\delta^{(3)}({\bf y}_1-{\bf x}_1)\delta^{(3)}({\bf y}_2-{\bf x}_2) =
E_n^{(0)}(y) \delta^{(3)}({\bf y}_1-{\bf x}_1)\delta^{(3)}({\bf y}_2-{\bf x}_2)
\nn\\
& & ~~ +\, ^{(0)}\langle \underbar{n}; {\bf y}_1 ,{\bf y}_2 \vert H_I
\vert \underbar{n}; {\bf x}_1,{\bf x}_2 \rangle^{(0)} 
\nn \\
& &
~~ - \, {1\over 2}\sum_{k\neq n} \int\!\! d^3z_1 \, d^3z_2 \,
^{(0)}\langle \underbar{n}; {\bf y}_1 ,{\bf y}_2 \vert H_I  
\vert \underbar{k}; {\bf z}_1 ,{\bf z}_2\rangle^{(0)}\,
^{(0)}\langle \underbar{k}; {\bf z}_1,{\bf z}_2 \vert H_I
\vert \underbar{n}; {\bf x}_1,{\bf x}_2 \rangle^{(0)}
\nn\\
& & \qquad \qquad \times 
\left( {1\over E_k^{(0)}(z) - E_n^{(0)}(y)} + {1\over E_k^{(0)}(z) -
    E_n^{(0)}(x)} \right)
\nn\\
& &
~~ - \, {1\over 2}\sum_{k\neq n} \int \!\! d^3z_1 \, d^3z_2 \,\int\!\! d^3\xi_1 \, d^3\xi_2 \,
^{(0)}\langle \underbar{n}; {\bf y}_1 ,{\bf y}_2 \vert H_I  
\vert \underbar{k}; {\bf z}_1 ,{\bf z}_2\rangle^{(0)}\,
\nn\\
& &  
\qquad \qquad \qquad \qquad \quad \times
^{(0)}\langle \underbar{k}; {\bf z}_1,{\bf z}_2 \vert H_I
\vert \underbar{n}; {\bfxi}_1,{\bfxi}_2 \rangle^{(0)} 
\,^{(0)}\langle \underbar{n}; {\bfxi}_1,{\bfxi}_2 \vert H_I
\vert \underbar{n}; {\bf x}_1,{\bf x}_2 \rangle^{(0)} 
\nn\\
& & \qquad \qquad \times 
{1\over E_k^{(0)}(z) - E_n^{(0)}(x)} {1\over E_k^{(0)}(z) - E_n^{(0)}(\xi)} 
\nn\\
& &
~~ - \, {1\over 2}\sum_{k\neq n} \int \!\! d^3z_1 \, d^3z_2 \,\int\!\!
d^3\xi_1 \, d^3\xi_2 \,
^{(0)}\langle \underbar{n}; {\bf y}_1,{\bf y}_2 \vert H_I
\vert \underbar{n}; {\bfxi}_1,{\bfxi}_2 \rangle^{(0)} 
\nn\\
& &  \qquad \qquad \qquad \qquad \quad \times
^{(0)}\langle \underbar{n}; {\bfxi}_1 ,{\bfxi}_2 \vert H_I  
\vert \underbar{k}; {\bf z}_1 ,{\bf z}_2\rangle^{(0)}\,
^{(0)}\langle \underbar{k}; {\bf z}_1,{\bf z}_2 \vert H_I
\vert \underbar{n}; {\bf x}_1,{\bf x}_2 \rangle^{(0)} 
\nn\\
& & \qquad \qquad \times 
{1\over E_k^{(0)}(z) - E_n^{(0)}(y)} {1\over E_k^{(0)}(z) - E_n^{(0)}(\xi)} 
\nn\\
& &
~~ + \, {1\over 2}\sum_{k,k'\neq n} \int \!\! d^3z_1 \, d^3z_2 \,\int\!\! d^3\xi_1 \, d^3\xi_2 \,
^{(0)}\langle \underbar{n}; {\bf y}_1,{\bf y}_2 \vert H_I
\vert \underbar{k}'; {\bfxi}_1,{\bfxi}_2 \rangle^{(0)} 
\nn\\
& &  \qquad \qquad \qquad \qquad \quad \times
^{(0)}\langle \underbar{k}'; {\bfxi}_1 ,{\bfxi}_2 \vert H_I  
\vert \underbar{k}; {\bf z}_1 ,{\bf z}_2\rangle^{(0)}\,
^{(0)}\langle \underbar{k}; {\bf z}_1,{\bf z}_2 \vert H_I
\vert \underbar{n}; {\bf x}_1,{\bf x}_2 \rangle^{(0)} 
\nn\\
& & \qquad \qquad \times 
\left( {1\over E_k^{(0)}(z) - E_n^{(0)}(y)} {1\over E_{k'}^{(0)}(\xi) - E_n^{(0)}(y)} 
+ {1\over E_k^{(0)}(z) - E_n^{(0)}(x)} {1\over E_{k'}^{(0)}(\xi) -
  E_n^{(0)}(x)} \right)
\nn\\
& & ~~ + {\cal O} (H_I^4). 
\label{En3}
\eea
The expansion up to ${\cal O}(H_I)$ was considered 
in \cite{M1} in order to obtain the $1/m$ potential. 
The ${\cal O}(H_I^2)$ term was obtained in \cite{M2}. 
The ${\cal O}(H_I^3)$ expression is new.
A detailed derivation of Eq. (\ref{En3}) will be given in Sec. \ref{estruc}.

\subsection{The NRQCD states at ${\cal O}(1/m^2)$}
\label{sec3c}
The states can also be formally expanded in $1/m$:
\be
|\underbar{n};{\bf x}_1, {\bf x}_2\rangle=|\underbar{n};{\bf x}_1,
    {\bf x}_2\rangle^{(0)}+{1 \over m}| \underbar{n};{\bf x}_1, {\bf
        x}_2\rangle^{(1)}+{1 \over m^2}|\underbar{n};{\bf x}_1, {\bf
        x}_2\rangle^{(2)}+\cdots \, .
\ee
It is convenient to write the above states in terms of some new states 
$|\underline{\rm \tilde n};{\bf x}_1, {\bf x}_2\rangle$, defined recursively as  
(see also Ref. \cite{M2}):
\begin{eqnarray}
|\underbar{$\tilde n$}; {\bf x}_1 ,{\bf x}_2\rangle &=& 
|\underbar{n}; {\bf x}_1 ,{\bf x}_2\rangle^{(0)} + {1\over E_n^{(0)}(x)-H^{(0)}}
\sum_{m\not=n}\int d^3x_1^\prime d^3x_2^\prime
|\underbar{m}; {\bf x}_1^\prime ,{\bf x}_2^\prime\rangle^{(0)}
{}^{(0)}\langle \underbar{m}; {\bf x}_1^\prime ,{\bf x}_2^\prime| 
\nn
\\
& &\times 
\bigg\{H_I |\underbar{$\tilde n$}; {\bf x}_1 ,{\bf x}_2\rangle  
- \int d^3x_1^\prime d^3x_2^\prime
|\underbar{$\tilde n$}; {\bf x}_1^\prime ,{\bf x}_2^\prime \rangle 
{}^{(0)}\langle \underbar{n}; {\bf x}_1^\prime ,{\bf x}_2^\prime| H_I 
|\underbar{$\tilde n$}; {\bf x}_1 ,{\bf x}_2\rangle \bigg\}
\nn
\\
&\equiv& |\underbar{$\tilde n$};{\bf x}_1,{\bf x}_2\rangle^{(0)}
 +{1 \over m}|\underbar{$\tilde n$};{\bf x}_1, {\bf x}_2\rangle^{(1)}
 +{1 \over m^2}|\underbar{$\tilde n$};{\bf x}_1, {\bf x}_2\rangle^{(2)}+\cdots \, .
\label{ntilde}
\end{eqnarray}
As a consequence of Eq. (\ref{ntilde}), it holds that 
\be
^{(0)}\langle \underbar{n}; {\bf x}_1 ,{\bf x}_2|\underbar{$\tilde n$}; {\bf y}_1 ,{\bf y}_2\rangle = 
\delta^{(3)} ({\bf x}_1 -{\bf y}_1)\delta^{(3)} ({\bf x}_2 -{\bf y}_2)
\ee 
or equivalently (this equation will become crucial in later sections
to simplify some calculations)
\be
\label{ortho}
^{(0)}\langle \underbar{n}; {\bf x}_1 ,{\bf x}_2|\underbar{$\tilde
n$}; {\bf y}_1 ,{\bf y}_2\rangle^{(i)} =0 \qquad \forall i \not=0
\,.
\ee

At ${\cal O}(1/m)$, we obtain
\be
|\underbar{n};{\bf x}_1, {\bf x}_2\rangle^{(1)} 
=
|\underbar{$\tilde n$};{\bf x}_1, {\bf x}_2\rangle^{(1)} 
=
- \sum_{k\neq n} \int \!\! d^3z_1 \, d^3z_2 
\vert \underbar{k}; {\bf z}_1 ,{\bf z}_2\rangle^{(0)}\,
{ ^{(0)}\langle \underbar{k}; {\bf z}_1 ,{\bf z}_2 \vert  H^{(1)}   
\vert \underbar{n}; {\bf x}_1 ,{\bf x}_2\rangle^{(0)} 
\over E_k^{(0)}(z) - E_n^{(0)}(x)}.
\label{n1}
\ee
At ${\cal O}(1/m^2)$, we obtain
\be
|\underbar{n};{\bf x}_1, {\bf x}_2\rangle^{(2)}
=
|\underbar{$\tilde n$};{\bf x}_1, {\bf x}_2\rangle^{(2)}+
|\underbar{n};{\bf x}_1, {\bf x}_2\rangle^{(2)}_{\rm norm.},
\label{n2}
\ee
where 
\bea 
& & |\underbar{$\tilde n$};{\bf x}_1, {\bf x}_2\rangle^{(2)}= 
~~ -  \sum_{k\neq n} \int \!\! d^3z_1 \, d^3z_2 
\vert \underbar{k}; {\bf z}_1 ,{\bf z}_2\rangle^{(0)}\,
{ ^{(0)}\langle \underbar{k}; {\bf z}_1 ,{\bf z}_2 \vert  H^{(2)}   
\vert \underbar{n}; {\bf x}_1 ,{\bf x}_2\rangle^{(0)} 
\over E_k^{(0)}(z) - E_n^{(0)}(x)}
\nn\\
& &  
~~ 
+ 
\sum_{k\neq n} 
\int \!\! d^3z_1 \, d^3z_2 
\vert \underbar{k}; {\bf z}_1 ,{\bf z}_2\rangle^{(0)}
\nn\\
& &
~~~~~~~~~~
\times 
\Bigg( 
- \int\!\! d^3\xi_1 \, d^3\xi_2 \, 
{ ^{(0)}\langle \underbar{k}; {\bf z}_1 ,{\bf z}_2 \vert  H^{(1)} 
\vert \underbar{n}; {\bfxi}_1 ,{\bfxi}_2\rangle^{(0)}\,
^{(0)}\langle \underbar{n}; {\bfxi}_1 ,{\bfxi}_2 \vert  H^{(1)}
\vert \underbar{n}; {\bf x}_1 ,{\bf x}_2\rangle^{(0)} 
\over (E_k^{(0)}(z) - E_n^{(0)}(x)) (E_k^{(0)}(z) - E_n^{(0)}(\xi))} 
\nn\\
& &
~~~~~~~~~~~~~~~
+ 
\sum_{j\neq n} 
\int\!\! d^3\xi_1 \, d^3\xi_2 \, 
{^{(0)}\langle \underbar{k}; {\bf z}_1 ,{\bf z}_2 \vert  H^{(1)} 
\vert \underbar{j}; {\bfxi}_1 ,{\bfxi}_2\rangle^{(0)}\,
^{(0)}\langle \underbar{j}; {\bfxi}_1 ,{\bfxi}_2 \vert  H^{(1)}   
\vert \underbar{n}; {\bf x}_1 ,{\bf x}_2\rangle^{(0)}\,
\over (E_k^{(0)}(z) - E_n^{(0)}(x)) (E_j^{(0)}(\xi) - E_n^{(0)}(x))}\Bigg), 
\label{state2tilde}
\eea
and the second term, due to the normalization
of the state, reads (note that $N_0=1+N_0^{(2)}/m^2+\dots$ is Hermitian):
\bea
&&
|\underbar{n};{\bf x}_1, {\bf x}_2\rangle^{(2)}_{\rm norm.}=
-{1 \over 2}\int d^3x_1^\prime d^3x_2^\prime |\underbar{$\tilde 0$}; 
{\bf x}_1^\prime ,{\bf x}_2^\prime\rangle^{(0)}
N_0^{(2)}({\bf x}_1^\prime,{\bf x}_2^\prime; {\bf p}_1^\prime, {\bf
p}_2^\prime) \delta^{(3)}({\bf x}_1^\prime-{\bf x}_1)\delta^{(3)}({\bf
x}_2^\prime-{\bf x}_2) 
\nn
\\
&&
\quad
=
~~ - \int \!\! d^3z_1 \, d^3z_2 
\vert \underbar{n}; {\bf z}_1 ,{\bf z}_2\rangle^{(0)}
\nn\\
& &
~~~~~~~~~~
\times 
\sum_{k\neq n}  
\,\int\!\! d^3\xi_1 \, d^3\xi_2 \, 
{
^{(0)}\langle \underbar{n}; {\bf z}_1 ,{\bf z}_2 \vert  H^{(1)}  
\vert \underbar{k}; {\bfxi}_1 ,{\bfxi}_2\rangle^{(0)}\,
^{(0)}\langle \underbar{k}; {\bfxi}_1 ,{\bfxi}_2 \vert  H^{(1)}
\vert \underbar{n}; {\bf x}_1 ,{\bf x}_2\rangle^{(0)}\,
\over (E_k^{(0)}(\xi) - E_n^{(0)}(x)) (E_k^{(0)}(\xi) - E_n^{(0)}(z))} .
\label{state2nor}
\eea

By using Eq. (\ref{H1nrqcd}) and the identities 
obtained in Refs. \cite{M1,M2}, explicit expressions for
the above  Eqs. (\ref{n1}) and (\ref{n2}) can be worked out. 
In particular, at order $1/m$ we obtain (the spin-independent part 
was first obtained in \cite{M1}):
\bea
& & |\underbar{n};{\bf x}_1, {\bf x}_2\rangle ^{(1)}= 
\nn\\
& & 
~~ -  \sum_{k\neq n} 
\left(
- {1\over 2} 
{ {}^{\,\,(0)} \langle k | [{\bf D}_1\cdot,g{\bf E}_1] | n \rangle^{(0)} \over
(E_n^{(0)} - E_k^{(0)})^2} \right.
+
\sum_{j\neq n} 
{ {}^{\,\,(0)} \langle k | g{\bf E}_1 | j \rangle^{(0)} \cdot
{}^{\,\,(0)} \langle j | g{\bf E}_1 | n \rangle^{(0)} 
\over (E_n^{(0)} - E_k^{(0)})^2 (E_n^{(0)} - E_j^{(0)})}  
\nn\\
& &~~~~~~~~
+ 2 \, (\bfnabla_1 E_n^{(0)})
\cdot{ {}^{\,\,(0)} \langle k | g{\bf E}_1 | n \rangle^{(0)} \over
(E_n^{(0)} - E_k^{(0)})^3}
+ \bfnabla_1 \cdot 
{ {}^{\,\,(0)} \langle k | g{\bf E}_1 | n \rangle^{(0)} \over
(E_n^{(0)} - E_k^{(0)})^2} 
\nn\\
& &~~~~~~~~
\left.
+
{c_F \over 2}\bfsigma_1\cdot
{ {}^{\,\,(0)} \langle k | g{\bf B}_1 | n \rangle^{(0)}
\over E_n^{(0)} - E_k^{(0)}}
\right)
\vert \underbar{k}; {\bf x}_1 ,{\bf x}_2\rangle^{(0)}
\nn \\
& & ~~
+ [g{\bf E}_1 \rightarrow -g {\bf E}_2^T, 
g{\bf B}_1 \rightarrow -g {\bf B}_2^T,
\bfsigma_1 \rightarrow \bfsigma_2, 
\bfnabla_1 \rightarrow \bfnabla_2, 
{\bf D}_1 \rightarrow {\bf D}_{c\,2}], 
\label{state1}
\eea
where $|n\rangle^{(0)}$ stands for a shorthand
notation of $|n;{\bf x}_1 ,{\bf x}_2\rangle^{(0)}$, the state
that encodes the gluonic content of the state $|\underbar{n};{\bf x}_1
,{\bf x}_2\rangle^{(0)}$ and is normalized as $^{(0)}\langle
n|m\rangle^{(0)} = \delta_{nm}$ (for a precise definition, see Eq.
(\ref{basis0M1}) and the following discussion).
We will use expression (\ref{state1}) in the subsequent sections.

\subsection{${\rm Im}\,E_0$ with relative accuracy ${\cal O}(1/m^2)$: 
             structure of the calculation}
\label{estruc}
In this paper, we are interested in computing ${\rm Im}\,E_n$ (actually 
${\rm Im}\,E_0$) with relative accuracy ${\cal O}(1/m^2)$. We will now explain 
in detail how the different terms of Eq. (\ref{En3}) appear 
within the quantum-mechanical calculation.

Equations. (\ref{norm1})--(\ref{norm}), as well as the analogous equations in
Ref. \cite{M2}, implicitly assume that the Hamiltonian is
Hermitian. This is not true at arbitrary orders and the iteration of
imaginary-dependent terms may lead to problems.
Nevertheless, at the relative ${\cal O}(1/m^2)$ accuracy we are aiming at in
this paper for the imaginary terms and for the $n=0$ state, such effects are zero.
Therefore, effectively, we only have to compute the expectation value of the
imaginary part of the NRQCD Hamiltonian in terms of the ${\cal O}(1/m^2)$
eigenstates of the Hermitian part of the NRQCD Hamiltonian.\footnote{However, a
systematic method to work with unstable particles should be worked out if a higher precision is
warranted.}  The reason is that the only imaginary contribution to the states  
up to ${\cal O}(1/m^2)$ comes from the first line of Eq. (\ref{state2tilde}) 
and this term is zero for $n=0$ because of the subsequent Eq. (\ref{herm}).

The imaginary terms in the NRQCD Lagrangian only appear in the matching coefficients 
of the 4-fermion operators, i.e. in ${\cal L}_{4-f}$. Therefore, the imaginary part of 
the NRQCD Hamiltonian has the structure of Eq. (\ref{H24}).
Profiting from this structure of the imaginary terms and since the
iteration of the leading imaginary terms gives zero, ${\rm Im}\,E_0$ can be
computed from 
\be
{\rm Im} \, E_0 \, \delta^{(3)}({\bf x}_1-{\bf x}'_1)\delta^{(3)}({\bf x}_2-{\bf x}'_2)  
=  
\langle {\underline 0};{\bf x}_1, {\bf x}_2|{\rm Im}\,H|{\underline 0};{\bf x}'_1,
    {\bf x}'_2\rangle.
\ee
Expanding in $1/m$ the states and ${\rm Im}\,H$, we can identify the
different terms of ${\rm Im}\,E_0$ in the $1/m$ expansion:
\be
{\rm Im}\, E_0= {1 \over m^2} {\rm Im} \,E_0^{(2)}
+{1 \over m^3} {\rm Im} \,E_0^{(3)}+{1 \over m^4} {\rm Im} \,E_0^{(4)}+ \cdots
\label{en234}
\,.
\ee
They read 
\be
\label{m2matching} 
{\rm Im} \, E_0^{(2)} \, 
\delta^{(3)}({\bf x}_1-{\bf x}'_1)\delta^{(3)}({\bf x}_2-{\bf x}'_2) 
=  
{}^{(0)}\langle {\underline 0};{\bf x}_1, {\bf x}_2|{\rm
Im} \, H^{(2)}|{\underline 0};{\bf x}'_1,
    {\bf x}'_2\rangle^{(0)},
\ee
\bea
\label{m3matching} 
&&
{\rm Im} \, E_0^{(3)} \, 
\delta^{(3)}({\bf x}_1-{\bf x}'_1)\delta^{(3)}({\bf x}_2-{\bf x}'_2) 
\\
\nn
&&
\qquad
=
{}^{(1)}\langle {\tilde {\underline 0}};{\bf x}_1, {\bf x}_2|{\rm
Im}\,H^{(2)}| {\underline 0};{\bf x}'_1,
    {\bf x}'_2\rangle^{(0)}
+
{}^{(0)}\langle {\underline 0};{\bf x}_1, {\bf x}_2|{\rm
Im}\,H^{(2)}|{\tilde {\underline 0}};{\bf x}'_1,
    {\bf x}'_2\rangle^{(1)},
\eea
\bea
\label{m4matching} 
&&
{\rm Im} \, E_0^{(4)} \, 
\delta^{(3)}({\bf x}_1-{\bf x}'_1)\delta^{(3)}({\bf x}_2-{\bf x}'_2) 
\\
\nn
&&
\qquad
=
{}^{(0)}\langle {\underline 0};{\bf x}_1, {\bf x}_2|{\rm
Im}\,H^{(4)}|{\underline 0};{\bf x}'_1, {\bf x}'_2\rangle^{(0)}
+
{}^{(1)}\langle {\underline 0};{\bf x}_1, {\bf x}_2|{\rm
Im}\,H^{(2)}|{\underline 0};{\bf x}'_1, {\bf x}'_2\rangle^{(1)} 
\\
\nn
&&
\qquad
+
{}^{(2)}\langle {\tilde {\underline 0}};{\bf x}_1, {\bf x}_2|{\rm
Im}\,H^{(2)}| {\underline 0};{\bf x}'_1,
    {\bf x}'_2\rangle^{(0)}
+
{}^{(0)}\langle {\underline 0};{\bf x}_1, {\bf x}_2|{\rm
Im}\,H^{(2)}|{\tilde {\underline 0}};{\bf x}'_1,
    {\bf x}'_2\rangle^{(2)}
\\
\nn
&&
\qquad
+
{}^{(0)}\langle {\underline 0};{\bf x}_1, {\bf x}_2|{\rm
Im}\,H^{(2)}| {\underline 0};{\bf x}'_1,
    {\bf x}'_2\rangle^{(2)}_{\rm norm.}+   
{}^{(2)}_{\rm norm.}\langle  {\underline 0};{\bf x}_1, {\bf x}_2|{\rm
Im}\,H^{(2)}|{\underline 0};{\bf x}'_1,
    {\bf x}'_2\rangle^{(0)}.
\eea

\medskip

After an explicit calculation, we have 
\be
{\rm Im} \, E_0^{(3)}=0, 
\ee
since 
\be
\label{sta1}
{}^{(1)}\langle {\tilde {\underline 0}};{\bf x}_1, {\bf x}_2|{\rm
Im}\,H^{(2)}|{\underline 0};{\bf x}'_1,
    {\bf x}'_2\rangle^{(0)}
=
{}^{(0)}\langle {\underline 0};{\bf x}_1, {\bf x}_2|{\rm
Im}\,H^{(2)}|{\tilde {\underline 0}};{\bf x}'_1,
    {\bf x}'_2\rangle^{(1)}
=0\,.
\ee
Moreover, we have 
\be
\label{state2}
{}^{(2)}\langle {\tilde {\underline 0}};{\bf x}_1, {\bf x}_2|{\rm
Im}\,H^{(2)}|{\underline 0};{\bf x}'_1,
    {\bf x}'_2\rangle^{(0)}
=
{}^{(0)}\langle {\underline 0};{\bf x}_1, {\bf x}_2|{\rm
Im}\,H^{(2)}|{\tilde {\underline 0}};{\bf x}'_1,
    {\bf x}'_2\rangle^{(2)}
=0
\,.
\ee
These results follow from Eq. (\ref{ortho}), supplemented by the 
following argument. The colour structure of ${\rm Im}\,H^{(2)}$ is such that, at the
gluonic level, the following matrix elements are produced 
within the total expression:
\be
{}^{(0)}\langle n| \one_1 \otimes \one_2 | 0 \rangle^{(0)}
={}^{(0)}\langle n| 0 \rangle^{(0)}=\delta_{n0} 
\ee
(by definition) and
\be
{}^{(0)}\langle n| T^a_1 \otimes T^{\dagger a}_2 | 0 \rangle^{(0)}.
\ee
In order to deal with this second expression, we note that the lowest
excitation, in the limit ${\bf x}_1 \to {\bf x}_2$,  
has no gluonic content and behaves like 
$|0; {\bf x}_1 ,{\bf x}_2 \rangle^{(0)} = \onec/\sqrt{N_c}|{\rm vac}\rangle$, so that:
\be 
{}^{(0)}\langle n| T^a_1 \otimes T^{\dagger a}_2 | 0 \rangle ^{(0)}
\delta^{(3)}({\bf x}_1-{\bf x}_2) 
= C_f \delta_{n0} \delta^{(3)}({\bf x}_1-{\bf x}_2),
\label{eq1}
\ee
where $C_f=(N_c^2-1)/(2N_c)$. 
The above expressions may appear problematic since they involve
the behaviour of the state in the limit ${\bf x}_1 \to {\bf
x}_2$. and some regularization could be required in this case. 
However, we actually only need a weaker condition to ensure that
Eq. (\ref{state2}) is zero. What we have is an expression like 
\be
\sum_{n\not=0}\sum_{k\not=0}{}^{(0)}\langle 0|O_1|n \rangle^{(0)}(\cdots)
^{(0)}\langle k|T^a_1 \otimes T^{\dagger a}_2 | 0 \rangle^{(0)}
\delta^{(3)}({\bf x}_1-{\bf x}_2),
\ee
where $O_1$ is some unspecified operator. Following Ref. \cite{M1},
this expression is the spectral decomposition of the Wilson loop
(for the definition of a Wilson loop with a number $n$ of operator
insertions, see Ref. \cite{M2}):
\be
\int dt_1\, ... \, dt_n \; \langle \!\langle O_1(t_1)(\cdots)T^a_1 \otimes T^{\dagger
a}_2(t_n) \rangle \!\rangle_c \; 
\delta^{(3)}({\bf x}_1-{\bf x}_2),
\label{wilop}
\ee
where $\langle \!\langle O  \rangle \!\rangle$ stays for the insertion of the 
operator $O$ on a static Wilson loop of spatial extension ${\bf x}_1 - {\bf
x}_2$. In the presence of more operators, the symbol $\langle \!\langle \cdots 
\rangle \!\rangle_c$ indicates the connected part (see in particular the
erratum of Ref. \cite{M2}).
One can see that the operator (\ref{wilop}) is zero in the limit ${\bf x}_1 \to {\bf
x}_2$. In order to obtain this result, it is very important that
the delta acts directly on the states\footnote{We may have situations
where the Wilson-loop operator has the structure
\be
\int dt_1\, ... \,dt_n
\left(\bfnabla_1\langle \!\langle O_1(t_1)(\cdots)T^a_1 \otimes T^{\dagger
    a}_2(t_n) 
\rangle \!\rangle_c \right)
\delta^{(3)}({\bf x}_1-{\bf x}_2).
\ee
In this case the argument does not apply since the delta does not act 
directly on the Wilson loop.}. In this situation, and in the
limit ${\bf x}_1 \to {\bf x}_2$, one can see that the disconnected
piece of the Wilson loop cancels with the connected piece, 
proving Eq. (\ref{state2}). 

For the other terms, we have
\bea 
& & {}^{(1)}\langle {\underline 0};{\bf x}_1, {\bf x}_2|{\rm
Im}\,H^{(2)}|{\underline 0};{\bf x}'_1, {\bf x}'_2\rangle^{(1)}  = 
\nn\\
& &
\quad 
~~ {1 \over 2} \sum_{k,k'\neq 0} \int \!\! d^3z_1 \, d^3z_2 \,\int\!\! d^3\xi_1 \, d^3\xi_2 \,
^{(0)}\langle \underbar{0}; {\bf y}_1,{\bf y}_2 \vert H^{(1)} 
\vert \underbar{k}'; {\bfxi}_1,{\bfxi}_2 \rangle^{(0)} 
\nn\\
& &  
\quad\qquad \qquad \qquad \times
  ^{(0)}\langle \underbar{k}'; {\bfxi}_1 ,{\bfxi}_2 \vert  
{\rm Im}\,H_{4-f}^{(2)} 
\vert \underbar{k}; {\bf z}_1 ,{\bf z}_2\rangle^{(0)} \,
^{(0)}\langle \underbar{k}; {\bf z}_1,{\bf z}_2 \vert H^{(1)} 
\vert \underbar{0}; {\bf x}_1,{\bf x}_2 \rangle^{(0)} 
\nn\\
& & 
\quad\qquad \times 
\left( {1\over E_k^{(0)}(z) - E_0^{(0)}(y)} {1\over E_{k^\prime}^{(0)}(\xi) - E_0^{(0)}(y)} 
+ {1\over E_k^{(0)}(z) - E_0^{(0)}(x)} {1\over E_{k'}^{(0)}(\xi) -
  E_0^{(0)}(x)} \right)\!,
\label{ImH211} \\
& & \nn \\
& & \nn \\
& & {}^{(0)}\langle {\underline 0};{\bf x}_1, {\bf x}_2|{\rm
Im}\,H^{(2)}| {\underline 0};{\bf x}'_1,
    {\bf x}'_2\rangle^{(2)}_{\rm norm.}
+
{}^{(2)}_{\rm norm.}\langle  {\underline 0};{\bf x}_1, {\bf x}_2|{\rm
Im}\,H^{(2)}| {\underline 0};{\bf x}'_1, {\bf x}'_2\rangle^{(0)} = 
\nn\\
& &
\quad
-{1 \over 2} \sum_{k\neq 0} \int \!\! d^3z_1 \, d^3z_2 \,\int\!\! d^3\xi_1 \, d^3\xi_2 \,
^{(0)}\langle \underbar{0}; {\bf y}_1 ,{\bf y}_2 \vert  H^{(1)}   
\vert \underbar{k}; {\bf z}_1 ,{\bf z}_2\rangle^{(0)}\,
\nn\\
& &  
\quad\qquad \qquad \qquad \times
^{(0)}\langle \underbar{k}; {\bf z}_1,{\bf z}_2 \vert H^{(1)} 
\vert \underbar{0}; {\bfxi}_1,{\bfxi}_2 \rangle^{(0)} 
 {}^{(0)}\langle \underbar{0}; {\bfxi}_1,{\bfxi}_2 \vert 
{\rm Im}\,H_{4-f}^{(2)}
\vert \underbar{0}; {\bf x}_1,{\bf x}_2 \rangle^{(0)} 
\nn\\
& & \quad\qquad \times 
{1\over E_k^{(0)}(z) - E_0^{(0)}(x)} {1\over E_k^{(0)}(z) - E_0^{(0)}(\xi)} 
\nn\\
& &
~~ -{1 \over 2} \sum_{k\neq 0} \int \!\! d^3z_1 \, d^3z_2 \,\int\!\! d^3\xi_1 \, d^3\xi_2 \,
  ^{(0)}\langle \underbar{0}; {\bf y}_1,{\bf y}_2 \vert 
{\rm Im} \, H_{4-f}^{(2)}
\vert \underbar{0}; {\bfxi}_1,{\bfxi}_2 \rangle^{(0)}  
\nn\\
& &  \quad\qquad \qquad \qquad \times
^{(0)}\langle \underbar{0}; {\bfxi}_1 ,{\bfxi}_2 \vert H^{(1)} 
\vert \underbar{k}; {\bf z}_1 ,{\bf z}_2\rangle^{(0)}\,
^{(0)}\langle \underbar{k}; {\bf z}_1,{\bf z}_2 \vert H^{(1)} 
\vert \underbar{0}; {\bf x}_1,{\bf x}_2 \rangle^{(0)} 
\nn\\
& & \quad\qquad \times 
{1\over E_k^{(0)}(z) - E_0^{(0)}(y)} {1\over E_k^{(0)}(z) - E_0^{(0)}(\xi)} 
.
\label{ImH2nor}
\eea
Indeed, the last two equations hold as well for an arbitrary $n$ and
not only for the state $n=0$, for which we have explicitly displayed them. 
It can be easily checked that the imaginary part of Eq. (\ref{En3}) for $n=0$ 
coincides with the above expression (\ref{en234}) supplemented by 
Eqs. (\ref{m2matching})--(\ref{m4matching}), (\ref{sta1}), 
(\ref{state2}), (\ref{ImH211}) and (\ref{ImH2nor}).

\subsection{${\rm Im}\,E_0$ with relative accuracy ${\cal O}(1/m^2)$: explicit
  expressions in terms of gluonic fields}
\label{sec3e}
The expressions obtained in the previous section can be rearranged in
terms of the pure gluonic content (see Refs. \cite{M1,M2}). 
In order to achieve this we have to make the quark field content of the
states explicit and use the Wick theorem. There is some freedom in choosing the
specific realization of the quark fields under spin
transformations. In \cite{M1}, the following state was chosen
\begin{equation}
\vert \underbar{n}; {\bf x}_1 ,{\bf x}_2  \rangle^{(0)}
\equiv  \psi^{\dagger}({\bf x}_1) \chi({\bf x}_2)
|n;{\bf x}_1 ,{\bf x}_2\rangle^{(0)} \qquad \forall {\bf x}_1,{\bf x}_2\,.
\label{basis0M1}
\end{equation}
In the basis of 4-fermion operators that we are using in this
paper (see Appendix \ref{appA}) and in the above basis,  
the quantum-mechanical operators that naturally appear are 
$\ones\otimes\ones$ and $\bfsigma^i\otimes\bfsigma^j$, 
where $\ones$ ($\bfsigma^i$) is the identity (sigma-matrix) 
in spin space acting either on the final- or the initial- spin quark--antiquark
state. Analogous definitions can be made for the operators 
acting on the colour subspace. 

Another possibility is the state 
\begin{equation}
\vert \underbar{n}; {\bf x}_1 ,{\bf x}_2  \rangle^{(0)}
\equiv  \psi^{\dagger}({\bf x}_1) \chi_c^{\dagger} ({\bf x}_2)
|n;{\bf x}_1 ,{\bf x}_2\rangle^{(0)} \qquad \forall {\bf x}_1,{\bf x}_2\,,
\label{basis0M2}
\end{equation}
which has been used in Ref. \cite{M2}. The quantum-mechanical
operators, which naturally appear in this way, are $\one_{1,2}$,
$\bfsigma_{1,2}^i$, and they represent the operators acting either
on the particle 1 or 2 (in this case we have always a particle 
interpretation). Analogous definitions can be made 
for the operators acting on the colour subspace. This representation
appears to be more convenient for the calculations of the
quantum-mechanical matching. In principle, one could also
write the local 4-fermion operators in a basis convenient for these
states by using Fierz transformations \cite{4f}. 

In both cases, we assume the state to be properly normalized in the
spin sector. Depending on the calculation, one definition turns out to
be more useful than the other. In any case, at the end, we are
interested to write the quantum-mechanical Hamiltonian relevant to the
Schr\"odinger equation. A way of avoiding ambiguities is to write
everything in terms of a definite set of spin operators. We will adopt
the operators ${\bf S}^i$ and $\one$ acting on a generic $1/2\otimes
1/2$ spin space and defined as
\be
{\bf S}^i={\bfsigma_{1}^i\over 2}\otimes\one_{2}+ \one_{1} \otimes {\bfsigma_{2}^i\over 2},
\qquad
\one=\one_{1}\otimes\one_{2}. 
\ee
It is possible to transform them in the operators $\ones\otimes\ones$ and 
$\bfsigma^i\otimes\bfsigma^j$ by using the identities:
\be
\chi_c {\bf S}^i\,{\bf S}^j \chi_c^\dagger 
=\chi^\dagger \bfsigma^i\otimes\bfsigma^j \chi, 
\qquad
\chi_c \left( 2\, \one-{\bf S}^2 \right)\chi_c^\dagger 
= \chi^\dagger \one_s\otimes\one_s \chi. 
\ee

\medskip

Let us now compute the different matrix elements that appear in
Eq. (\ref{m4matching}). The contribution due to the dimension-8 
4-fermion operators reads
\bea
\nn
& & ^{(0)}\langle \underbar{0}; {\bf x}_1 ,{\bf x}_2 \vert  
{\rm Im}\,H_{4-f}^{(4)}
\vert \underbar{0}; {\bf y}_1 ,{\bf y}_2\rangle^{(0)} = 
\Bigg(
C_A {\rm Im}\,f_1({}^{2S+1}P_J){\cal T}^{ij}_{SJ}
\bfnabla^i\delta^{(3)}({\bf r})\bfnabla^j
\\
\nn
& &~~
+{C_A\over 2}{\rm Im}\,g_1({}^{2S+1}S_J)\Omega^{ij}_{SJ}\bigg\{
\bfnabla^i\bfnabla^j 
+ { \delta_{ij} \over 3}{\cal E}_1,\delta^{(3)}({\bf r}) \bigg\}
+{T_F\over 3} {\rm Im}\,f_8({}^{2S+1}P_J){\cal T}^{ii}_{SJ}{\cal
  E}_1 \delta^{(3)}({\bf r}) \Bigg)
 \\
 &&
 \qquad\qquad
 \times
  \delta^{(3)}({\bf x}_1-{\bf y}_1)\delta^{(3)}({\bf x}_2-{\bf y}_2), 
\label{H4}
\eea
where $C_A=N_c$, $\bfnabla \equiv \bfnabla_{\bf r}$, ${\bf r} \equiv {\bf x}_1 - {\bf
  x}_2$ and (${\cal T}_S$ will be used in Sec. \ref{twostepmatching})
\bea 
{\cal T}^{ij}_{01} &=& \delta_{ij}(2\one-{\bf S}^2), \label{defT1}\\
{\cal T}^{ij}_{10} &=& {1\over 3} {\bf S}^i \, {\bf S}^j, \\
{\cal T}^{ij}_{11} &=& {1\over 2}\epsilon_{ki\ell}\, \epsilon_{kj\ell'} \,
{\bf S}^\ell \, {\bf S}^{\ell'}, \\
{\cal T}^{ij}_{12} &=& 
\left({\delta_{ik}{\bf S}^\ell + \delta_{i\ell}{\bf S}^k \over 2} 
- {{\bf S}^i\delta_{k\ell}\over 3}\right) 
\left({\delta_{jk}{\bf S}^\ell + \delta_{j\ell}{\bf S}^k \over 2} 
- {{\bf S}^j\delta_{k\ell}\over 3}\right), \label{defT2} \\
\Omega^{ij}_{00} &=& \delta_{ij}(2\one-{\bf S}^2), \\
\Omega^{ij}_{11} &=& \delta_{ij} \, {\bf S}^2,  \\
\Omega^{ij}_{11}({}^3S_1,{}^3D_{1}) &=& {\bf S}^i \, {\bf S}^j 
- {\delta_{ij}\over 3} {\bf S}^2, \\ 
{\cal T}_S &=& {1 \over 3} \Omega^{ii}_{SS} \label{defT}.
\eea
Equations. (\ref{defT1})--(\ref{defT2}) and (\ref{defT}) provide the explicit
expressions of the operators ${\cal T}_S$ and ${\cal T}_{SJ}^{ij}$ 
first used in Ref. \cite{pw}. 
The non-perturbative constant ${\cal E}_1$ (as well as all the other 
constants ${\cal E}_3$, ${\cal B}_1$, ${\cal E}_3^{(2)}$, ${\cal E}_3^{(2,c)}$
and ${\cal E}_3^{(2,{\rm norm.})}$ appearing 
in this section) will be defined in Sec. \ref{corr}. 
If we consider the electromagnetic contribution due to $H_{4-f}^{(4)}$, we obtain 
(in this case there are no octet operators)  
\bea
\label{H4EM}
& & ^{(0)}\langle \underbar{0}; {\bf x}_1 ,{\bf x}_2 \vert  
{\rm Im} \,H_{4-f}^{(4,{\rm EM})}
\vert \underbar{0}; {\bf y}_1 ,{\bf y}_2\rangle^{(0)} = 
\Bigg(
C_A {\rm Im} \,f_{\rm EM}({}^{2S+1}P_J){\cal T}^{ij}_{SJ}
\bfnabla^i\delta^{(3)}({\bf r})\bfnabla^j
\\
& &~~
+{C_A\over 2}{\rm Im} \,g_{\rm EM}({}^{2S+1}S_J)\Omega^{ij}_{SJ}\bigg\{
\bfnabla^i\bfnabla^j 
+ { \delta_{ij}  \over 3}{\cal E}_1,\delta^{(3)}({\bf r})  \bigg\}
 \Bigg)
  \delta^{(3)}({\bf x}_1-{\bf y}_1)\delta^{(3)}({\bf x}_2-{\bf y}_2). 
\nn
\eea

\medskip

In order to calculate the contribution due to the $1/m$ correction to the
state, we need to know (a $\one$ is understood where no spin-operator is displayed):
\bea 
& & 
^{(0)}\langle \underbar{n}; {\bf x}_1 ,{\bf x}_2 \vert H^{(1)} 
\vert \underbar{m}; {\bf y}_1 ,{\bf y}_2\rangle^{(0)} = \nn 
\\
& & ~~
\left(
{1\over 2} 
{ {}^{\,\,(0)} \langle n | [{\bf D}_1\cdot,g{\bf E}_1] | m \rangle^{(0)} \over
E_n^{(0)} - E_m^{(0)}}
-
\sum_{j\neq n} 
{ {}^{\,\,(0)} \langle n | g{\bf E}_1 | j \rangle^{(0)} \cdot
{}^{\,\,(0)} \langle j | g{\bf E}_1 | m \rangle^{(0)} 
\over (E_n^{(0)} - E_m^{(0)}) (E_n^{(0)} - E_j^{(0)})} \right. 
\nn\\
& &~~~~~~~~~~~~~~~~~~~~~~
-
(\bfnabla_1 E_n^{(0)})
\cdot{ {}^{\,\,(0)} \langle n | g{\bf E}_1 | m \rangle^{(0)} \over
(E_n^{(0)} - E_m^{(0)})^2}
- \bfnabla_1 \cdot 
{ {}^{\,\,(0)} \langle n | g{\bf E}_1 | m \rangle^{(0)} \over
E_n^{(0)} - E_m^{(0)}} 
\nn\\
& &~~~~~~~~~~~~~~~~~~~~~~
\left.
-
{c_F \over 2}\bfsigma_1\cdot
{}^{\,\,(0)} \langle n | g{\bf B}_1 | m \rangle^{(0)}
\right)
\delta^{(3)}({\bf x}_1-{\bf y}_1)\delta^{(3)}({\bf x}_2-{\bf y}_2) 
\nn \\
& & ~~
+ [g{\bf E}_1 \rightarrow -g {\bf E}_2^T, 
g{\bf B}_1 \rightarrow -g {\bf B}_2^T,
\bfsigma_1 \rightarrow \bfsigma_2, 
\bfnabla_1 \rightarrow \bfnabla_2, 
{\bf D}_1 \rightarrow {\bf D}_{c\,2}]
~~~~~~~~~~~ \forall n\neq m,
\label{H1}
\eea
\bea
& & 
^{(0)}\langle \underbar{n}; {\bf x}_1 ,{\bf x}_2 \vert  
{\rm Im}\, H_{4-f}^{(2)}
\vert \underbar{m}; {\bf y}_1 ,{\bf y}_2\rangle^{(0)} = \nn 
\\
\nn
& & ~~
-\Bigg(
\left[ 2 \left(
{\rm Im}\,f_1(^1 S_0) 
-  {T_F\over N_c} {\rm Im}\,f_8(^1 S_0) \right) 
\right.
\\
& &~~
\left.
+  \left(
{\rm Im}\,f_1(^3 S_1)   -{\rm Im}\,f_1(^1 S_0)  
+  {T_F\over N_c} \left({\rm Im}\,f_8(^1 S_0) - {\rm Im}\,f_8(^3 S_1) \right)
\right)
{\bf S}^2  \right]  
\delta^{(3)}({\bf r}) 
\langle n| \onec\otimes\onec |m\rangle 
\nn\\
& & ~~ 
+ \left[ 2\,{\rm Im}\,f_8(^1 S_0) 
+ \left( {\rm Im}\,f_8(^3 S_1) - {\rm Im}\,f_8(^1 S_0) \right)
{\bf S}^2\right]  
\delta^{(3)}({\bf r}) \, T_F \, \delta_{nm}
\Bigg)
\nn\\
& & ~~~~~~~~~~~~~~~~~~~~~~~~~~~ \times
\delta^{(3)}({\bf x}_1-{\bf y}_1)\delta^{(3)}({\bf x}_2-{\bf y}_2),
\label{H2}
\eea
where $F_j \equiv F({\bf x}_j)$, $F$ being a generic gluonic operator.
In particular from the last equation it follows that 
\be
^{(0)}\langle \underbar{n}; {\bf x}_1 ,{\bf x}_2 \vert  
{\rm Im}\, H_{4-f}^{(2)}
\vert \underbar{0}; {\bf y}_1 ,{\bf y}_2\rangle^{(0)} = 0  \qquad \forall
n\neq 0.
\label{herm}
\ee
It is this equation that guarantees that, for the $n=0$ quarkonium
state, no imaginary contribution is carried by the state (see the
discussion at the beginning of Sec. \ref{estruc}).  Finally, from the
above equations it follows that the contributions due to the $1/m$
correction to the state read:
\bea
&&
{}^{(1)}\langle {\tilde {\underline 0}};{\bf x}_1, {\bf x}_2|{\rm 
Im}\,H^{(2)}|{\tilde {\underline 0}};{\bf y}_1, {\bf y}_2 \rangle^{(1)} = 
\nn
\\
&&~~
\Bigg(
{T_F \over 9}{\cal E}_3
{\bfnabla}\delta^{(3)}({\bf r}){\bfnabla}
\left[
4\,{\rm Im} \,f_8(^1 S_0)
-2{\bf S}^2 \left({\rm Im} \,f_8(^1 S_0)-{\rm Im}
\,f_8(^3 S_1)\right) 
\right]
\nn
\\
&&~~
+2\,T_F\,c_F^2\,{\cal B}_1\,\delta^{(3)}({\bf r})
\left[
{\rm Im} \,f_8(^3 S_1)+
{1 \over 6}{\bf S}^2
({\rm Im} \,f_8(^1 S_0)-3{\rm Im} \,f_8(^3 S_1))
\right]
\nn
\\
&&~~
+{T_F \over 3}{\cal E}_3^{(2)}\delta^{(3)}({\bf r})
\left[
4\,{\rm Im} f_8(^1 S_0)-2{\bf S}^2\left({\rm Im} f_8(^1 S_0)-{\rm Im} f_8(^3 S_1)\right)
\right]
\nn
\\
&&~~
-{C_A \over 3}({\cal E}_3^{(2)}-{\cal E}_3^{(2,c)})\delta^{(3)}({\bf r})
\left[
4\,{\rm Im} f_1(^1 S_0)-2{\bf S}^2\left({\rm Im} f_1(^1 S_0)-{\rm Im} f_1(^3 S_1)\right)
\right]
\Bigg)
\nn
\\
&& ~~~~~~~~~~~~~~~~~~~~~~~~~~~ \times
\delta^{(3)}({\bf x}_1-{\bf y}_1)\delta^{(3)}({\bf x}_2-{\bf y}_2)
.
\eea
For the electromagnetic contribution we have the intermediate vacuum, which does not allow
an intermediate emission of gluons. This means that 
\be
{}^{(1)}\langle {\underline 0};{\bf x}_1, {\bf x}_2|{\rm 
Im}\,H^{(2)}_{\rm EM}|{\underline 0};{\bf y}_1, {\bf y}_2\rangle^{(1)}=0.
\ee

The contributions due to the normalization of the state read
\bea
&&
{}^{(0)}\langle {\underline 0};{\bf x}_1, 
{\bf x}_2|{\rm Im}\,H^{(2)}|{\underline 0};{\bf y}_1,
    {\bf y}_2\rangle^{(2)}_{\rm norm.} +
{}^{(2)}_{\rm norm.}\langle  {\underline 0};{\bf x}_1, {\bf x}_2|{\rm
Im}\,H^{(2)}| {\underline 0};{\bf y}_1, {\bf y}_2\rangle^{(0)} =
\nn
\\
&&~~
\bigg(
- {2 \over 9}C_A {\cal E}_3
\left\{ \bfnabla^2, \delta^{(3)}({\bf r})\right\}
\left[
{\rm Im} \,f_1(^1 S_0)+
{{\bf S}^2 \over 2}\left({\rm Im} \,f_1(^3 S_1)-{\rm Im} \,f_1(^1 S_0)\right)
\right] 
\nn
\\
\nn
&&~~
-2\,C_A\,c_F^2\,{\cal B}_1\,\delta^{(3)}({\bf r})
\left[ {\rm Im} \,f_1(^1 S_0)+
{{\bf S}^2 \over 6}\left({\rm Im} \,f_1(^3 S_1)-3{\rm Im} \,f_1(^1 S_0)\right)
\right]
\\
\nn
&&~~
-{C_A \over 3}\left({\cal E}_3^{(2,c)}+{\cal E}_{3}^{(2,{\rm norm.})}\right)
\delta^{(3)}({\bf r})
\left[
4\,{\rm Im} f_1(^1 S_0)-2\,{\bf S}^2
\left({\rm Im} f_1(^1 S_0)-{\rm Im} f_1(^3 S_1)\right)
\right]
\bigg)
\\
&& ~~~~~~~~~~~~~~~~~~~~~~~~~~~ \times
\delta^{(3)}({\bf x}_1-{\bf y}_1)\delta^{(3)}({\bf x}_2-{\bf y}_2).
\eea
Exactly the same contribution is obtained from the electromagnetic
terms if we change the subscript 1 in the matching coefficients by EM.

\subsection{Gluonic correlators}
\label{corr}
The non-perturbative constants ${\cal E}_n$, ${\cal B}_n$, 
${\cal E}_3^{(2)}$, ${\cal E}_3^{(2,c)}$ and ${\cal E}_3^{(2,{\rm norm.})}$,
which appeared in the previous section, are pure gluonic 
quantities, since the fermionic fields have been integrated out. 
Within the quantum-mechanical matching, they are first obtained in
terms of gluonic states. For instance, we obtain the expressions 
\bea
{\cal E}_n {\delta_{ij} \over 3}&=&  (-i)^{n+1} n! \sum_{k\neq 0}
{\langle 0|g{\bf E}^i|k\rangle\langle k|g{\bf E}^j|0\rangle \over 
(E_k^{(0)} - E_0^{(0)})^{n+1}},
\label{corrEn}
\\
{\cal B}_n {\delta_{ij} \over 3}&=& 
 (-i)^{n+1} n! \sum_{k\neq 0}
{\langle 0|g{\bf B}^i|k\rangle\langle k|g{\bf B}^j|0\rangle \over 
(E_k^{(0)} - E_0^{(0)})^{n+1}},
\label{corrBn}
\\
\nn
{\cal E}_3^{(2,c)} &=& - {3! \over 4} \sum_{n,r,s\neq 0} 
\left\{
{\langle 0|g{\bf E}_1|r\rangle \cdot \langle r|g{\bf E}_1|n\rangle
\langle n|g{\bf E}_1|s\rangle \cdot \langle s|g{\bf E}_1|0\rangle \over 
(E_0^{(0)} - E_r^{(0)})(E_0^{(0)} - E_m^{(0)})^{4}(E_0^{(0)} - E_s^{(0)})}
\right.
\\
\nn
\qquad\qquad
&&
+{\langle 0|g{\bf E}^T_2|r\rangle \cdot \langle r|g{\bf E}^T_2|n\rangle
\langle n|g{\bf E}^T_2|s\rangle \cdot \langle s|g{\bf E}^T_2|0\rangle \over 
(E_0^{(0)} - E_r^{(0)})(E_0^{(0)} - E_m^{(0)})^{4}(E_0^{(0)} - E_s^{(0)})}
\\
\nn
\qquad\qquad
&&
+{\langle 0|g{\bf E}_1|r\rangle \cdot \langle r|g{\bf E}_1|n\rangle
\langle n|g{\bf E}^T_2|s\rangle \cdot \langle s|g{\bf E}^T_2|0\rangle \over 
(E_0^{(0)} - E_r^{(0)})(E_0^{(0)} - E_m^{(0)})^{4}(E_0^{(0)} - E_s^{(0)})}
\\
\qquad\qquad
&&
\left.
+{\langle 0|g{\bf E}^T_2|r\rangle \cdot \langle r|g{\bf E}^T_2|n\rangle
\langle n|g{\bf E}_1|s\rangle \cdot \langle s|g{\bf E}_1|0\rangle \over 
(E_0^{(0)} - E_r^{(0)})(E_0^{(0)} - E_m^{(0)})^{4}(E_0^{(0)} - E_s^{(0)})}
\right\}, \\
\cdots 
&& \qquad\qquad\qquad\qquad \cdots \qquad\qquad\qquad \cdots \qquad\qquad\qquad\qquad.
\nn
\eea
For the first two equations, there is no need to specify whether 
the gluonic fields are inserted on the particle or on the antiparticle line 
since they give the same contribution. We do not give here the complete list of expressions 
at the quantum-mechanical level, since this section does this in terms of Wilson loop operators. 
The former may be derived straightforwardly from the latter by spectral decomposition.

\medskip

Using the techniques of Refs. \cite{M1,M2}, it is possible  
to express ${\cal E}_n$, ${\cal B}_n$, 
${\cal E}_3^{(2)}$, ${\cal E}_3^{(2,c)}$ and ${\cal E}_3^{(2,{\rm norm.})}$
in terms of the more familiar gluonic field correlators. 
We obtain (traces as well as suitable Schwinger lines connecting the 
gluon fields are understood if not explicitly displayed)\footnote{
Note that the quantity $\cal E$ used in Ref. \cite{pw} corresponds here 
to $N_c \,{\cal E}_3$.}
\bea
{\cal E}_n = 
{1 \over N_c}\int_0^\infty dt \, t^n \langle g{\bf E}(t)\cdot g{\bf E}(0)\rangle,
\eea
\bea
{\cal B}_n = 
{1 \over N_c}\int_0^\infty dt \, t^n \langle g{\bf B}(t)\cdot g{\bf B}(0)\rangle,
\eea
\be
{\cal E}^{(2,c)}_3 = {1 \over 4N_c}
\int_0^\infty dt_1\int_0^{t_1}dt_2\int_0^{t_2}dt_3 \, (t_2-t_3)^3 \, 
\langle \{g{\bf E}(t_1)\cdot, g{\bf E}(t_2)\}\, \{g{\bf E}(t_3)\cdot, g{\bf E}(0)\}\rangle_c, 
\ee
\bea
{\cal E}^{(2,{\rm norm.})}_{3}&=& -{1 \over 4N_c}
\left\{
\int_0^\infty dt_1\int_0^{t_1}dt_2\int_0^{t_2}dt_3 
\right.
\nn
\\
\nn
&&
\times
\bigg(
\left((t_2-t_3)^3+(t_1-t_3)^3\right)
\langle \{g{\bf E}(t_1)\cdot, g{\bf E}(t_2)\}\, \{g{\bf E}(t_3)\cdot,
g{\bf E}(0)\}\rangle_c 
\\
\nn
&&
\qquad\qquad\quad
+ (t_1-t_2)^3
\langle \{g{\bf E}^i(t_1), g{\bf E}^j(t_2)\}\, \{g{\bf E}^i(t_3),
g{\bf E}^j(0)\}\rangle_c 
\\
\nn
&&
\qquad\qquad\quad
+ 4(t_1-t_2)^3
\langle g{\bf E}^i(t_1) g{\bf E}^j(t_2)\, g{\bf E}^j(t_3)
g{\bf E}^i(0) \rangle_c 
\bigg)
\\
\nn
&&
- 2 \int_0^\infty dt_1\int_0^{t_1}dt_2
(t_1-t_2)^3
\bigg(
\langle g{\bf E}^i(t_1) [i{\bf D}^i,g{\bf E}^j](t_2) g{\bf E}^j(0) \rangle
\\
\nn
&&
\qquad\qquad\quad
+
\langle g{\bf E}^i(t_1)  g{\bf E}^j(t_2) [i{\bf D}^i,g{\bf E}^j](0) \rangle
+
\langle g{\bf E}^i(t_1) [i{\bf D}^j,g{\bf E}^j](t_2) g{\bf E}^i(0) \rangle
\bigg)
\\
&&
\left.
+
\int_0^\infty dt_1 \, t_1^3 \, 
\langle g{\bf E}^i(t_1) [i{\bf D}^i,[i{\bf D}^j,g{\bf E}^j]](0)\rangle
\right\}
+{1 \over 4}{\cal E}_0{\cal E}_4+{1 \over 3}{\cal E}_1{\cal E}_3,
\label{E32norm}
\eea
\bea
{\cal E}^{(2)}_3 &\equiv& {1 \over 4 N_c}
\int_0^\infty dt_1\int_0^{t_1}dt_2\int_0^{t_2}dt_3 \,(t_2-t_3)^3 
\bigg\{
\langle \{g{\bf E}(t_1)\cdot, g{\bf E}(t_2)\}\, \{g{\bf E}(t_3)\cdot, g{\bf 
E}(0)\}\rangle_c
\nn
\\
&&  
~~~~~~~~~~~~~~~~~~~~~~~~~
- {4 \over N_c}\langle {\rm Tr}(g{\bf E}(t_1)\cdot g{\bf E}(t_2))\, {\rm
Tr}(g{\bf E}(t_3)\cdot g{\bf E}(0))\rangle_c
\bigg\} ,
\label{E32}
\eea
where
\bea
\langle \{g{\bf E}(t_1)\cdot, g{\bf E}(t_2)\}\, \{g{\bf E}(t_3)\cdot g{\bf
E}(0)\}\rangle_c &\equiv&
\langle \{g{\bf E}(t_1)\cdot g{\bf E}(t_2)\}\, \{g{\bf E}(t_3)\cdot g{\bf
E}(0)\}\rangle
\nn
\\
&&-{1 \over N_c} \langle g{\bf E}(t_1)\cdot g{\bf E}(t_2)\rangle\langle
g{\bf E}(t_3)\cdot g{\bf E}(0)\rangle, 
\eea
and similarly for the other structures with four chromoelectric fields that
appear in Eqs. (\ref{E32norm}) and (\ref{E32}).

For further use, we also define:
\bea
{\cal E}^{(2,t)}_3&=&{\cal E}^{(2)}_3+{\cal E}^{(2,{\rm norm.})}_3,\\
{\cal E}^{(2,{\rm EM})}_3&=&{\cal E}^{(2,c)}_3+{\cal E}^{(2,{\rm norm.})}_3.
\eea

\section{pNRQCD}
\label{pNRQCD}

\subsection{Matching to pNRQCD}
\label{sec3g}
Expressions (\ref{En3}) and alike are no more than formal expansions
in $H_I$, i.e. in $1/m$, until some dynamical assumption is made.  We
will assume a mass gap of order $\lQ \gg mv^2$ between the lowest-lying 
excitation and the higher ones.  Under this assumption all the
excitations ($\underbar{n}\not= {\underline 0}$) decouple from the
ground state ($\underbar{n}= {\underline 0}$), which is identified
as the only degree of freedom of pNRQCD.  It corresponds to the
singlet state $\rm S$ in the pNRQCD Lagrangian (\ref{lpnrqcd}).
Moreover, the above expansion acquires a dynamical meaning, becoming an
expansion in $\lQ/m$ and $v$ in the effective field theory.

The above assumption is the same as was made in Refs. \cite{M1,M2} in the
situation without massless fermions. In this work, we are including
light fermions. Nevertheless, at least in this paper, we will assume
that this does not change the structure of the leading order solution
(this was also assumed in Ref. \cite{pw}).  In other words, we will
assume that the size of the typical splittings between the ground
state (heavy quarkonium) and the gluonic excitations (hybrids) is much
larger than the typical splittings produced by the solutions of the
Schr\"odinger equation for the heavy quarkonium. This is, indeed,
supported by lattice simulations where the plots of the static
potentials for the heavy quarkonium and hybrids show the same pattern
after the inclusion of light fermions \cite{lattice}. Nevertheless, in
principle, a new problem may arise. Once light fermions have been
incorporated into the spectrum, new gauge-invariant states appear
besides the heavy quarkonium, hybrids and glueballs. On the one hand,
we have the states with no heavy quark content. Due to chiral
symmetry, there is a mass gap, of ${\cal O}(\Lambda_{\chi})$, between
the Goldstone bosons, which are massless in the chiral limit, and the
rest of the spectrum. We will consider that the Goldstone bosons are
ultrasoft degrees of freedom and that $\Lambda_{\chi} \sim \lQ$, so
that the rest of the spectrum should be integrated out. Besides these,
we also have bound states made of one heavy quark and light quarks. In
practice, we are considering the $Q \bar q$--$\bar Q q $ system. The
energy of this system is, according to the HQET counting rules
\cite{Neubert}: 
\be 
m_{Q \bar q}+m_{\bar Q q}=2m+2\,{\bar \Lambda} .
\ee 
Therefore, since ${\bar \Lambda} \sim \lQ$, we will assume that
they also have to be integrated out.  Problems may appear if we try to
study the heavy quarkonium near threshold.  In this case there is no
mass gap between the heavy quarkonium and the creation of a $Q
\bar q$--$\bar Q q $ pair. Thus, if we want to study the heavy quarkonium
near threshold, we should include these degrees of freedom in the
spectrum (for a model-dependent approach to this situation see, for
instance, \cite{CHARMONIUM}). We will not do so in this paper. It may
happen, however, that the mixing between the heavy quarkonium and the
$Q \bar q$--$\bar Q q $ is small. Indeed, such a mixing is suppressed
in the large $N_c$ counting.

Summarizing, light fermions contribute within this picture in three ways:

\noindent
1) hard light fermions, they are encoded into the matching coefficients of 
the NRQCD Lagrangian and obtained from the computation of perturbative
Feynman diagrams at the scale $m$;

\noindent
2) soft light fermions, a term that denotes, in a generic way, all the
fermions that are incorporated in the potentials; it is expected that
their main effects can be simulated by a variation of the value of the
parameters in the potentials;

\noindent
3) ultrasoft light fermions, these are the ones that will become pions
and, since they are also ultrasoft degrees of freedom, they should be
incorporated in the effective Lagrangian together with the heavy
quarkonium. However, we will not consider them in the present paper,
even if we do not expect to find conceptual problems in an eventual
incorporation.

\medskip

In conclusion, the matching condition to pNRQCD for the real part reads
\be
{\rm Re}\,E_0={\rm Re}\,h=-{\bfnabla^2 \over m}+V^{(0)}+{V^{(1)} \over m}
+{ V^{(2)} \over m^2}+\cdots \,.
\ee
At ${\cal O}(1/m)$ the matching has been performed in Ref. \cite{M1} and 
at ${\cal O}(1/m^2)$ in Ref. \cite{M2} (for the case without 
light fermions). We refer to those articles for further details 
about the structure of the potentials. For the imaginary piece,  
we have the analogous matching condition:
\be
{\rm Im}\, E_0={\rm Im}\, h \,.
\ee
Using the results of the previous sections, we can now write 
the first two terms in the $1/m$ expansion of ${\rm Im}\, h$
(the $P$-wave-dependent terms were obtained in Ref. \cite{pw}):
\be
\label{imh}
{\rm Im} \, h= 
{{\rm Im} \, h^{(2)} \over m^2} +{{\rm Im} \, h^{(4)} \over m^4}  + \cdots ,
\ee
where
\bea
\label{imh2}
{\rm Im} \, h^{(2)}&=&-{C_A \over 2}\delta^{(3)}({\bf r})
\Bigg(
4\, {\rm Im} \, f_1(^1 S_0)
-2\,{\bf S}^2\left({\rm Im}\, f_1(^1 S_0)-{\rm Im}\, f_1(^3 S_1)\right)
\\
\nn
&&
\qquad\qquad
+ 4\,{\rm Im}\, f_{\rm EM}(^1 S_0)
-2\,{\bf S}^2\left({\rm Im} \, f_{\rm EM}(^1 S_0)-{\rm Im}\, f_{\rm EM}(^3 S_1)\right)
\Bigg),
\eea
\bea
\label{imh4}
&&
{\rm Im} \, h^{(4)}
=
C_A \, {\cal T}^{ij}_{SJ} \bfnabla^i\delta^{(3)}({\bf r})\bfnabla^j \, 
\left({\rm Im}\,f_1({}^{2S+1}P_J)+{\rm Im}\,f_{\rm
    EM}({}^{2S+1}P_J)\right) 
\\
\nn
&&\qquad\qquad
+{C_A\over 2}\,
\Omega^{ij}_{SJ}\bigg\{ \bfnabla^i\bfnabla^j 
+ {\delta_{ij}  \over 3}{\cal E}_1, \delta^{(3)}({\bf r}) \bigg\} \,
\left({\rm Im}\,g_1({}^{2S+1}S_J)
+{\rm Im}\,g_{\rm EM}({}^{2S+1}S_J)\right)
\\
\nn
&&\qquad\qquad
+{T_F\over 3}\, {\cal T}^{ii}_{SJ}{\cal E}_1\delta^{(3)}({\bf r})\,
{\rm Im}\,f_8({}^{2S+1}P_J)
\\
\nn
&&\qquad\qquad
+{T_F \over 9}{\cal E}_3
{\bfnabla}\delta^{(3)}({\bf r}){\bfnabla}
\Bigg(
4\,{\rm Im} \,f_8(^1 S_0)
-2\,{\bf S}^2 \left({\rm Im} \,f_8(^1 S_0)-{\rm Im}\,f_8(^3 S_1)\right) 
\Bigg)
\\
\nn
&&\qquad\qquad
+2\,T_F\,c_F^2\,{\cal B}_1\,\delta^{(3)}({\bf r})
\Bigg(
{\rm Im} \,f_8(^3 S_1)+
{1 \over 6}{\bf S}^2
({\rm Im} \,f_8(^1 S_0)-3\,{\rm Im} \,f_8(^3 S_1))
\Bigg)
\nn
\\
\nn
&&\qquad\qquad
+{T_F \over 3}{\cal E}_3^{(2)}\delta^{(3)}({\bf r})
\Bigg(
4\,{\rm Im} \, f_8(^1 S_0)-2\,{\bf S}^2
\left({\rm Im} \,f_8(^1 S_0)-{\rm Im} \,f_8(^3 S_1)\right)
\Bigg)
\\
\nn
&&\qquad\qquad
-{C_A \over 3}{\cal E}_3^{(2,t)}\delta^{(3)}({\bf r})
\Bigg(
4\,{\rm Im} f_1(^1 S_0)-2\,{\bf S}^2\left({\rm Im} f_1(^1 S_0)-{\rm Im} f_1(^3 S_1)\right)
\Bigg)
\\
\nn
&&\qquad\qquad
- C_A{2 \over 9}{\cal E}_3
\left\{ \bfnabla^2, \delta^{(3)}({\bf r})\right\}
\Bigg(
{\rm Im} \,f_1(^1 S_0)+{\rm Im} \,f_{\rm EM}(^1 S_0)
\\
\nn
&&\qquad\qquad\qquad\qquad
+{{\bf S}^2 \over 2}
\left( {\rm Im} \,f_1(^3 S_1)-{\rm Im} \,f_1(^1 S_0) 
+{\rm Im} \,f_{\rm EM}(^3 S_1)-{\rm Im} \,f_{\rm EM}(^1 S_0)\right)
\Bigg) 
\nn
\\
\nn
&&\qquad\qquad
-2\,C_A\,c_F^2\,{\cal B}_1\,\delta^{(3)}({\bf r})
\Bigg(
{\rm Im} \,f_1(^1 S_0)+{\rm Im} \,f_{\rm EM}(^1 S_0)
\\
\nn
&&\qquad\qquad\qquad\qquad
+{{\bf S}^2 \over 6}\left({\rm Im} \,f_1(^3 S_1)-3\,{\rm Im} \,f_1(^1 S_0)
+{\rm Im} \,f_{\rm EM}(^3 S_1)-3{\rm Im} \,f_{\rm EM}(^1 S_0)\right)
\Bigg)
\\
\nn
&&\qquad\qquad
-{C_A \over 3}{\cal E}_3^{(2,{\rm EM})}\delta^{(3)}({\bf r})
\Bigg(
4\,{\rm Im}\, f_{\rm EM}(^1 S_0)-2\,{\bf S}^2\left({\rm Im}\, f_{\rm EM}(^1 S_0)-
{\rm Im} \,f_{\rm EM}(^3 S_1)\right)
\Bigg).
\eea

The above expressions have been given in 4 dimensions. Therefore, they should
be generalized to $d$ dimensions if we want to work in an $\MS$-like scheme in
order to use the same scheme as for the NRQCD matching coefficients
computation. This becomes relevant when logarithmic ultraviolet divergences 
appear in the non-perturbative constants.  
Hence, eventual lattice calculations must be converted to $\MS$ in this case. 
Nevertheless, in several situations, it is not necessary to work in an $\MS$
scheme if we only want to obtain the non-perturbative objects from experiment, since
the scheme dependence simply goes into a redefinition of the non-perturbative 
constants. Finally, note also that in addition to the divergences in the 
non-perturbative constants, which are due to large momentum transfers $k$, 
at some point there will also be ultraviolet divergences 
arising in quantum-mechanical perturbation theory, which are due to large 
relative momenta $p$. These must also be regulated in dimensional
regularization and $\MS$-subtracted, along the lines worked out in Ref. \cite{CMY}.

\subsection{Power counting in pNRQCD}
\label{sec3h}
With the above results, we are in a position to compute the inclusive 
decays of heavy quarkonium into light particles by using Eq. (\ref{imag}). 
Before doing so, we have to specify some power-counting 
rules in order to estimate the importance of the different
terms of the pNRQCD Hamiltonian. Previous discussions on this subject, 
some of which we will repeat here, can be found in Refs. \cite{M1,M2}. 

With the results of Sec. \ref{sec3g} and using Eq. (\ref{imag}),  
the decay width of $S$-wave quarkonium has schematically the following structure: 
\bea
\Gamma &\sim& 
{\rm Im}\,c_{4-f}^{d=6}{|R_{ns0s}(0)|^2 \over m^2}
\left(1+{\lQ^2 \over m^2}+\cdots \right)
\nn
\\
&+&
{\rm Im}\,c_{4-f}^{d=8}\left({R_{ns0s}(0)(\bfnabla^2 R_{ns0s}(0))\over m^4}
+{|R_{ns0s}(0)|^2 \over m^2}{\lQ^2 \over m^2}+\cdots \right)
+ \cdots \,,
\label{gammas}
\eea
where $c_{4-f}$ stands for the NRQCD 4-fermion matching
coefficients and $R_{ns0s}$ is the $S$-wave radial component of the solution of
the real piece of the Schr\"odinger equation:
\be
\label{Schr}
({\rm Re}\, h)\,\phi_{njls}({\bf r})=E_{njls}\,\phi_{njls}({\bf r}),
\ee 
with the normalization ($|s\rangle_{\rm spin}$ denotes the normalized spin component): 
\be
\phi_{ns0s}({\bf r})=R_{ns0s}(r){1 \over \sqrt{4\pi}}|s\rangle_{\rm spin}.
\ee
Although $E_{njls}$ coincides with the binding energy of the system
at the order we are working at, it will no longer be so when
iterations of imaginary parts start playing a role.

\medskip

From Eq. (\ref{gammas}), we can see how the power counting has to be
organized. On the one hand, we have an explicit expansion in $\lQ/m$,
independent of the details of the bound state. In the most
conservative situation ($\lQ \sim mv$), it would correspond to having
the power counting $\lQ/m \sim v$. We can also find derivatives of the
wave function divided by $m$. They typically scale like $\bfnabla/m
\sim v$. On the other hand, the normalization condition of the
wave function sets the scaling $|R_{njls}|^2 \sim (mv)^3$. This means
that a formal ${\cal O}(mv^5)$ accuracy (leaving aside possible $\als(m)$
suppressions due to the NRQCD matching coefficients) is achieved with
Eq. (\ref{gammas}).  At the same order of accuracy, the decay width of
$P$-wave quarkonium has the structure: 
\be
\label{simPwave}
\Gamma \sim
{\rm Im}\,c_{4-f}^{d=8}{|\bfnabla R_{nj1s}(0)|^2 \over m^4} + \cdots \,.
\ee

In the above discussion, we have only considered the leading order
power counting of the wave function at the origin $\sim (mv)^3$. This
accuracy is sufficient for the $P$-wave function of Eq. (\ref{simPwave}), 
as well as for the wave functions multiplying $\lQ^2/m^2$ terms or
with two $\bfnabla$ in Eq. (\ref{gammas}) but not for the leading
order term.  In this case, one has to take into account that the wave
function at the origin also has subleading contributions in $v$:
$|R_{njls}(0)|^2 \sim (mv)^3(1+av+bv^2+\cdots)$. Therefore, we have to
further specify the solution of Eq. (\ref{Schr}), for which we have to
set the power counting of the potentials in the Schr\"odinger
equation. Since we do not know the specific dynamics of the different
potentials, the only thing we can do is to require consistency of the
theory and allow, in principle, the most conservative counting.  This
would correspond to setting the counting by the largest scale that has
been integrated out, i.e. the potentials would scale like $(mv)^d$,
$d$ being their dimension.\footnote{Notice that our power counting
rules are different from those of \cite{nrqcd,pcnrqcd}. Whereas ours
are meant to apply in the situation $\lQ \gg mv^2$, the power counting
rules in Refs. \cite{nrqcd,pcnrqcd} rather follow the counting $\lQ
\sim mv^2$. Indeed, if we take $\lQ \sim mv^2$ in our results we
obtain a similar power counting for the NRQCD matrix elements.} For
definiteness, we will also assume $\als(m) \sim v^q$ with $q > 1$.

\medskip

{\bf Leading order.} Consistency of the theory requires the virial theorem 
to be fulfilled. In other words, the potential at leading order needs to fulfil
\be
h^{(0)}\phi^{(0)}_{njls}({\bf r})
=
\left({{\bf p}^2 \over m}+V_{\rm LO}\right)\phi^{(0)}_{njls}({\bf r})
=
E^{(0)}_{njls}\,\phi^{(0)}_{njls}({\bf r}),
\ee
with the power counting
\be
{\bf p}^2/m \sim V_{\rm LO} \sim E^{(0)}_{njls} \sim mv^2.
\ee
It follows that $V^{(0)} \sim mv^2$ (even if, using the most conservative power counting, 
we would have obtained $V^{(0)} \sim mv$). Moreover, in our power counting 
we have $V^{(1)}/m \sim mv^2$.\footnote{As a consequence, if the potential  
$V^{(1)}$ is non-perturbative, we have no general argument to 
consider $V^{(1)}/m$ subleading with respect to $V^{(0)}$. 
A lattice simulation or some model-dependent studies are, therefore, 
highly desirable to discern the issue. Whereas it is difficult to obtain this
information from the spectrum structure, the study of the decays may perhaps 
shed some light on this problem. Finally, we note that, in the perturbative situation, 
$V^{(1)}$ has an extra $\als^2$ suppression. Further discussions can be found 
in Ref. \cite{M1}.}
Therefore, in the most conservative situation, we would have
\be
V_{\rm LO}=V^{(0)}+{V^{(1)} \over m}.
\ee
The important point here is that, at this order, the potential is
spin-independent ($E_{njls}^{(0)} \equiv E_{nl}^{(0)}$ 
and $R^{(0)}_{njls} \equiv R^{(0)}_{nl}$).
Therefore, the leading-order $P$-wave function reads
\be
\phi^{(0)}_{n1s}({\bf r})=R^{(0)}_{n1}(r) \;
\langle {\bf \hat{r}}|js\rangle,
\ee
where $|{\bf \hat r}\rangle$ is the normalized eigenstate of the position 
and $|js\rangle$ stay for $J$ (total angular
momentum) and $S$ eigenstates such that
\be
\langle {\bf \hat{r}}|j0\rangle=Y_j^m({\bf \hat{r}})|0\rangle_{\rm spin} \quad (j=l=1)\,,
\qquad 
\langle {\bf \hat{r}}|j1\rangle={\cal Y}^1_{jm}({\bf \hat{r}})
\,,
\ee 
where $m$ denotes the third component of the angular momentum and 
detailed expressions of ${\cal Y}^1_{jm}({\bf \hat{r}})$ can be found 
in Ref. \cite{GP}, Appendix B.

\medskip

{\bf Next-to-leading order.} The ${\cal O}(1/m^2)$ potential scales utmost as $V^{(2)}/m^2 
\sim mv^3$. Therefore, in the most conservative situation, we would have
\be
V_{\rm NLO}={V^{(2)} \over m^2} .
\ee
At this order, spin-dependent contributions start to appear.
In particular, the spin-dependent potential contributing to the $S$-wave function 
at the origin reads
\be
\delta V= {{\bf S}_1\cdot{\bf S}_2 \over m^2}\,{\rm Re}\,V_{S^2}^{(1,1)}(r),
\label{vss}
\ee
where \cite{M2}
\be
{\rm Re}\,V_{S^2}^{(1,1)}(r)={2 \,c_F^{2} \over 3}\, i \, \int_0^{\infty} dt \,  
\lla g{\bf B}_1(t) \cdot g{\bf B}_2 (0) \rra
+2\,C_A\left({\rm Re}f_1(^1 S_0)-{\rm Re}f_1(^3 S_1) \right) \,
\delta^{(3)}({\bf r}).
\ee
This potential produces the following correction to the $S$-wave function:
\be
{R_{ns0s}({0})\over \sqrt{4\pi}}
={R_{n0}^{(0)}({0})\over \sqrt{4\pi}}
+{1 \over 2m^2}\left(s(s+1)-{3 \over 2}\right)
\langle {\bf r}={\bf 0}|{1 \over E_{n0}^{(0)}-h^{(0)}}\,P_n
\,{\rm Re}\,V_{S^2}^{(1,1)}|n0\rangle,
\label{rss}
\ee
where $P_n\equiv I-|n0\rangle\langle n0|$ and 
$\langle {\bf r}|njls\rangle = \phi^{(0)}_{njls}({\bf r})$ 
($\langle {\bf r}|n0\rangle = \phi^{(0)}_{n0}({\bf r})$).

If the spin-dependent potential (\ref{vss}) is ${\cal O}(mv^3)$, 
it just provides the leading order spin-dependent correction to the $S$-wave function
at the origin and one can use the difference between vector and pseudoscalar
decays to fix the value of the correction. 
If the spin-dependent potential is ${\cal O}(mv^4)$, it provides a correction 
to the $S$-wave function squared at the origin, which is of the same order as the 
${\cal O}(v^2)$ corrections to the decay width that we have already evaluated.
Therefore, in this last situation, Eq. (\ref{rss}) would account for the full difference 
between the vector and pseudoscalar wave functions at the origin at 
relative order ${\cal O}(v^2)$, which is the precision we are aiming at in
this work. This last counting seems to be supported by the size of 
the spin-dependent splittings in the bottomonium and charmonium spectra.

For the spin-independent contributions, we will make no assumption at this 
or higher orders, as their effects will be encoded into the wave functions,  
which will be left unevaluated. Our results
allow for the most conservative counting where $V^{(1)}/m \sim mv^2$ and
$V^{(2)}({\hbox{spin-independent}})/m^2 \sim mv^3$.
We note that, in this power counting, potentials with imaginary part arise
in the pNRQCD Hamiltonian at order $m \als(m)^2v^3$  (where the powers
in $\als(m)$ come from the imaginary part of the 4-fermion matching 
coefficients in NRQCD). Therefore, corrections due
to the iteration of imaginary terms, which could affect the 
validity of Eq. (\ref{imag}), are far beyond the accuracy of this paper.
In fact, the general factorization formula put forward in \cite{nrqcd} may 
not hold beyond a certain order.

In any case, we do not rule out that a different power counting may
also lead to consistent equations in the non-perturbative regime for
some specific ratios of $\lQ$ versus $m$ and versus $p$ and $k$. This point
deserves further investigation and may lead to a different
implementation of the matching procedure. We recall that the issue 
of assessing the power counting in the non-perturbative
situation has been addressed before by Beneke
\cite{Beneke} and by Fleming et al. \cite{FRL}. In both cases, the authors 
have given some freedom to the possible size of the NRQCD matrix
elements by introducing a parameter $\lambda$ that interpolates
between the power counting in the perturbative limit and other
possible power countings according to the value of $\lambda$. In this 
respect, our formalism may shed more light to clarify this problem, 
since it incorporates the factorization between
the soft and the ultrasoft scales, allowing us to write the
NRQCD matrix elements in terms of the wave function at the origin and of some 
bound-state-independent constants. Another point of concern is whether there 
are non-perturbative effects that are not accounted for in the $1/m$ matching. 

\medskip
We conclude this section by giving a useful equality, 
valid in dimensional regularization:
\be
{R_{nl}^{(0)}(0)(\bfnabla^2 R_{nl}^{(0)}(0))\over m^4}=-{|R_{nl}^{(0)}(0)|^2 \over m^2}
{E_{nl}^{(0)} \over m}, 
\label{eqgk}
\ee 
which follows from the fact that we know the behaviour of the potential 
and the wave function (up to a constant) at short distances and that 
(see Appendix \ref{Appreg})
\be
\langle n,j,l,s|V^{(0)}|{\bf r}={\bf 0} \rangle = 
\langle n,j,l,s|V^{(1)}|{\bf r}={\bf 0}\rangle  =0
\qquad \hbox{(in dim. regularization)}.
\ee

With this we have discussed the relative importance of the different
terms that will appear in our evaluation of the decay widths. The
results can be found in Sec. \ref{results}.

\section{The matching in the case $mv \gg \lQ \gg mv^2$}
\label{twostepmatching}
Although it is not clear whether quarkonia states fulfilling $mv \gg \lQ \gg
mv^2$ exist in nature ($mv \sim k \sim p$ and $mv^2 \sim E$ will always be 
understood in the present section), this situation is worth investigating for
several reasons. First of all, the calculation in the general case of the 
Sec. \ref{qmmatching} is non-standard and, hence, any independent check of it, even
if it is in a particular case, is welcome. Secondly, the calculation in this
case can be divided into two steps. The first step can be carried out by a perturbative calculation 
in $\als$, which involves far more familiar techniques. The second step, even
if it is non-perturbative in $\als$, admits a diagrammatic representation,
which makes the calculation somewhat more intuitive. Third, the more detailed information on the 
potential allows us to make important tests on how the terms in the potential 
can be consistently reshuffled by means of unitary transformations \cite{M1}, 
as is illustrated in the example provided in Appendix \ref{Appunit}.

\subsection{pNRQCD$^\prime$}
As mentioned in the Introduction, we shall call pNRQCD$^\prime$ the EFT for energies below $mv$.
Since $mv \gg \lQ$, the integration of the energy scale $mv$, namely the 
matching between NRQCD and pNRQCD$^\prime$ can be carried out perturbatively
in $\als$. This is done following 
Refs. \cite{Mont,long}. A tree-level matching is sufficient, but higher orders
in the multipole expansion will be needed. 
We only display below the terms eventually required in the calculation:
\bea
{\mathcal L}_{\rm pNRQCD'} &=& 
{\rm Tr} \,\Big\{ {\rm S}^\dagger \left( i\partial_0 - h_s  \right) {\rm S} 
+ {\rm O}^\dagger \left( iD_0 - h_o   \right) {\rm O} \Big\} 
\nonumber \\ 
& & + {\rm Tr} \left\{\!   
{\rm O}^\dagger {\bf r} \cdot g{\bf E}\,{\rm S} 
+ \hbox{H.c.} + 
{{\rm O}^\dagger {\bf r} \cdot g{\bf E} \, {\rm O} \over 2} +
{{\rm O}^\dagger {\rm O} {\bf r} \cdot g{\bf E} \over 2} \!\right\} 
\nn \\
& & +  {1 \over 8} {\rm Tr} \left\{\! {\rm O}^\dagger {\bf r}^i {\bf r}^j \,  
g {\bf D}^i {\bf E}^j \, {\rm O} - {\rm O}^\dagger {\rm O} {\bf r}^i {\bf r}^j
\, g {\bf D}^i {\bf E}^j \!\right\} 
\nn \\
& & +  {1 \over 24} {\rm Tr} \left\{\! {\rm O}^\dagger {\bf r}^i {\bf r}^j
  {\bf r}^k \, g {\bf 
    D}^i{\bf D}^j {\bf E}^k \,{\rm S} 
+ \hbox{H.c.} \!\right\} 
\nn \\
& &+ {c_F \over 2m} {\rm Tr} \left \{ {\rm O}^\dagger (\bfsigma_1 - \bfsigma_2) \cdot g{\bf B}\,{\rm S} 
+ \hbox{H.c.} \right \} - {1\over 4} G_{\mu \nu}^{a} G^{\mu \nu \, a}\, ,
\label{pnrqcd0}
\eea
where the traces are over colour space only. S and O are chosen here to
transform as a $1/2\otimes 1/2$ representation in spin space 
(hence ${\bfsigma}_1 - {\bfsigma}_2 = {\bfsigma}_1\otimes \one_2 - \one_1
\otimes{\bfsigma}_2$); $h_s$ and $h_o$ read as follows 
(again we only display terms eventually required in the calculation):
\bea 
h_s &=&  -{{\bf \nabla}^2  \over m}-C_f {\als \over r}
-i \, {C_A \over  2}{\delta^{(3)}({\bf r})\over m^2}\Bigg ( 4\, {\rm Im} \, f_1(^1 S_0)
-2\,{\bf S}^2\left({\rm Im}\, f_1(^1 S_0)-{\rm Im}\, f_1(^3 S_1) \right) 
\nn \\
&& + 4\,{\rm Im} \, f_{\rm EM}(^1 S_0) -2\,{\bf S}^2\left({\rm Im} \, f_{\rm
    EM}(^1 S_0)-{\rm Im}\, f_{\rm EM}(^3 S_1)\right) \Bigg) 
\nn \\
&& + i \, {C_A\over 2}\,
{\Omega^{ij}_{SJ}\over m^4}\bigg\{ \bfnabla^i\bfnabla^j, \delta^{(3)}({\bf r}) \bigg\} \,
\left({\rm Im}\,g_1({}^{2S+1}S_J)
+{\rm Im}\,g_{\rm EM}({}^{2S+1}S_J)\right)\, , \\
\label{octp}
h_o & = &  -{{\bf \nabla}^2  \over m} 
+  \left ( {C_A \over 2} - C_f \right ){\als \over r} 
\nn\\
&&
- i\, {T_F \over 2}{\delta^{(3)}({\bf r})\over m^2}\Bigg ( 4\, {\rm
  Im} \, f_8(^1 S_0) -2\,{\bf S}^2\left({\rm Im}\, f_8(^1 S_0)-{\rm Im}\,
  f_8(^3 S_1) \right) \Bigg) 
\nn \\
&& + i \, T_F \, {\cal T}^{ij}_{SJ} \bfnabla^i{\delta^{(3)}({\bf r})\over m^4}\bfnabla^j \, 
{\rm Im}\,f_8({}^{2S+1}P_J) \, . 
\eea

The Feynman rules associated to this Lagrangian are displayed in Fig. 1.

\subsection{Matching pNRQCD to pNRQCD$^\prime$} 
The matching of pNRQCD$^\prime$ to pNRQCD can no longer be done perturbatively in $\als$, 
but it can indeed be done perturbatively in the following ratios of scales:
$\lQ/mv$ (multipole expansion), $\lQ/m$ and $mv^2/ \lQ$. The diagrams contributing 
to the calculation are displayed in Figs. 2 to 9.
 
\medskip

We have focused on contributions to $S$-wave states involving imaginary parts. 
Since the imaginary parts, which are inherited from NRQCD, sit on local 
($\delta^{(3)} ({\bf r})$, $\bfnabla \delta^{(3)} ({\bf r}) \bfnabla$, 
etc.) terms in the pNRQCD$^\prime$ Lagrangian, they tend 
to cancel when multiplied by the ${\bf r}$'{\small s} arising from the multipole expansion.
 Hence, for an imaginary part to contribute, it is necessary to have a sufficient number of derivatives 
(usually arising from the $mv^2/ \lQ$ expansion) as to kill all the ${\bf
  r}$'s. Since derivatives are always 
accompanied by powers of $1/m$, it implies that at a given order of $1/m$,
only a finite number of terms in the multipole expansion contribute. 
In our case a fourth order in the multipole expansion is sufficient. 
The natural way to organize the calculation in our case would be to assign a size $mv^{p}$ 
to $\lQ$, $1<p<2$ and $v^{q}$ to $\als$, $1<q<2$, 
and to carry out the calculation at the desired order in $v$. However, our
main goal here is not the phenomenological relevance 
of the situation $mv \gg \lQ \gg mv^2$, but providing an independent
calculation to support the results of Sec. \ref{sec3g}. 
Hence, irrespectively of what $p$ and $q$ may be, we will only be 
interested in fishing up the imaginary pieces that  
contribute to $S$-wave states up to order $1/m^4$.

The two diagrams in Fig. 2 correspond to the leading contribution in the $\lQ/mv$ and $ \lQ/m$ 
expansion, respectively. Figure 3 displays the evaluation of each of them in the
$mv^2/ \lQ$ expansion. The diagrams in Fig. 4 correspond to the
next-to-leading order contributions in the $ \lQ/mv$ expansion, and Figs. 5--9 display 
their evaluation in the $mv^2/ \lQ$ expansion. 
It is then clear that the basic skeleton of the calculation consists of the 
$x= (\lQ/mv)^2$ and $y= (\lQ/m)^2$ expansions, which suggests writing the pNRQCD Hamiltonian as: 
\be
h = h_s+  h_x + h_{2x} + h_{y}+... \, .
\ee
The interpolating fields of pNRQCD$^\prime$ and pNRQCD will be related by:
\bea
S\vert_{\rm pNRQCD'}&=&Z^{1\over 2} S\vert_{\rm pNRQCD}=\left( 1+ Z_x +Z_{2x}
  + Z_y + ... \, \right)^{{1 \over 2}} S\vert_{\rm pNRQCD} 
\nonumber \\
&=& \left ( 1+ {1 \over 2} \left (Z_x +Z_{2x} + Z_y - {1 \over 4} Z_x^2
  \right ) + ... \, \right ) S\vert_{\rm pNRQCD}\, .
\eea
The matching calculation reads:
\bea
 & &
\int^{\infty}_{-\infty} dt \, e^{-iEt}\int d^3 {\bf R} \, \langle {\rm vac}\vert
T\{S({\bf R}, {\bf x}, t) S({\bf 0}, {\bf x}^{\prime}, 0)\}\vert {\rm vac}
\rangle\vert_{\rm pNRQCD'} 
\cr & &
= \int^{\infty}_{-\infty} dt \,  e^{-iEt} \int d^3 {\bf R} \, 
Z^{1\over 2}\langle {\rm vac}\vert T\{S({\bf R}, {\bf x}, t) S({\bf 0}, 
{\bf x}^{\prime}, 0)\}\vert {\rm vac} \rangle
\vert_{\rm pNRQCD}{Z^{1\over 2}}^{\dagger} 
\, .
\label{mtcal}
\eea
The right-hand side of the matching calculation has the following structure 
(up to a global $i$ factor, which is dropped):
\bea 
& &
{1\over E-h_s}+{1\over E-h_s}( h_x + h_{2x} +h_{y} ) {1\over E-h_s}
+ {1 \over 2} \left( Z_x +Z_{2x} +Z_y - { Z_x^2 \over 4} \right)
{1\over E-h_s}
\cr 
&+& {1\over E-h_s}{1 \over 2} \left( Z_x +Z_{2x} +Z_y 
- {Z_x^2 \over 4} \right)^{\dagger}+
\left( {Z_x \over 2} \right) {1\over E-h_s}\left( {Z_x \over 2}
\right)^{\dagger} 
\cr 
&+& {1\over E-h_s}h_x{1\over E-h_s}h_x{1\over E-h_s}+
\left( {Z_x \over 2} \right){1\over E-h_s}h_x{1\over E-h_s}+{1\over
  E-h_s}h_x{1\over E-h_s}\left( {Z_x \over 2} \right)^{\dagger}  .
\label{wfpt}
\eea
Hence, once we have made sure that, up to contact terms, the left-hand side of
Eq. (\ref{mtcal}) has exactly this structure, we can easily identify the 
contributions to the pNRQCD Hamiltonian from the second term of the expression 
(\ref{wfpt}).

\subsection{Calculation}
Let us then proceed to the calculation of the left-hand side of Eq. (\ref{mtcal}) 
(in order to match Eq. (\ref{wfpt}) a global $i$ factor will also be dropped).

Diagram (a) of Fig. 2 gives:
\bea
{1 \over E-h_s} \, {i \over N_c}  \int_0^{\infty} dt \, \langle i {\bf r}
\cdot g{\bf E} (t) \, e^{-i(h_o-E)t} \, i {\bf r} \cdot g{\bf E} (0) \rangle \, {1\over
  E-h_s} \, .
\eea
The fact that $mv^2/\lQ$ is small is implemented by expanding the
exponential. This guarantees that we will eventually get usual, 
energy-independent, potentials.

\medskip

The first contributions arise at 
${\cal O}\left(mv^2/\lQ\right)$ from the ${\cal O}(1/m^4)$ 
$P$-wave (Fig. 3a) and $S$-wave (Fig. 3e) terms in the octet potential of (\ref{octp}): 
\bea
& & {\rm (a)} ~~ {i \over E-h_s} \, 
{T_F {\cal T}_{S J}^{ii}\,{\rm Im}\,f_8 (^{2S+1} P_J) \over 3 N_c m^4 }
\int_0^{\infty} dt \, t \, \langle g{\bf E}(t) \cdot g{\bf E} (0) \rangle \,
{\delta^{(3)} ({\bf r}) \over E-h_s} \, , \
\\
& & {\rm (e)} ~~ {i \over E-h_s} \, { {\cal T}_{S}\,{\rm Im}\,g_1 (^{2S+1} S_S) \over
  m^4 }  \int_0^{\infty} dt \, t \, \langle g{\bf E}(t) \cdot g{\bf E} (0)
\rangle \, {\delta^{(3)} ({\bf r}) \over E-h_s} \, ,
\eea
where ${\cal T}_{SJ}^{ij}$ are defined in Eqs. (\ref{defT1})--(\ref{defT2}) and 
${\cal T}_S$ in Eq. (\ref{defT}).

\medskip

At ${\cal O}(m^2v^4/\lQ^2)$ and higher, it is convenient to write $E-h_o=E-h_s + (V_o -V_s)$. 
Ill-defined expressions arise in the calculation, from products of
distributions (both products of two delta functions and products of 
delta functions with non-local potentials, which explode as ${\bf r}\rightarrow
0$). It is most convenient to use dimensional regularization in this case, 
which sets all these terms to zero. This is shown in Appendix \ref{Appreg}, 
where the relation with other regularizations is also discussed.
Having this in mind, it is clear that, at the order we are interested in, 
$\rm{Im}\, (V_o -V_s)^2 =0$ and ${\rm Im}\,(V_o -V_s){\bf r} =0$. 
Hence, we only have to consider:
\be
{1\over E-h_s}{\bf r} (E-h_s)^2 {\bf r}{1\over E-h_s}\label{m2e2} \, .
\ee
If we decide to take one power $(E-h_s)$ to the right and one to the left we have:
\be
{\bf r}^2 +{\bf r}[{\bf r}, h_s]{1\over E-h_s}+{1\over E-h_s}[h_s,{\bf r}]
{\bf r}+{1\over E-h_s}[h_s,{\bf r}][{\bf r}, h_s]{1\over E-h_s} \, , 
\label{m2e2sym}
\ee
which does not produce any imaginary part. However, an equally acceptable expression is:
\be
{\bf r}^2+{1\over 2}[{\bf r},[{\bf r}, h_s]]{1\over E-h_s}
+{1\over E-h_s}{1\over 2}[[h_s,{\bf r}], {\bf r}]
+{1\over E-h_s}{1\over 2}\{[[{\bf r}, h_s],h_s],{\bf r}\}{1\over E-h_s} \, ,
\label{m2e2asym}
\ee
which does produce an imaginary part. This apparent paradox only reflects 
the fact that expression (\ref{m2e2}) by itself (as well as some of the
expressions we will find below) does not determine uniquely its contribution 
to the potentials. 
This expression always leads to contact terms, wave-function normalization and
potential, as is apparent in (\ref{m2e2sym}) 
and (\ref{m2e2asym}), but depending on how we decide to organize the
calculation, the terms associated to each of these pieces 
change. For instance, when matched to (\ref{wfpt}), (\ref{m2e2sym}) gives:
\be
h_x= [h_s,{\bf r}][{\bf r}, h_s] \quad \quad Z_x={\bf r}[{\bf r}, h_s]\, , 
\ee
whereas (\ref{m2e2asym}) gives:
\be
h_x= {1\over 2}\{[[{\bf r}, h_s],h_s],{\bf r}\} \quad \quad Z_x
   = {1\over 2}[{\bf r},[{\bf r}, h_s]] \, .
\ee
This should not be a surprise. It has already been discussed in Ref. \cite{M1}
that this ambiguity exactly corresponds to the freedom of making unitary
transformations in a quantum-mechanical Hamiltonian, and does not affect
physical observables. This is discussed in detail in Appendix \ref{Appunit} 
for the decay widths of the $S$-wave states we are concerned with. 
In order to fix the contribution to the potential of any term once forever,  
we will use the following prescription. If we have an expression 
with singlet propagators $1/(E-h_s)$ only in the
external legs, and an even number of powers of $(E-h_s)$, 
we will take the one closest to the left propagator to the left and 
the one closest to the right propagator to the right, and repeat until no power
is left except in contact terms. Accordingly, in the intermediate steps, when
terms with a single external leg $1/(E-h_s)$ and several powers of $(E-h_s)$
are produced, one should take these powers towards the $1/(E-h_s)$ leg until
no power is left except in contact terms. If the number of powers of $(E-h_s)$
is odd, we use the same prescription until a single power is left.
We then write  $ (E-h_s)=(E-h_s)/2+(E-h_s)/2$ and take one half 
to the right and one half to the left. 
Expressions with an internal singlet propagator also appear, which require a
more careful treatment as will be discussed after (\ref{128}) below.
Note that this prescription to organize the calculation needs not coincide with the prescription 
for fixing the wave-function normalization in Sec. \ref{sec3g}.
Hence, we only expect to agree with the results of that section up to a unitary transformation. 
Anyway, with this prescription, (\ref{m2e2}) gives rise to the potential
obtained in (\ref{m2e2sym}) 
and hence to no imaginary part.

At ${\cal O}(m^3v^6/\lQ^3)$ only the following two terms in $(E-h_o)^3=(E-h_s)^3 +
(E-h_s)(V_s-V_o)(E-h_s)+ ... \, $ contribute, giving rise to:
\bea
&&{1 \over 2} \left ( (E-h_s) {\bf r}^2 + {\bf r}^2 (E-h_s) \right ) + {3
  \over 2} \left ( [h_s,{\bf r} ] {\bf r} + {\bf r} [ {\bf r},h_s] \right ) 
\nn\\
&& +{1 \over 2} \left ({1\over E-h_s}[h_s,[h_s,{\bf r} ]] {\bf r} + {\bf r} [ [
  {\bf r},h_s],h_s]{1 \over E-h_s} \right ) 
\nn \\
&&+ {1 \over 2} \left ([{\bf r} [{\bf r},h_s], h_s ] {1 \over E-h_s} + {1
    \over E-h_s} [h_s, [h_s,{\bf r}] {\bf r}] \right ) 
\nn\\
&& + [h_s, {\bf r} ] [{\bf r},h_s] {1 \over E-h_s} + {1 \over E-h_s} [h_s, {\bf r} ] [{\bf r},h_s] 
\nn \\
&&+ {1 \over 2} \left ( {1 \over E-h_s} [h_s,[h_s,{\bf r}]] [{\bf r},h_s] {1
    \over E-h_s}+ {1 \over E-h_s} [h_s, {\bf r} ]  [ [ {\bf r},h_s],h_s] {1
    \over E-h_s} \right )  
\nn \\
&& + {\bf r}\,  (V_s - V_o)\, {\bf r}
+ {1\over E-h_s}[ h_s,{\bf r} ] \, (V_s - V_o) \, {\bf r} 
\nn\\
&& + {\bf r} \,(V_s-V_o) \, [{\bf r}, h_s] {1\over E-h_s} + {1\over E-h_s} [ h_s,{\bf r} ] \,
(V_s - V_o) \, [{\bf r}, h_s] {1\over E-h_s} \, . 
\nn
\eea
It is the first term in the fifth line that renders the contribution depicted in Fig. 3d:
\bea
&& {\rm (d)} ~~ {-i \over E-h_s} \, {1 \over 9} { {\cal T}_{S}{\rm Im}f_1 (^{2S+1}
  S_S) \over m^2 }  \int_0^{\infty} dt \, t^3  \langle g{\bf E}(t) \cdot g{\bf E}
(0) \rangle \, \left \{ \delta^{(3)} ({\bf r}) \, , {\bfnabla^2 \over m^2}
\right \} {1 \over E-h_s}  \, .
\eea

At ${\cal O}(m^4v^8/\lQ^4)$ and higher, only imaginary parts beyond $1/m^4$ are produced.

Consider next the diagram Fig. 2b. Since the chromomagnetic moment already
provides two powers of $1/m$, only the linear term in the expansion 
of the exponential contributes (Figs. 3b and 3c). 
This gives:
\bea
& & {\rm (b)}~~ {i \over E-h_s} \, {T_F c_F^2\over N_c} { {\cal T}_{S}{\rm Im}f_8
  (^{3-2S} S_{1-S}) \over 3^S m^4 }  \int_0^{\infty} dt \, t  \langle g{\bf
  B}(t) \cdot g{\bf B} (0) \rangle \, {\delta^{(3)} ({\bf r}) \over E-h_s} \, , 
\nn \\
& & {\rm (c)}~~ {-i \over E-h_s} \, { c_F^2{\cal T}_{S}{\rm Im}f_1 (^{2S+1} S_S)
  \over 3^Sm^4 }  \int_0^{\infty} dt \, t  \langle g{\bf B}(t) \cdot g{\bf B} (0)
\rangle \, {\delta^{(3)} ({\bf r}) \over E-h_s} \, .
\eea

Consider next Fig. 4a. Because of the four ${\bf r}$'s in the expression, only the
following term in the expansion $(E-h_o)^3=(E-h_s)^3 + ... \,$ contributes. 
We obtain:
\be
{- i \over E-h_s} {{\cal T}_{S}{\rm Im}f_1 (^{2S+1} S_S) \over 72 \, m^4 }
(\delta_{ij} \delta_{kl} + \delta_{ik} \delta_{jl}+ \delta_{il}
\delta_{jk})\int_0^{\infty} dt \, t^3  \langle g {\bf E}^i (t)  [ {\bf D}^j ,
[{\bf D}^k , g{\bf E}^l (0)]]  \rangle  {\delta^{(3)} ({\bf r}) \over E-h_s} \, .
\ee
For the symmetric diagram, we have:
\be
{- i \over E-h_s} {{\cal T}_{S}{\rm Im}f_1 (^{2S+1} S_S) \over 72 \, m^4 }
(\delta_{ij} \delta_{kl} + \delta_{ik} \delta_{jl}+ \delta_{il}
\delta_{jk})\int_0^{\infty} dt \, t^3 \langle [{\bf D}^i, [{\bf D}^j, g{\bf
  E}^k (t)]] 
g{\bf E}^l (0)\rangle  {\delta^{(3)} ({\bf r}) \over E-h_s} \, .
\ee
In fact both contributions are the same, adding up to (see formula i) above 
Eq. (15) of Ref. \cite{M1}):
\be
{i \over E-h_s} {{\cal T}_{S}{\rm Im}f_1 (^{2S+1} S_S) \over 12 \, m^4 }
\int_0^{\infty} dt \, t^3 \langle [{\bf D},\cdot g{\bf E} (t)] [ {\bf D},
\cdot g{\bf E} (0)] \rangle  {\delta^{(3)} ({\bf r}) \over E-h_s} \, .
\ee

Consider next Fig. 4b. The only contributions come from $(E-h_o)^3=(E-h_s)^3
+... \, $ in one octet propagator and $1$ in the other. 
We obtain (Fig. 5b):
\bea
&& {1 \over E-h_s} \, {{\cal T}_{S}{\rm Im}f_1 (^{2S+1} S_S) \over 12 \, m^4
  }(\delta_{ij} \delta_{kl} + \delta_{ik} \delta_{jl}+ \delta_{il}
\delta_{jk}) \int_0^{\infty} dt_1 \, \int_0^{t_1} dt_2 \, 
\Big[ (t_1-t_2)^3+ t_2^3 \Big] 
\nn \\
&& \qquad\qquad\qquad\qquad\qquad
\times \langle g {\bf E}^i (t_1) [ {\bf D}^j , g{\bf E}^k ](t_2) g{\bf E}^l
(0) \rangle  {\delta^{(3)} ({\bf r}) \over E-h_s} \, .
\eea

Then consider Fig. 4c. From here we get several contributions. Because of the
four ${\bf r}$'s we need a total of three powers of $(E-h_o)$. When all the
powers come from the octet propagator in the middle, we get contributions from
$(E-h_o)^3=(E-h_s)^3 + (E-h_s)(V_s-V_o)(E-h_s)+\cdots$. 
The ones from the second term read (Fig. 6):
\bea
&& {i \over E-h_s} {{\cal T}_{S}\over 6 N_c m^4} 
\Big(T_F \, {\rm Im} \, f_8 (^{2S+1} S_S) - N_c \, {\rm Im} \, f_1 (^{2S+1} S_S)\Big) 
\nn\\
&&\qquad  
\times 
\int_0^\infty
dt_1\int_0^{t_1}dt_2\int_0^{t_2} dt_3 \, (t_2-t_3)^3 
\nn\\
&& \qquad 
\times \Bigl \{ \langle \{ g{\bf E}(t_1), \cdot g{\bf E}(t_2) \} \, \{
g{\bf E}(t_3), \cdot g{\bf E}(0) \} \rangle 
\nn\\
&& \qquad\qquad\qquad\qquad
- {4 \over N_c}\langle {\rm
  Tr}(g{\bf E}(t_1)\cdot g{\bf E}(t_2))\, {\rm Tr}(g{\bf E}(t_3)\cdot g{\bf
  E}(0))\rangle \Bigr \} {\delta^{(3)} ({\bf r}) \over E-h_s} \, . 
\label{xx3xx}  
\eea
When a  power of $(E-h_o)$ does not come from the octet propagator in the
middle, all the powers can be substituted by $(E-h_s)$. If we put these
contributions together with the first term before (\ref{xx3xx}), we obtain (Fig. 7):
\bea
&& {i \over E-h_s} {1\over 12 m^4}{\cal T}_{S} \, {\rm Im} \, f_1 (^{2S+1} S_S)
\left \{ \int_0^\infty dt_1\int_0^{t_1}dt_2\int_0^{t_2} dt_3 \, \Big[ (t_1-t_3)^3 +
  t_2^3 \Big] \right. 
\nonumber \\
&& ~~\left. \times \left [ \langle \{ g{\bf E} (t_1)\cdot , g{\bf E} (t_2) \} \{
    g{\bf E} (t_3)\cdot , g{\bf E} (0) \}  \rangle - {4 \over N_c} \langle
    {\rm Tr} [ g{\bf E} (t_1)\cdot g{\bf E} (t_2) ] \, {\rm Tr} [ g{\bf E}
    (t_3) \cdot g{\bf E} (0) ] \rangle \right ] \, \, \right. 
\nonumber \\
&& \left. + \, \int_0^{\infty} dt_1 \, \int_0^{t_1} dt_2 \, \int_0^{t_2} dt_3
  \, \Big[ (t_1-t_2)^3 + t_3^3 \Big] \, \right.  
\nonumber \\
&&~~ \left. \times \Bigg ( \left [ \langle \{ g{\bf E}^i (t_1), g{\bf E}^j (t_2)
    \} \{ g{\bf E}^i (t_3) , g{\bf E}^j (0) \}  \rangle - {4 \over N_c}
    \langle {\rm Tr} [ g{\bf E}^i (t_1) g{\bf E}^j (t_2) ] \, {\rm Tr} [
    g{\bf E}^i (t_3) g{\bf E}^j (0) ] \rangle \right] \right.  
\nonumber \\
&& ~~ \left.  + \left [ \, \langle \{ g{\bf E}^i (t_1), g{\bf E}^j (t_2) \} \{
    g{\bf E}^j (t_3) , g{\bf E}^i (0) \} \rangle - {4 \over N_c} \langle {\rm
      Tr} [ g{\bf E}^i (t_1) g{\bf E}^j (t_2) ] \, {\rm Tr} [ g{\bf E}^j (t_3)
    g{\bf E}^i (0) ] \rangle \right ] \Bigg ) \! \right \} 
\nonumber  \\
&& \times {\delta^{(3)} ({\bf r}) \over E-h_s} \, . 
\eea

Consider next Fig. 4d. Clearly this diagram contains the iteration of lower
order potentials, which must be isolated. This is achieved by adding and
subtracting the projection operator into the gluonic ground state $1=(1-\vert
0\rangle\langle 0\vert) +\vert 0\rangle\langle 0\vert$. The piece $ (1-\vert
0\rangle\langle 0\vert)$ contains new contributions to the potential only,
whereas the piece $\vert 0\rangle\langle 0\vert$ contains both 
the iteration of lower order potentials and new contributions to the potential.  
Consider first the piece $(1-\vert 0\rangle\langle 0\vert)$. It is identical to Fig. 4c by taking 
$V_o\rightarrow V_s$ in the expression before (\ref{xx3xx}) 
and changing the chromoelectric field correlators accordingly. We then have (Fig. 8):
\bea
\label{128}
&& {i \over E-h_s} {1\over 3 N_c m^4}\, {\cal T}_{S} \, {\rm Im} \, f_1 (^{2S+1} S_S) 
\Bigg\{ \int_0^{\infty} dt_1\int_0^{t_1} dt_2\int_0^{t_2} dt_3 \, \Big[ (t_1-t_3)^3 + t_2^3 \Big] 
\nonumber \\
&& ~~\left. \times \Big[ \langle {\rm Tr} [ g{\bf E} (t_1)\cdot g{\bf E} (t_2)
    ] \, {\rm Tr} [ g{\bf E} (t_3) \cdot g{\bf E} (0) ] \rangle - \langle
    g{\bf E} (t_1)\cdot g{\bf E} (t_2) \rangle \langle g{\bf E} (t_3) \cdot
    g{\bf E} (0) \rangle \Big] \, \, \right. 
\nonumber \\
&& \left. + \, \int_0^{\infty} dt_1 \, \int_0^{t_1} dt_2 \, \int_0^{t_2} dt_3
  \, \Big[ (t_1-t_2)^3 + t_3^3 \Big] \, \right.  
\nonumber \\
&&~~ \times \left( \Big[ \langle {\rm Tr} [ g{\bf E}^i (t_1) g{\bf 
      E}^j (t_2) ] \,  {\rm Tr} [ g{\bf E}^i (t_3) g{\bf E}^j (0) ] \rangle -
    \langle g{\bf E}^i (t_1) g{\bf E}^j (t_2) \rangle \langle g{\bf E}^i
    (t_3) g{\bf E}^j (0) \rangle \Big] \right.  
\nonumber \\
&&~~  \left. + \Bigg[ \, \langle {\rm Tr} [ g{\bf E}^i (t_1) g{\bf E}^j (t_2) ]
    \, {\rm Tr} [ g{\bf E}^j (t_3) g{\bf E}^i (0) ] \rangle - \langle g{\bf 
      E}^i (t_1) g{\bf E}^j (t_2) \rangle \langle g{\bf E}^j (t_3) g{\bf E}^i
    (0) \rangle \Bigg] \right) \Bigg\}
 \nonumber  \\
&& \times {\delta^{(3)} ({\bf r}) \over E-h_s} \, . 
\eea

Consider next the contribution from $\vert 0\rangle\langle 0\vert$. 
The vacuum insertion leads to an internal singlet propagator. To be specific, we have:
\bea
&& {-i \over E-h_s} {1 \over N_c^2} \int_0^{\infty} dt \, \langle i{\bf r}
\cdot  g{\bf E} (t) e^{-i(h_o-E) \, t} i{\bf r} \cdot g{\bf E} (0) \rangle {1 \over E-h_s} 
\nonumber \\
&& 
\qquad\qquad \times \int_0^{\infty} dt^{\prime} \, \langle i{\bf r} \cdot g{\bf E}
(t^{\prime}) e^{-i(h_o-E)\, t^{\prime}} i{\bf r} \cdot g{\bf E} (0) \rangle {1 \over
  E-h_s} \, .
\eea
The exponentials of ($E- h_o$) will be expanded. In order to be consistent
with the calculation of the lower order potentials and subtract only their 
iteration, we must treat the powers of ($E-h_o$) at each side of the
internal singlet propagator exactly as we did in the calculation of the lower
order potentials. Let us illustrate how it works when we have two powers of
($E-h_o$) on each side. The only contributions occur when $(E-h_o) \sim (E-h_s) $. 
If we write the propagator in the middle as 
$1 /(E-h_s) = \left ( 1 /(E-h_s) \right ) (E-h_s) \left ( 1 /(E-h_s)\right )$ 
we can use (\ref{m2e2}) and (\ref{m2e2sym}) in order to obtain:
\bea
&& \left( 
{\bf r}^2 +{\bf r}[{\bf r}, h_s]{1\over E-h_s}+{1\over E-h_s}[h_s,{\bf r}]
{\bf r}+{1\over E-h_s}[h_s,{\bf r}][{\bf r}, h_s]{1\over E-h_s} \right) 
\nn \\
&\times& (E-h_s) 
\label{it}\\
&\times& \left ( {\bf r}^2 +{\bf r}[{\bf r}, h_s]{1\over E-h_s}+{1\over
    E-h_s}[h_s,{\bf r}] {\bf r}+{1\over E-h_s}[h_s,{\bf r}][{\bf r},
  h_s]{1\over E-h_s} \right ) \, . 
\nn 
\eea
We can easily identify the contributions that match the following terms in (\ref{wfpt}):
\bea
&& \left( {Z_x \over 2} \right) 
{1\over E-h_s}\left( {Z_x \over 2} \right)^{\dagger}+{1\over E-h_s}h_x{1\over
  E-h_s}h_x{1\over E-h_s} 
\nonumber \\
&& + \left( {Z_x \over 2} \right)
{1\over E-h_s}h_x{1\over E-h_s}
+{1\over E-h_s}h_x{1\over E-h_s}\left( {Z_x \over 2} \right)^{\dagger} \, .
\eea
We also see that, apart from the terms above, there are additional terms in
(\ref{it}) that may (and do) eventually lead to new contributions to the
potential (none of them with imaginary parts for this example). For them we
use the same prescription as stated at the beginning of the section. 
The contributions to the imaginary parts come from the following terms in the 
expansion only: (i) an $(E-h_o)^4$ from an octet propagator and a $1$ from the other one 
(Fig. 9, first diagram), and (ii) an $(E-h_o)^3$ from an octet propagator and
an $(E-h_o)$ from the other one (Fig. 9, all of them).
They read:
\bea
&& {\rm (i)} ~~ {7\,i \over 9 \, N_c} {{\rm Im} \, f_1 (^{2S+1}S_S) \over m^4}{{\cal T}_S
  \over h_s - E} \left ( \int_0^{\infty} dt \, t^4 \langle g {\bf E} (t) \cdot
  g {\bf E} (0) \rangle \right ) 
\nn\\
&&
\qquad\qquad\qquad\qquad\qquad\qquad\qquad
\times
\left ( \int_0^{\infty} dt^{\prime} \, 
\langle g {\bf E} (t^{\prime}) \cdot g {\bf E} (0) \rangle \right )
{\delta^{(3)} ({\bf r}) \over h_s - E} \, , 
\nonumber \\
&& {\rm (ii)} ~~  { 4\, i \over 27 \, N_c} {{\rm Im}\, f_1 (^{2S+1}S_S) \over m^4}{{\cal
    T}_S \over h_s - E} \left ( \int_0^{\infty} dt \, t^3 \langle g {\bf E}
  (t) \cdot g {\bf E} (0) \rangle \right ) 
\nn\\
&&
\qquad\qquad\qquad\qquad\qquad\qquad\qquad
\times
\left ( \int_0^{\infty} dt^{\prime}
  \, t^{\prime} \langle g {\bf E} (t') \cdot g {\bf E} (0) \rangle \right )
{\delta^{(3)} ({\bf r}) \over h_s - E}  \, . 
\nonumber 
\eea

\subsection{Results}
\label{sec5d}
Combining all the above calculations we obtain the same result as in 
Sec. \ref{sec3g}, except for the terms proportional to ${\rm Im}(^{2S+1}
S_S)$. With the mere replacement:
\bea
&& {\mathcal E}_3^{(2,t)} \longrightarrow \, \overline{\mathcal E}_3^{(2,t)}
\, , 
\nonumber \\
&& {\mathcal E}_3^{(2,{\rm EM})} \longrightarrow \, \overline{\mathcal E}_3^{(2,{\rm EM})} \, , 
\eea 
where we have defined:
\bea
&& \overline{\cal E}_3^{(2,t)} = -{1 \over 8 N_c} 
\left\{ \int_0^\infty dt_1\int_0^{t_1}dt_2\int_0^{t_2} dt_3 \, \Big[ (t_1-t_3)^3 +
  t_2^3 \Big] \right. 
\nonumber \\
&& ~~ \left. \times \left [ \langle \{ g{\bf E} (t_1)\cdot , g{\bf E} (t_2) \} \{
    g{\bf E} (t_3)\cdot , g{\bf E} (0) \} \rangle - {4 \over N_c} \langle
    g{\bf E} (t_1)\cdot g{\bf E} (t_2) \rangle 
\langle g{\bf E} (t_3) \cdot g{\bf E} (0) \rangle \right ] \, \, \right. 
\nonumber \\
&& \left. + \, \int_0^{\infty} dt_1 \, \int_0^{t_1} dt_2 \, 
\int_0^{t_2} dt_3 \, \Big[ (t_1-t_2)^3 + t_3^3 \Big] \, \right.  
\nonumber \\ 
&& ~~ \times \left( \left [ \langle \{ g{\bf E}^i (t_1), g{\bf E}^j (t_2) \} 
\{ g{\bf E}^i (t_3) , g{\bf E}^j (0) \} \rangle 
- {4 \over N_c} \langle g{\bf E}^i (t_1) g{\bf E}^j (t_2) 
\rangle \langle g{\bf E}^i (t_3) g{\bf E}^j (0) \rangle \right] \right.  
\nonumber \\
&& ~~ \left.  + \left[ \, \langle \{ g{\bf E}^i (t_1), g{\bf E}^j (t_2) \} 
\{ g{\bf E}^j (t_3) , g{\bf E}^i (0) \} \rangle - {4 \over N_c} 
\langle g{\bf E}^i (t_1) g{\bf E}^j (t_2)  \rangle 
\langle g{\bf E}^j (t_3) g{\bf E}^i (0) \rangle \right] \right)
\nonumber  \\
&& \left. - i (\delta_{ij} \delta_{kl} + \delta_{ik} \delta_{jl}
+ \delta_{il} \delta_{jk}) \int_0^{\infty} dt_1 \, 
\int_0^{t_1} dt_2 \, \Big[ (t_1-t_2)^3+ t_2^3 \Big] 
\langle g {\bf E}^i (t_1) [ {\bf D}^j , g{\bf E}^k ](t_2) g{\bf E}^l (0)
\rangle \right. 
\nonumber \\
&& \left. + \, \int_0^{\infty} dt \, t^3  \langle [ {\bf D}\cdot ,g {\bf E}
  (t) ]  [{\bf D} \cdot , g{\bf E} (0)] \rangle + {7 \over 6} {\mathcal E}_4
  {\mathcal E}_0 + {2 \over 9} {\mathcal E}_3 {\mathcal E}_1 \right\} \, ,
\eea
and 
\bea
&& \overline{\cal E}_3^{(2,{\rm EM})} = -{1 \over 2 N_c^2} 
\left\{ \int_0^\infty dt_1\int_0^{t_1}dt_2\int_0^{t_2} 
dt_3 \, \Big[ (t_1-t_3)^3 + t_2^3 \Big] \right. 
\nonumber \\
&& ~~\left. \times \Bigg[ \langle {\rm Tr} [ g{\bf E} (t_1)\cdot g{\bf E} (t_2)
    ] {\rm Tr} [ g{\bf E} (t_3) \cdot g{\bf E} (0) ] \rangle 
- \langle g{\bf E} (t_1)\cdot g{\bf E} (t_2) \rangle \langle g{\bf E} (t_3) 
\cdot g{\bf E} (0) \rangle \Bigg] \, \, \right. 
\nonumber \\
&& \left. + \, \int_0^{\infty} dt_1 \, 
\int_0^{t_1} dt_2 \, \int_0^{t_2} dt_3 \, \Big[ (t_1-t_2)^3 + t_3^3 \Big] \, \right.  
\nonumber \\ 
&&~~ \times  \left[ \left ( \langle 
 \{ g{\bf E}^i (t_1), g{\bf E}^j (t_2) \} \{ g{\bf E}^i (t_3) ,
g{\bf E}^j (0) \}  \rangle 
- {4 \over N_c} \langle g{\bf E}^i (t_1) g{\bf E}^j (t_2)  \rangle 
\langle g{\bf E}^i (t_3) g{\bf E}^j (0) \rangle \right ) \right.  
\nonumber \\
&&~~ \left.  + \left ( \, \langle \{ g{\bf E}^i (t_1), g{\bf E}^j
    (t_2) \} \{ g{\bf E}^j (t_3) , g{\bf E}^i (0) \}  \rangle 
- {4 \over N_c} \langle g{\bf E}^i (t_1) g{\bf E}^j (t_2) \rangle 
\langle g{\bf E}^j (t_3) g{\bf E}^i (0) \rangle \right ) \right]  
\nonumber  \\
&& \left. - i (\delta_{ij} \delta_{kl} + \delta_{ik} \delta_{jl}
+ \delta_{il} \delta_{jk}) \int_0^{\infty} dt_1 \, \int_0^{t_1} dt_2 \, \Big[
(t_1-t_2)^3+ t_2^3 \Big] \langle g {\bf E}^i (t_1) [ {\bf D}^j , g{\bf E}^k ](t_2)
g{\bf E}^l (0) \rangle \right. 
\nonumber \\
&& \left. + \, \int_0^{\infty} dt \, t^3  \langle [ {\bf D} \cdot , g {\bf
    E}(t)] [{\bf D}\cdot , g{\bf E} (0)] \rangle 
+ {7 \over 6} {\mathcal E}_4 {\mathcal E}_0 + {2 \over 9} {\mathcal E}_3 {\mathcal E}_1 \right\} \, ,
\eea
the same expressions apply.

As mentioned before, the difference is due to the different prescription to
fix the wave-function normalization in Sec. \ref{sec3b}. In Appendix
\ref{Appunit} we show that there exists an unitary transformation such that our
results can be taken to the form of Sec. \ref{sec3g}, and hence they are equivalent for all purposes.
 
In fact, it is somewhat surprising  that the two calculations lead to
identical results (up to a unitary transformation).  On general grounds, one
could only expect that the result in this section be a particular case of the
general results of Sec. \ref{qmmatching}. In fact the real parts of the
potentials in $h$ are indeed particular cases of the potentials in \cite{M2}. 
However, since we did not need their specific form at any stage we have not 
lost generality in our final expressions. More surprising is 
the fact that the matching coefficients of the terms in the multipole
expansion in pNRQCD$^\prime$ (\ref{pnrqcd0}) were only calculated at tree 
level here, whereas the expressions in Sec. \ref{qmmatching} correspond to an
all-loop result. This indicates that there must be a symmetry protecting these 
terms against higher-loop corrections, which may (or may not) be an extension
of reparametrization invariance \cite{LukeManohar} or Poincar\'e 
invariance itself \cite{BGV}.\footnote{For the leading order term, 
the non-renormalization was verified at one loop in \cite{RGstatic}.} 

In summary, we have presented in this section an alternative derivation 
of (\ref{hadrV})--(\ref{electrchi}), which does not rely so heavily on the
$1/m$ expansion. The matching from NRQCD to pNRQCD$^\prime$, which can be done 
perturbatively in $\als$, can indeed be implemented in the $1/m$ expansion, as originally proposed
\cite{Mont}, but it can also be done entirely in the framework of 
the threshold expansion \cite{BV,KPSS}, where the kinetic term is kept in 
the denominator for potential loop contributions and the on-shell condition 
is used (the results obtained in either way are related by local field
redefinitions). The matching between pNRQCD$^\prime$ and pNRQCD is done 
in the $\lQ/mv$, $\lQ/m$ and $mv^2/ \lQ$ expansions. 
The approaches taken in these two steps are quite different from 
the strict $1/m$ expansion of Sec. \ref{qmmatching}, and the coincidence 
of the results strongly supports their correctness.

\section{Results}
\label{results}
In this section we list our expressions for $S$-wave decays up to 
${\cal O}(c(\als(m))mv^3\times (\lQ^2/m^2, E/m))$ and for $P$-wave decays 
up to ${\cal O}(c(\als(m))mv^5)$. The $P$-wave decay widths were first obtained in \cite{pw} 
and are given here for completeness. The $S$-wave decay widths are new.
In order to help the reader and for further convenience, we will start 
by recalling, at the same level of accuracy, the expressions of the 
decay widths as they are known from NRQCD.
In the following we define the radial part of the vector $S$-wave function as 
$R_{n101}\equiv R^V_{n0}=R_{n0}^{(0)}(1+{\cal O}(v))$ 
and the radial part of the pseudoscalar $S$-wave function 
as $R_{n000} \equiv R^P_{n0}=R_{n0}^{(0)}(1+{\cal O}(v))$. 
The quantity $R^{(0)\,\prime}_{n1}$ is 
the derivative of the leading order $P$-wave function.  
The symbols $V$ and $P$ stand for the vector 
and pseudoscalar $S$-wave heavy quarkonium and the symbol $\chi$ for  
the generic $P$-wave quarkonium (the states $\chi(n10)$ and $\chi(nJ1)$
are usually called $h((n-1)P)$ and $\chi_J((n-1)P)$, respectively).

\subsection{Decay Widths in NRQCD}
Including up to the NRQCD 4-fermion operators of
dimension 8, the inclusive decays of heavy quarkonia are given by:
\bea
&&\Gamma(V_Q (nS) \rightarrow LH) = {2\over m^2}\Bigg( 
{\rm Im\,}f_1(^3 S_1)   \langle V_Q(nS)|O_1(^3S_1)|V_Q(nS)\rangle
\nn
\\
&&\quad
+ {\rm Im\,}f_8(^3 S_1) \langle V_Q(nS)|O_8(^3S_1)|V_Q(nS)\rangle
+ {\rm Im\,}f_8(^1 S_0) \langle V_Q(nS)|O_8(^1S_0)|V_Q(nS)\rangle 
\nn
\\
&&\quad 
+ {\rm Im\,}g_1(^3 S_1)
{\langle V_Q(nS)|{\cal P}_1(^3S_1)|V_Q(nS)\rangle \over m^2}
+ {\rm Im\,}f_8(^3 P_0)
{\langle V_Q(nS)|O_8(^3P_0)|V_Q(nS)\rangle \over m^2}
\nn
\\
&&\quad
+ {\rm Im\,}f_8(^3 P_1)
{\langle V_Q(nS)|O_8(^3P_1)|V_Q(nS)\rangle \over m^2}
+ {\rm Im\,}f_8(^3 P_2)
{\langle V_Q(nS)|O_8(^3P_2)|V_Q(nS)\rangle \over m^2}\Bigg),
\\
&&
\nn 
\\
&&\Gamma(P_Q (nS) \rightarrow LH) = {2\over m^2}\Bigg( 
{\rm Im\,}f_1(^1 S_0)   \langle P_Q(nS)|O_1(^1S_0)|P_Q(nS)\rangle
\nn
\\
&&\quad
+ {\rm Im\,}f_8(^1 S_0) \langle P_Q(nS)|O_8(^1S_0)|P_Q(nS)\rangle
+ {\rm Im\,}f_8(^3 S_1) \langle P_Q(nS)|O_8(^3S_1)|P_Q(nS)\rangle 
\nn
\\
&&\quad 
+ {\rm Im\,}g_1(^1 S_0)
{\langle P_Q(nS)|{\cal P}_1(^1S_0)|P_Q(nS)\rangle \over m^2}
+ {\rm Im\,}f_8(^1 P_1)
{\langle P_Q(nS)|O_8(^1P_1)|P_Q(nS)\rangle \over m^2} \Bigg),
\\
&&
\nn
\\
&&\Gamma(\chi_Q(nJS)  \rightarrow LH)= 
{2\over m^2}\Bigg( {\rm Im \,}  f_1(^{2S+1}P_J) 
{\langle \chi_Q(nJS) | O_1(^{2S+1}P_J ) | \chi_Q(nJS) \rangle \over m^2}
\nn
\\
&&\quad
+ f_8(^{2S+1}S_S) \langle \chi_Q(nJS) | O_8(^1S_0 ) | \chi_Q(nJS) \rangle\Bigg).
\eea
At the same order the electromagnetic decays are given by:
\bea 
&&\Gamma(V_Q (nS) \rightarrow e^+e^-)= {2\over m^2}\Bigg( 
{\rm Im\,}f_{ee}(^3 S_1)   \langle V_Q(nS)|O_{\rm EM}(^3S_1)|V_Q(nS)\rangle
\nn
\\
&&\quad
+ {\rm Im\,}g_{ee}(^3 S_1)
{\langle V_Q(nS)|{\cal P}_{\rm EM}(^3S_1)|V_Q(nS)\rangle \over m^2}\Bigg),
\\
&&
\nn
\\
&&\Gamma(P_Q (nS) \rightarrow \gamma\gamma)= {2\over m^2}\Bigg( 
{\rm Im\,}f_{\gamma\gamma}(^1 S_0)   
\langle P_Q(nS)|O_{\rm EM}(^1S_0)|P_Q(nS)\rangle
\nn
\\
&&\quad
+ {\rm Im\,}g_{\gamma\gamma}(^1 S_0)
{\langle P_Q(nS)|{\cal P}_{\rm EM}(^1S_0)|P_Q(nS)\rangle \over m^2} \Bigg),
\\
&&
\nn
\\
&&\Gamma(\chi_Q(nJ1)  \rightarrow \gamma\gamma)= 
2 {\rm Im \,}  f_{\gamma\gamma}(^3P_J) 
{\langle \chi_Q(nJ1) | O_{\rm EM}(^3P_J ) | \chi_Q(nJ1) \rangle \over m^4}
\qquad {\rm for} \; J=0,2\,.
\eea

\subsection{Decay Widths in pNRQCD}
\label{sec4b}
Up to ${\mathcal O}(c(\als(m))mv^3\times (\lQ^2/m^2, E/m))$ for $S$-wave
and ${\mathcal O}(c(\als(m))mv^5)$ for $P$-wave, the inclusive decays of
heavy quarkonia are given in pNRQCD by:
\bea
&&\Gamma(V_Q (nS) \rightarrow LH) ={C_A \over \pi}{|R^V_{n0}({0})|^2 \over m^2}
\left[
{\rm Im\,}f_1(^3 S_1)\left(1-{E_{n0}^{(0)} \over m}{2{\cal E}_3 \over 9}
+{2{\cal E}^{(2,t)}_3 \over 3 m^2}+{c_F^2{\cal B}_1 \over 3 m^2}\right)
\right.
\nn
\\
&& \qquad
-{\rm Im\,}f_8(^3 S_1){2 (C_A/2-C_f) {\cal E}^{(2)}_3 \over 3 m^2}
-{\rm Im\,}f_8(^1 S_0){ (C_A/2-C_f) c_F^2{\cal B}_1 \over 3 m^2}
\nn
\\
&& \qquad 
+{\rm Im\,}g_1(^3 S_1)\left( {E_{n0}^{(0)} \over m} - {{\cal E}_1 \over m^2 } \right) 
\nn
\\
&& \qquad
\left.
-\left({\rm Im\,}f_8(^3 P_0)+3{\rm Im\,}f_8(^3 P_1)+5{\rm Im\,}f_8(^3 P_2) \right)
{(C_A/2-C_f) {\cal E}_1 \over 9 m^2} 
\right], \label{hadrV}
\\
&&
\nn 
\\
&&\Gamma(P_Q (nS) \rightarrow LH) ={C_A \over \pi}{|R^P_{n0}({0})|^2 \over m^2}
\left[
{\rm Im\,}f_1(^1 S_0)\left(1-{E_{n0}^{(0)} \over m}{2{\cal E}_3 \over 9}
+{2{\cal E}^{(2,t)}_3 \over 3 m^2}+{c_F^2{\cal B}_1 \over m^2}\right)
\right.
\nn
\\
&& \qquad
-{\rm Im\,}f_8(^1 S_0){2 (C_A/2-C_f) {\cal E}^{(2)}_3 \over 3 m^2}
-{\rm Im\,}f_8(^3 S_1){(C_A/2-C_f) c_F^2{\cal B}_1 \over m^2 }
\nn
\\
&& \qquad
\left.
+{\rm Im\,}g_1(^1 S_0)\left( {E_{n0}^{(0)} \over m} - {{\cal E}_1 \over m^2  }\right)
-{\rm Im\,}f_8(^1 P_1){(C_A/2-C_f) {\cal E}_1 \over m^2 }
\right],
\\
\nn
&&
\\
&&\Gamma(\chi_Q(nJS)  \rightarrow LH)= 
{C_A\over \pi}{| R^{(0)\,\prime}_{n1} ({0}) |^2 \over m^4}
\Bigg[ 3 \, {\rm Im\,}\,   f_1(^{2S+1}P_J) 
+ {2 T_F\over 3 C_A } {\rm Im\,} \,  f_8(^{2S+1}{\rm{S}}_S) \, {\cal E}_3 \Bigg]. 
\eea
At the same order the electromagnetic decays are given by:
\bea
&&\Gamma(V_Q (nS) \rightarrow e^+e^-) ={C_A \over \pi}{|R^V_{n0}({0})|^2 \over m^2}
\left[
{\rm Im\,}f_{ee}(^3 S_1)\left(1-{E_{n0}^{(0)} \over m}{2{\cal E}_3 \over 9}
+{2{\cal E}^{(2,{\rm EM})}_3 \over 3 m^2}+{c_F^2{\cal B}_1 \over 3 m^2}\right) \right.
\nn
\\
&& \qquad \left.
+{\rm Im\,}g_{ee}(^3 S_1)\left({E_{n0}^{(0)} \over m}-{{\cal E}_1 \over
    m^2}\right)
\right],
\\
&&
\nn
\\
&&\Gamma(P_Q (nS) \rightarrow \gamma\gamma) =
{C_A \over \pi}{|R^P_{n0}({0})|^2 \over m^2}
\left[
{\rm Im\,}f_{\gamma\gamma}(^1 S_0)\left(1-{E_{n0}^{(0)} \over m}{2{\cal E}_3 \over 9}
+{2{\cal E}^{(2,{\rm EM})}_3 \over 3 m^2}+{c_F^2{\cal B}_1 \over m^2}\right) \right.
\nn
\\
&& \qquad \left.
+{\rm Im\,}g_{\gamma\gamma}(^1 S_0)\left({E_{n0}^{(0)} \over m}
-{{\cal E}_1 \over m^2}\right)
\right],
\\
&&
\nn
\\
&&\Gamma(\chi_Q(nJ1)  \rightarrow \gamma\gamma)= 
3 {C_A \over \pi}{|R^{(0)\,\prime}_{n1}({0})|^2 \over m^4} f_{\gamma\gamma}(^3P_J) 
\qquad {\rm for} \; J=0,2\,. \label{electrchi}
\eea

\subsection{NRQCD Matrix elements}
\label{nrqcdmatrix}
By comparing the decay widths in NRQCD and pNRQCD
we obtain the following {\it dictionary} between the matrix elements of
NRQCD and the non-perturbative constants of pNRQCD, valid up to (once
normalized to $m$) ${\mathcal O}(v^3\times (\lQ^2/m^2, E/m))$ for the
$S$-wave matrix elements and up to ${\mathcal O}(v^5)$ for $P$-wave
matrix elements:
\bea
\label{O13S1}
&&\langle V_Q(nS)|O_1(^3S_1)|V_Q(nS)\rangle=
C_A {|R^V_{n0}({0})|^2 \over 2\pi}
\left(1-{E_{n0}^{(0)} \over m}{2{\cal E}_3 \over 9}
+{2{\cal E}^{(2,t)}_3 \over 3 m^2 }+{c_F^2{\cal B}_1 \over 3 m^2 }\right),
\\
&&\langle P_Q(nS)|O_1(^1S_0)|P_Q(nS)\rangle=
C_A {|R^P_{n0}({0})|^2 \over 2\pi}
\left(1-{E_{n0}^{(0)} \over m}{2{\cal E}_3 \over 9}
+{2{\cal E}^{(2,t)}_3 \over 3 m^2}+{c_F^2{\cal B}_1 \over m^2}\right),
\\
&&\langle V_Q(nS)|O_{\rm EM}(^3S_1)|V_Q(nS)\rangle=
C_A {|R^V_{n0}({0})|^2 \over 2\pi}
\left(1-{E_{n0}^{(0)} \over m}{2{\cal E}_3 \over 9}
+{2{\cal E}^{(2,{\rm EM})}_3 \over 3 m^2}+{c_F^2{\cal B}_1 \over 3 m^2}\right),
\\
\label{OEM1S0}
&&\langle P_Q(nS)|O_{\rm EM}(^1S_0)|P_Q(nS)\rangle=
C_A {|R^P_{n0}({0})|^2 \over 2\pi}
\left(1-{E_{n0}^{(0)} \over m}{2{\cal E}_3 \over 9}
+{2{\cal E}^{(2,{\rm EM})}_3 \over 3 m^2}+{c_F^2{\cal B}_1 \over m^2}\right),
\\
&&
\langle \chi_Q(nJS) | O_1(^{2S+1}P_J ) | \chi_Q(nJS) \rangle = 
\langle \chi_Q(nJS) | O_{\rm EM}(^{2S+1}P_J ) | \chi_Q(nJS) \rangle  
\nn
\\
&&\qquad\qquad\qquad\qquad\qquad\qquad\qquad\quad
={3 \over 2}{C_A \over \pi} |R^{(0)\,\prime}_{n1}({0})|^2,
\label{chio1}
\\
&&
\langle V_Q(nS)|{\cal P}_1(^3S_1)|V_Q(nS)\rangle=
\langle P_Q(nS)|{\cal P}_1(^1S_0)|P_Q(nS)\rangle
\nn\\
&&
\qquad\qquad\qquad
=\langle V_Q(nS)|{\cal P}_{\rm EM}(^3S_1)|V_Q(nS)\rangle=
\langle P_Q(nS)|{\cal P}_{\rm EM}(^1S_0)|P_Q(nS)\rangle
\nn\\
&&\qquad\qquad\qquad\qquad\qquad\qquad\qquad\quad
=C_A {|R^{(0)}_{n0}({0})|^2 \over 2\pi}
\left(m E_{n0}^{(0)} -{\cal E}_1 \right),
\label{P13S1}
\\
&&
\langle V_Q(nS)|O_8(^3S_1)|V_Q(nS)\rangle=
\langle P_Q(nS)|O_8(^1S_0)|P_Q(nS)\rangle
\nn\\
&&\qquad\qquad\qquad\qquad\qquad\qquad\qquad\quad
=C_A {|R^{(0)}_{n0}({0})|^2 \over 2\pi}
\left(- {2 (C_A/2-C_f) {\cal E}^{(2)}_3 \over 3 m^2 }\right),
\\
&&
\langle V_Q(nS)|O_8(^1S_0)|V_Q(nS)\rangle=
{\langle P_Q(nS)|O_8(^3S_1)|P_Q(nS)\rangle \over 3}
\nn\\
&&\qquad\qquad\qquad\qquad\qquad\qquad\qquad\quad
=C_A {|R^{(0)}_{n0}({0})|^2 \over 2\pi}
\left(-{(C_A/2-C_f) c_F^2{\cal B}_1 \over 3 m^2 }\right),
\\
&&
\langle V_Q(nS)|O_8(^3P_J)|V_Q(nS)\rangle=
{\langle P_Q(nS)|O_8(^1P_1)|P_Q(nS)\rangle \over 3}
\nn\\
&&\qquad\qquad\qquad\qquad\qquad\qquad\qquad\quad
=(2J+1)\,C_A {|R^{(0)}_{n0}({0})|^2 \over 2\pi}
\left(-{(C_A/2-C_f) {\cal E}_1 \over 9 }\right),
\\
&&
\langle \chi_Q(nJS)\vert O_8(^1S_0)\vert \chi_Q(nJS) \rangle
= {T_F\over 3}
{\vert R^{(0)\,\prime}_{n1}({0})\vert^2 \over \pi m^2} {\cal E}_3.
\label{matoct}
\eea
Any other $S$-wave dimension-6 matrix element is 0 at NNLO and any other
$S$-wave dimension-8 matrix element is 0 at LO. 

\medskip

Equation (\ref{P13S1}) is worth emphasizing. 
It is of the {\it singlet} type but, because of the term proportional to
${\cal E}_1$, its leading contribution is not only
proportional to what one would expect from a pure singlet potential model. 
In Ref. \cite{GK} the authors have also elaborated on Eq. (\ref{P13S1}). Within the
context of NRQCD \cite{nrqcd}, they use the leading equations of
motion\footnote{We have also used the equations of motion in order to derive
Eq. (\ref{eqgk}). Nevertheless, we have done so in the context of pNRQCD.}, 
the power-counting rules of \cite{nrqcd,pcnrqcd} and some
arguments to neglect some mass-like terms, which could be generated
under regularization. They get 
\bea
&&
\langle V_Q(nS)|{\cal P}_1(^3S_1)|V_Q(nS)\rangle_{\hbox{Ref. \cite{GK}}} =
C_A {|R^{(0)}_{n0}({0})|^2 \over 2\pi} m E_{n0}^{(0)} ,
\\
&&
\langle P_Q(nS)|{\cal P}_1(^1S_0)|P_Q(nS)\rangle_{\hbox{Ref. \cite{GK}}} =
C_A {|R^{(0)}_{n0}({0})|^2 \over 2\pi} m E_{n0}^{(0)},
\eea
where the term proportional to ${\cal E}_1$ is missing. Nevertheless,
this does not necessarily reflect any inconsistency in any of the
derivations since, according to the (perturbative-like) power-counting
rules of \cite{nrqcd,pcnrqcd}, the term due to ${\cal E}_1$ would be subleading. In any case, it
would be very interesting to see how a term proportional to ${\cal E}_1$ would appear in the
derivation of Ref. \cite{GK}. Here, we only would like to point out the 
possibility that an ${\cal E}_1/m$ term may show up as a correction to the neglected
mass-like term in Ref. \cite{GK}.
Finally, let us note that in the dynamical situation $mv \sim \lQ$, where 
${\cal E}_1 \sim \lQ^2 \sim m^2v^2 \sim m E_{n0}^{(0)}$,  
both terms on the right-hand side of Eq. (\ref{P13S1}) are of the same order 
and contribute to the decay width at order $c(\als(m))mv^5$.
Phenomenologically this is particularly relevant 
to the case of pseudoscalar decays into light hadrons and to 
the electromagnetic decays. In the case of vector decays into light 
hadrons the contribution coming from the operator 
$\langle V_Q(nS)|{\cal P}_1(^3S_1)|V_Q(nS)\rangle$ may 
not be so important since the matching coefficient  ${\rm Im\,} g_1(^3S_1) \sim
\als(m)^3$ is suppressed by a factor $\als(m)$ 
with respect to the other ones  
(with the exception of ${\rm Im\,} f_1(^3S_1)$ and ${\rm Im\,} f_8(^3P_1)$, 
which are also of order $\als(m)^3$).

\subsection{Evolution equations}
\label{eveq}
In \cite{nrqcd} evolution equations for the 4-fermion operators have been
obtained. If we focus on the states that we are studying
in this paper, the following evolution equations for 
the NRQCD matrix elements are obtained:
\bea
\langle V_Q(nS)|\left(\nu {d\over d\nu}
O_1(^3S_1)\right)
|V_Q(nS)\rangle &=&
{8 \als \over 3\pi m^2} \left(
\langle V_Q(nS)|O_8(^3P_0)|V_Q(nS)\rangle \right.
\nn
\\
&&
+ \langle V_Q(nS)|O_8(^3P_1)|V_Q(nS)\rangle 
\nn
\\
&&
+ \langle V_Q(nS)|O_8(^3P_2)|V_Q(nS)\rangle 
\nn
\\
&&
\left.
- C_f \langle V_Q(nS)|{\cal P}_1(^3S_1)|V_Q(nS)\rangle \right),
\label{EE1}
\\
\langle P_Q(nS)|
\left( \nu {d\over d\nu}O_1(^1S_0)\right)
|P_Q(nS)\rangle &=&
{8 \als \over 3\pi m^2} \left(
\langle P_Q(nS)|O_8(^1P_1)|P_Q(nS)\rangle \right.
\nn
\\
&&
\left.
- C_f \langle P_Q(nS)|{\cal P}_1(^1S_0)|P_Q(nS)\rangle  \right),
\label{EE2}
\\
\label{EE3}
\langle V_Q(nS)|
\left(\nu {d\over d\nu}
O_{\rm EM}(^3S_1)\right)
|V_Q(nS)\rangle &=&
- {8 C_f \als \over 3\pi m^2} 
\langle V_Q(nS)|{\cal P}_{\rm EM}(^3S_1)|V_Q(nS)\rangle ,
\\
\label{EE4}
\langle P_Q(nS)|\left(
\nu {d\over d\nu}
O_{\rm EM}(^1S_0)\right)
|P_Q(nS)\rangle &=&
-{8 C_f \als \over 3\pi m^2} 
\langle P_Q(nS)|{\cal P}_{\rm EM}(^1S_0)|P_Q(nS)\rangle.  
\eea
Since we have, at ${\cal O}(\als)$ and leading log accuracy, 
\bea 
\nu {d\over d\nu} {\cal E}_3 &=&  12 \, C_f \, {\als \over \pi},
\label{e3run}
\\
\nu {d\over d\nu} {\cal E}^{(2)}_3 &=& 
\nu {d\over d\nu} {\cal E}^{(2,c)}_3 =
\nu {d\over d\nu} {\cal E}^{(2,t)}_3 = 0,
\eea
Eqs. (\ref{O13S1})--(\ref{OEM1S0}) are compatible with the evolution equations 
(\ref{EE1})--(\ref{EE4}) at leading log accuracy. Note that at this order
there is no $\nu$ dependence in the states, and hence the derivatives with
respect to $\nu$ can be taken out of the expectation values.
In Ref. \cite{pw} it was proved that Eq. (\ref{e3run}) 
gives the correct running for the octet operator of Eq. (\ref{matoct}).
In Appendix \ref{ARG}, the reader can find the evolution equations and their 
leading-order solutions for the imaginary parts of all the 
4-fermion matching coefficients needed in this work.

\section{Model-independent predictions}
\label{mind}
The inclusive decays of the heavy quarkonium (either hadronic or
electromagnetic) are usually considered up to, and including, NRQCD
matrix elements of 4-fermion operators of dimension 8. This means
to consider the ${\cal O}(1/m^2,1/m^4)$ local 4-fermion operators of the NRQCD
Lagrangian. With this accuracy, the decay into light hadrons of a vector
$S$-wave state is described in NRQCD by the matrix elements of two singlet
operators ($O_1(^3S_1)$ and ${\cal P}_1$), and three octet operators
($O_8(^3S_1)$, $O_8(^1S_0)$ and $O_8(P)$).  The corresponding
pseudoscalar $S$-wave state decay needs, at the same level of
accuracy, the additional knowledge of the matrix element of the
singlet operator $O_1(^1S_0)$. The electromagnetic decays of the same
$S$-states need the additional knowledge of the matrix elements of the
singlet electromagnetic operators $O_{\rm EM}(^3S_1)$ and $O_{\rm
EM}(^1S_0)$ respectively.  The decay of a $P$-wave quarkonium state
into light hadrons and the corresponding electromagnetic decay are
described in NRQCD with the above accuracy by the matrix element of a
singlet ($O_1(P)$) and an octet ($O_8(^1S_0)$) operator.  If we
consider that in the bottomonium system in principle 14 $S$- and
$P$-wave states lie below threshold ($\Upsilon(nS)$ and $\eta_b(nS)$
with $n=1,2,3$; $h_b(nP)$ and $\chi_{bJ}(nP)$ with $n=1,2$ and
$J=0,1,2$) and that in the charmonium system this is the case for 8
states ($\psi(nS)$ and $\eta_c(nS)$ with $n=1,2$; $h_c(1P)$ and
$\chi_{cJ}(1P)$ with $J=0,1,2$), all the bottomonium and charmonium
$S$- and $P$-wave decays into light hadrons and into photons or
$e^+e^-$ are described in NRQCD up to ${\cal O}(1/m^4)$ by 46 unknown
NRQCD matrix elements (40 for the $S$-wave decays and $6$ for the $P$-wave
decays). These matrix elements have to be fixed either by lattice
simulations \cite{latbo} or by fitting the data \cite{Maltoni}. Only
in the specific case of matrix elements of singlet operators does NRQCD
allow an interpretation in terms of quarkonium wave functions and one
can resort to potential models.

At the same level of accuracy $S$- and $P$-wave bottomonium and
charmonium decays are described in pNRQCD, under the dynamical
assumption $\lQ \gg mv^2$, by only 19 non-perturbative
parameters. These are the 13 wave functions (one for each of the 10
$S$-wave quarkonium states below threshold, for which we need to
distinguish different spin states, and a total number of 3 for the
$P$-wave quarkonium states) and 6 universal non-perturbative
parameters, which do not depend on the flavour and on the state
(${\cal E}_1$, ${\cal E}_3$, ${\cal B}_1$, ${\cal E}_3^{(2)}$, ${\cal
E}_3^{(2,t)}$ and ${\cal E}_3^{(2, {\rm EM})}$).

In the above discussion we have counted NRQCD matrix elements by their dimensionality only. 
A more refined discussion would require that a maybe less conservative 
power counting be assigned to the NRQCD matrix elements as well as that 
the $\als(m)$ suppression due to the short-distance NRQCD matching
coefficients be taken into account. 
As we have already mentioned throughout the paper, the power counting of
the NRQCD matrix elements is an open issue. To consider all the
possibilities and phenomenological consequences goes beyond the scope  of
the present paper, whose aim is to set the theoretical framework. 
However, we would like to  mention a few things. In the standard NRQCD power counting \cite{pcnrqcd},
the octet matrix elements are ${\cal O}(v^4)$ suppressed for $S$-wave decays if
compared with the leading order. This is not so within our framework where, assuming the
counting $\lQ \sim mv$, they would only be ${\cal O}(v^2)$-suppressed. This is
potentially relevant to $\Gamma(V \rightarrow LH)$ since ${\rm Im}\,
f_1( ^3S_1)$ is ${\cal O}(\als(m))$-suppressed with respect ${\rm Im} \,f_8( S)$. 
In other words, the octet matrix element effects could
potentially be much more important than usually thought for these decays. 
It would be interesting to analyse this possibility further.

The dramatic reduction in the number of parameters makes it possible, 
in the framework of pNRQCD, to formulate several new predictions with
respect to NRQCD. In particular it is possible to relate information
gained from decay widths of quarkonium with a specific flavour and
principal quantum number to decay widths of quarkonium with different
flavour and/or principal quantum number. Following this strategy in
\cite{pw} the non-perturbative parameter ${\cal E}_3$ has been fixed
on the charmonium $P$-wave decay data and used to predict
ratios of $P$-wave decay widths for the bottomonium system (in this
case and at leading order there is no ambiguity on the relative size
between the singlet and the octet contribution).  Here we
will concentrate on some exact model-independent relations valid for
$S$-wave decays.

Let us consider the ratios of hadronic and electromagnetic decay widths
for states with the same principal quantum number:
\bea
& &R_n^V = {\Gamma(V_Q (nS) \rightarrow LH) \over
\Gamma(V_Q (nS) \rightarrow e^+e^-)},
\label{rv}
\\
& &R_n^P = {\Gamma(P_Q (nS) \rightarrow LH) \over
\Gamma(P_Q (nS) \rightarrow \gamma\gamma)}.
\label{rp}
\eea
Ten of these ratios exist, ten being the number of bottomonium and charmonium
states below threshold. As we discussed above, in NRQCD, and if one
includes all the NRQCD operators up to ${\cal O}(1/m^4)$, these 10 ratios are
described by 40 non-perturbative
parameters. It is a specific prediction of pNRQCD that, for the states
for which the assumption $\lQ \gg mv^2$ holds, the wave-function
dependence drops out from the right-hand side of Eqs. (\ref{rv})
and (\ref{rp}). The residual flavour dependence is encoded in the
powers of $1/m$, in $E_{n0}^{(0)}$ and in the Wilson coefficients, while
the residual dependence on the principal quantum number is encoded in
the leading order binding energy $E_{n0}^{(0)}$.
In principle, if all the 10 bottomonium
and charmonium $S$-wave states below threshold belonged to the
dynamical regime  $\lQ \gg mv^2$, then, in the framework of
pNRQCD, the ratios of hadronic and electromagnetic decay widths
would be described by the 6 non-perturbative
universal parameters listed above only.

Particularly simple is, in pNRQCD, the expression of the ratios between
$R_n^V$ and $R_n^P$ with different principal quantum number.
We obtain up to order $v^2$ (with the counting $\lQ\sim mv$)
($M(nS) - 2 m = E_{n0}^{(0)}(1+{\cal O}(v))$, $M(nS)$ being the meson
mass):
\bea
& &{R_n^V\over R_m^V} = 1
+ \left({{\rm Im\,}g_1(^3 S_1) \over {\rm Im\,}f_1(^3 S_1)}
- {{\rm Im\,}g_{ee}(^3 S_1) \over {\rm Im\,}f_{ee}(^3 S_1)} \right)
{M(nS) - M(mS) \over m},
\label{rrv}
\\
& &{R_n^P\over R_m^P} = 1
+ \left({{\rm Im\,}g_1(^1 S_0) \over {\rm Im\,}f_1(^1 S_0)}
- {{\rm Im\,}g_{\gamma\gamma}(^1 S_0) \over {\rm Im\,}f_{\gamma\gamma}(^1
S_0)} \right)
{M(nS) - M(mS) \over m}.
\label{rrp}
\eea
It is to be stressed that the octet-type contributions cancel
(otherwise they would be $1/\als(m)$ enhanced in the vector case). This
prediction should be compared with the one expected in NRQCD. Within the
standard (perturbative-like) power counting, the same prediction is
obtained in NRQCD. However, if one counts $\als(m) \sim v^2$ as was
done in \cite{GK}, the contribution due to the octet matrix elements
is of the same order as the corrections obtained above and it should
be taken into account in the vector case. Therefore, in principle, one is able to check
the theory and/or the power counting.
As an example, taking $m_b = 5$ GeV we get for the $\Upsilon(2S)$ and
$\Upsilon(3S)$ states of the bottomonium system,
$R_2^\Upsilon/R_3^\Upsilon \simeq 1.3$, which is close
(within a 10\% accuracy) to the experimental central value of about $1.4$
that one can get from \cite{pdg}. Let us also notice that, since
${{\rm Im\,}g_1(^1 S_0) /{\rm Im\,}f_1(^1 S_0)}
- {{\rm Im\,}g_{\gamma\gamma}(^1 S_0) /{\rm Im\,}f_{\gamma\gamma}(^1 S_0)}
= {\cal O}(\als(m))$, up to corrections of order $v^3$ we find that
$R_n^P$, i.e. the ratio between hadronic and electromagnetic decay
widths for pseudoscalar quarkonium, is the same for all radial excitations.
However, it is not the purpose of this work to carry out
a comprehensive and detailed phenomenological analysis, 
which is left to a subsequent publication.

\section{Conclusions}
\label{conclusions}
We have obtained the imaginary part of the pNRQCD Hamiltonian up to
${\cal O}(1/m^4)$ in the non-perturbative regime ($k \gtrsim \lQ \gg mv^2$). 
The expressions are given in Eqs. (\ref{imh})--(\ref{imh4}).
As for any quantum-mechanical Hamiltonian, also the pNRQCD Hamiltonian 
is defined up to a unitary transformation. An alternative expression, related 
to the previous one by a unitary transformation, can be found in Sec. \ref{sec5d}.

We have applied our results to calculate the inclusive decay widths to light
hadrons, photons and leptons up to ${\mathcal O}(c(\als(m))mv^3\times (\lQ^2/m^2,E/m))$ 
for $S$-wave heavy quarkonium and up to ${\mathcal O}(c(\als(m))mv^5)$ for $P$-wave heavy quarkonium.
These are given in Eqs. (\ref{hadrV})--(\ref{electrchi}) and are the main 
result of the paper. An alternative way to present it is given in 
Sec. \ref{nrqcdmatrix}, where all the NRQCD matrix elements entering 
in quarkonium decays up to this order are expressed in terms of the quarkonium 
wave functions at the origin 
and 6 non-perturbative gluonic correlators, which are flavour- and 
state-independent, and for this reason may be called universal.
The wave-function dependence factorizes in all these expressions.
It is particularly remarkable that this is also true for the octet matrix elements.

We have derived our expressions in two different ways:
in Sec. \ref{qmmatching} under the general assumption $\lQ \siml k$ 
and in Sec. \ref{twostepmatching} under the particular assumption 
$k \gg \lQ $. In the first case, we have matched directly 
NRQCD to pNRQCD in an entirely non-perturbative one-step procedure,   
based on the Hamiltonian formulation of NRQCD. 
In the second case, we have matched NRQCD to pNRQCD 
in a two-step procedure, the first perturbative, the second non-perturbative, 
but still with a clear diagrammatic interpretation based on the multipole expansion.
The fact that these two completely different ways of deriving the pNRQCD Hamiltonian 
give the same answer up to a unitary transformation can be considered 
a stringent test on the correctness of the result.
In Sec. \ref{eveq} we have also checked that the evolution equations of 
our universal parameters are consistent at leading log accuracy with the 
known evolution equations of the NRQCD matrix elements.

In Sec. \ref{mind} we have considered the phenomenological implications 
of our results. There exist 14 charmonium and bottomonium states below threshold.
We expect our results to be applicable to most of these states. 
The exceptions are, on the one hand, the $\Upsilon (1 S)$, which is commonly 
understood as a weak coupling state (i.e. $k \gg E \gtrsim \lQ$),
and, on the other hand, states that are too close to the $D$-$\bar D$
threshold for charmonium or to the $B$-$\bar B$ threshold for bottomonium, like, maybe, the $\psi (2S)$.
Going from NRQCD to pNRQCD reduces the number of non-perturbative parameters needed 
to calculate the inclusive decay widths associated with these states 
by about a factor of 2. The situation is even better if we consider 
ratios of hadronic and electromagnetic decay widths. Since the wave-function 
dependence factorizes, it drops out in the ratios. It follows that only 6 universal 
parameters, which depend only on the light degrees of freedom of QCD, are needed. 
The already known data will be sufficient to fix all these parameters, to allow 
checks and to make new predictions. Moreover, suitable combinations of ratios
give rise to novel parameter-free, model-independent predictions. 
We have considered some of them in Sec. \ref{mind}.

The non-perturbative universal parameters that we have introduced do not 
necessarily need to be fitted to the experimental data. We have provided expressions 
for them in terms of correlators of gluonic fields. This allows for an eventual 
evaluation on the lattice. These parameters may also be obtained 
from QCD vacuum models \cite{Simonov}. We note that, once they become fixed, our results 
make the evaluation of NRQCD octet matrix elements possible from properties of
the wave functions at the origin. Hence, any potential model that leads to
definite wave functions \cite{EichtenQuigg} will provide definite results for
these matrix elements. Nevertheless, it should be pointed out that, 
if we wish to obtain the NRQCD matrix elements given 
in Eqs. (\ref{O13S1})--(\ref{OEM1S0}) with the aforementioned accuracy, 
any potential model to be used here must be consistent with the structure 
of the potential derived from NRQCD in terms of Wilson loops in
Refs. \cite{M1,M2}. In fact, the wave functions defined in this paper can also
be computed in a model-independent way without resorting to data fitting. 
This is so because our wave functions correspond to the solution of a
Schr\"odinger equation where the potentials are given in terms of expectation
values of Wilson loops with suitable field insertions. Therefore, once lattice simulations are 
provided for the potentials \cite{lattice2}, 
the wave function can be obtained unambiguously without any model dependence.

Since our method reduces the number of unknown parameters with respect to NRQCD, we expect 
it to 
become increasingly relevant as the number of needed NRQCD matrix elements increases. 
This seems to be necessary in the calculation of charmonium decay widths, where 
the non-relativistic expansion converges slowly. Indeed, higher order operators 
have been considered recently in Refs. \cite{mawang,bope}.
In Appendix \ref{appdir}, we give the general matching formula for the NRQCD matrix 
elements to the pNRQCD results without going 
through the whole matching procedure outlined in the main body of the paper.

We have also addressed, mainly in Sec. \ref{sec3h}, the issue of the power 
counting in NRQCD in the non-perturbative case. We believe that our formalism 
provides a suitable theoretical framework to study it. 
The power counting of NRQCD is not known a priori in the non-perturbative regime and 
it could, in principle, be different, depending on each dynamical system. This
is particularly transparent in pNRQCD. There, the potentials are functions of
$r$ and $\lQ$. Therefore, as the typical value of $r$ changes from system to
system, one should accordingly assign a different size to each given
potential. Moreover, having expressed the NRQCD matrix elements in terms of
wave functions and universal correlators, we have disentangled the soft scale
$k$, now entering in the wave function square, from the $\lQ / m$ and $E / m$
corrections. In fact, this is why we can construct ratios of convenient decay
rates where the $k$-dependence drops, providing a more constrained set of
relations. For these ratios the fixing of the power counting reduces 
to the evaluation of the correlators, while taking into account possible 
enhancement effects due to the NRQCD matching coefficients. 

Finally, although in the present paper we have focused on inclusive decays to
light hadrons, there should be no conceptual problem, a priori, in considering
the NRQCD matrix elements that appear in heavy quarkonium production. We also
expect there a significant reduction in the number of non-perturbative 
parameters. In particular, our formalism may shed some light on the power 
counting problems that appear in the heavy quarkonium polarization data \cite{FRL}.

\bigskip

{\bf Acknowledgements.}
D.E., A.P. and J.S. are supported by MCyT and Feder (Spain), FPA2001-3598,
and by  CIRIT (Catalonia), 2001SGR-00065. They thank the Benasque Center
of Science for hospitality while this work was being written up.
A.V. is supported by the European Community through a Marie-Curie fellowship, 
contract No. HPMF-CT-2000-00733.

\vfill\eject

\appendix

\section{4-fermion operators}
\label{appA}
Here we list the relevant 4-fermion operators of dimension 6 and 8, as 
taken from Ref. \cite{nrqcd},
\begin{eqnarray}
O_1({}^1S_0) &=& \psi^\dagger \chi \, \chi^\dagger \psi ,
\label{O1singS} \\
O_1({}^3S_1) &=& \psi^\dagger \mbox{\boldmath $\sigma$} \chi \cdot
\chi^\dagger \mbox{\boldmath $\sigma$} \psi ,
\label{O1tripS} \\
O_8({}^1S_0) &=& \psi^\dagger T^a \chi \, \chi^\dagger T^a \psi ,
\label{O8singS} \\
O_8({}^3S_1) &=& \psi^\dagger \mbox{\boldmath $\sigma$} T^a \chi \cdot
\chi^\dagger \mbox{\boldmath $\sigma$} T^a \psi,
\label{O8tripS}
\\
O_1({}^1P_1) &=& \psi^\dagger (-\mbox{$\frac{i}{2}$} \tensor{\bf D}) \chi
        \cdot \chi^\dagger (-\mbox{$\frac{i}{2}$} \tensor{\bf D}) \psi ,
\label{OsingP}
\\
O_1({}^3P_{0}) &=&  {1 \over 3} \;
\psi^\dagger (-\mbox{$\frac{i}{2}$} \tensor{\bf D} \cdot \mbox{\boldmath $\sigma$}) \chi
        \, \chi^\dagger (-\mbox{$\frac{i}{2}$} \tensor{\bf D} \cdot \mbox{\boldmath
$\sigma$}) \psi ,
\label{OtripP0}
\\
O_1({}^3P_{1}) &=&  {1 \over 2} \;
\psi^\dagger (-\mbox{$\frac{i}{2}$} \tensor{\bf D} \times \mbox{\boldmath
$\sigma$}) \chi
        \cdot \chi^\dagger (-\mbox{$\frac{i}{2}$} \tensor{\bf D} \times
\mbox{\boldmath $\sigma$}) \psi ,
\label{OtripP1}
\\
O_1({}^3P_{2}) &=& \psi^\dagger (-\mbox{$\frac{i}{2}$} \tensor{{\bf D}}{}^{(i}
\bfsigma^{j)}) \chi
        \, \chi^\dagger (-\mbox{$\frac{i}{2}$} \tensor{{\bf D}}{}^{(i} \bfsigma^{j)}) \psi ,
\label{OtripP2}
\\
{\cal P}_1({}^1S_0) &=& {1\over 2}
\left[\psi^\dagger \chi \, \chi^\dagger (-\mbox{$\frac{i}{2}$} \tensor{\bf
D})^2 \psi \;+\; {\rm H.c.}\right] ,
\label{PsingS}
\\
{\cal P}_1({}^3S_1) &=& {1\over 2}\left[\psi^\dagger \mbox{\boldmath $\sigma$}
\chi
        \cdot \chi^\dagger \mbox{\boldmath $\sigma$} (-\mbox{$\frac{i}{2}$}
\tensor{\bf D})^2 \psi \;+\; {\rm H.c.}\right] ,
\label{PtripS}
\\
{\cal P}_1({}^3S_1,{}^3D_{1}) &=& {1\over 2}\left[\psi^\dagger \bfsigma^i \chi \,
        \chi^\dagger \bfsigma^j (-\mbox{$\frac{i}{2}$})^2 \tensor{{\bf D}}{}^{(i}
\tensor{{\bf D}}{}^{j)} \psi
\;+\;{\rm H.c.}\right],
\label{PtripSD}
\\
O_8({}^1P_1) &=& \psi^\dagger (-\mbox{$\frac{i}{2}$} \tensor{\bf D}) T^a\chi
        \cdot \chi^\dagger (-\mbox{$\frac{i}{2}$} \tensor{\bf D}) T^a\psi ,
\label{OoctP}
\\
O_8({}^3P_{0}) &=&  {1 \over 3} \;
\psi^\dagger (-\mbox{$\frac{i}{2}$} \tensor{\bf D} \cdot \mbox{\boldmath $\sigma$}) T^a\chi
        \, \chi^\dagger (-\mbox{$\frac{i}{2}$} \tensor{\bf D} \cdot \mbox{\boldmath
$\sigma$}) T^a\psi ,
\label{OtripoctP0}
\\
O_8({}^3P_{1}) &=&  {1 \over 2} \;
\psi^\dagger (-\mbox{$\frac{i}{2}$} \tensor{\bf D} \times \mbox{\boldmath
$\sigma$}) T^a\chi
        \cdot \chi^\dagger (-\mbox{$\frac{i}{2}$} \tensor{\bf D} \times
\mbox{\boldmath $\sigma$}) T^a\psi ,
\label{OtripoctP1}
\\
O_8({}^3P_{2}) &=& \psi^\dagger (-\mbox{$\frac{i}{2}$} \tensor{{\bf D}}{}^{(i}
\bfsigma^{j)}) T^a\chi
        \, \chi^\dagger (-\mbox{$\frac{i}{2}$} \tensor{{\bf D}}{}^{(i} \bfsigma^{j)}) T^a\psi ,
\label{OtripoctP2}
\\
{\cal P}_8({}^1S_0) &=& {1\over 2}
\left[\psi^\dagger T^a\chi \, \chi^\dagger (-\mbox{$\frac{i}{2}$} \tensor{\bf
D})^2 T^a\psi \;+\; {\rm H.c.}\right] ,
\label{PoctS}
\\
{\cal P}_8({}^3S_1) &=& {1\over 2}\left[\psi^\dagger \mbox{\boldmath $\sigma$}
T^a\chi
        \cdot \chi^\dagger \mbox{\boldmath $\sigma$} (-\mbox{$\frac{i}{2}$}
\tensor{\bf D})^2 T^a\psi \;+\; {\rm H.c.}\right] ,
\label{PtripoctS}
\\
{\cal P}_8({}^3S_1,{}^3D_{1}) &=& {1\over 2}\left[\psi^\dagger \bfsigma^i T^a\chi \,
        \chi^\dagger \bfsigma^j (-\mbox{$\frac{i}{2}$})^2 \tensor{{\bf D}}{}^{(i}
\tensor{{\bf D}}{}^{j)} T^a\psi
\;+\;{\rm H.c.}\right],
\label{PtripoctSD}
\end{eqnarray}
where we use the conventional notation 
$T^{(ij)} \equiv (T^{ij}+T^{ji})/2 - T^{kk}\delta^{ij}/3$.
The electromagnetic operators are defined as follows:
\bea 
O_{\rm EM}(^1S_0) &=& 
\psi^\dagger \chi |{\rm vac}\rangle \langle {\rm vac}| \chi^\dagger \psi,
\\
O_{\rm EM}(^3S_1) &=& 
\psi^\dagger {\bfsigma} \chi |{\rm vac}\rangle \langle {\rm vac}|
\chi^\dagger {\bfsigma} \psi,
\\
O_{\rm EM}({}^1P_1) &=& \psi^\dagger (-\mbox{$\frac{i}{2}$} \tensor{\bf D})
\chi |{\rm vac}\rangle \cdot \langle {\rm vac}| 
\chi^\dagger (-\mbox{$\frac{i}{2}$} \tensor{\bf D}) \psi,
\\
O_{\rm EM}({}^3P_{0}) &=&  {1 \over 3} \;
\psi^\dagger (-\mbox{$\frac{i}{2}$} \tensor{\bf D} \cdot \bfsigma) \chi 
|{\rm vac}\rangle \langle {\rm vac}|
\chi^\dagger (-\mbox{$\frac{i}{2}$} \tensor{\bf D} \cdot \bfsigma) \psi,
\\
O_{\rm EM}({}^3P_{1}) &=&  {1 \over 2} \;
\psi^\dagger (-\mbox{$\frac{i}{2}$} \tensor{\bf D} \times \bfsigma) \chi 
|{\rm vac}\rangle  \cdot \langle {\rm vac}|
\chi^\dagger (-\mbox{$\frac{i}{2}$} \tensor{\bf D} \times \bfsigma) \psi,
\\
O_{\rm EM}({}^3P_{2}) &=& \psi^\dagger (-\mbox{$\frac{i}{2}$}
\tensor{\bf D}{}^{(i} \bfsigma^{j)}) \chi
|{\rm vac}\rangle \langle {\rm vac}|
\chi^\dagger (-\mbox{$\frac{i}{2}$} \tensor{\bf D}{}^{(i} \bfsigma^{j)}) \psi,
\\ 
{\cal P}_{\rm EM}(^1S_0) &=& {1\over 2} \left[ \psi^\dagger \chi 
|{\rm vac}\rangle \langle {\rm vac}|
\chi^\dagger \left( -{i\over 2} {\bf D}^2 \right)
\psi + {\rm H.c.} \right],
\\
\label{Pem3S1}
{\cal P}_{\rm EM}(^3S_1) &=& {1\over 2} \left[ \psi^\dagger {\bfsigma} \chi 
|{\rm vac}\rangle \langle {\rm vac}|
\chi^\dagger {\bfsigma} \left( -{i\over 2} {\bf D}^2 \right)
\psi + {\rm H.c.} \right],
\\
{\cal P}_{\rm EM}({}^3S_1,{}^3D_{1}) &=& {1\over 2}\left[\psi^\dagger \bfsigma^i \chi 
|{\rm vac}\rangle \langle {\rm vac}| 
\chi^\dagger \bfsigma^j (-\mbox{$\frac{i}{2}$})^2 \tensor{\bf D}{}^{(i}
\tensor{\bf D}{}^{j)} \psi  + {\rm H.c.}\right],
\eea
where $|{\rm vac}\rangle $ is the vacuum state of QCD.

\section{Direct Matching to pNRQCD of NRQCD matrix elements}
\label{appdir}
In principle, it is possible to match directly to pNRQCD matrix
elements of NRQCD that involve operators different from the
Hamiltonian $H$. In this way NRQCD matrix elements can be expressed in
terms of non-local correlators without going through the full matching
procedure outlined in the main body of the paper. This is 
useful if no iteration of these NRQCD operators are necessary in the
matching calculation.  In order to do this it is necessary to have an
explicit expression for the state $|{\underline 0};{\bf x}_1, {\bf
x}_2\rangle$.  Up to ${\cal O}(1/m)$ it can be found in
Eq. (\ref{state1}).  This way of doing will be particularly useful in
order to work out higher order operators that will appear in going
beyond ${\cal O}(mv^5)$ in the expansion of the heavy quarkonium decay
width. Higher order operators appear to be necessary for charmonium
decays, where the non-relativistic expansion converges slowly, assuming
$v^2_c \sim 0.3$.

The master equation is ($|H\rangle$ represents a generic heavy-quarkonium state at rest, 
${\bf P}=0$, with quantum numbers $n$, $j$, $l$ and $s$ as defined in Ref. \cite{nrqcd}):
\bea
&&\langle H|O|H\rangle =  {1\over \langle {\bf P}=0| {\bf P}=0 \rangle }
\int d^3{r}\int d^3{r}'\int d^3{R}\int d^3{R}' \,
\langle {\bf P}=0| {\bf R}\rangle \langle njls |{\bf r}\rangle
\\
\nn
&& 
\qquad\qquad\qquad\qquad\qquad
\times 
\bigg[
\langle \underbar{0}; {\bf x}_1 {\bf x}_2|
\int d^3\xi \; O(\bfxi) 
\;
| \underbar{0}; {\bf x}_1^\prime {\bf x}_2^\prime \rangle
\bigg]
\langle {\bf R}'|{\bf P}=0\rangle \langle {\bf r}'|njls \rangle
\,,
\eea
where ${\bf r} = {\bf x}_1-{\bf x}_2$, ${\bf r}' = {\bf x}'_1-{\bf x}'_2$, 
${\bf R} = ({\bf x}_1+{\bf x}_2)/2$ and ${\bf R}' = ({\bf x}'_1+{\bf
x}'_2)/2$ (note that $\langle  {\bf R}'|{\bf P}=0\rangle=1$ and
 $\langle {\bf P}=0| {\bf P}=0 \rangle=\int d^3x$). 
As an example, let us consider here the NRQCD matrix element 
\be
\langle \chi_Q(n01)|{\cal F}_{\rm EM}(^3P_0)|\chi_Q(n01)\rangle, 
\label{matrixma}
\ee
of the dimension-9 operator 
\be
{\cal F}_{\rm EM}(^3P_0) = {1\over 6}\psi^\dagger \bfsigma\cdot g{\bf E}
\chi |{\rm vac} \rangle \langle {\rm vac}| \chi^\dagger {\bfsigma}\cdot {\bf D}
  \psi + {\rm H.c.},
\ee
which is relevant to describe the electromagnetic decay 
$\chi_{c0} \rightarrow \gamma\gamma$ at order $mv^7$ accuracy.
Owing to spin symmetry, the same matrix element enters into the 
$\chi_{c2} \rightarrow \gamma\gamma$ decay.
These contributions have recently been considered in \cite{mawang}.
In the Hamiltonian formalism of Sec. \ref{qmmatching} the matrix 
element (\ref{matrixma}) is written as 
\bea
&&
\langle \chi_Q(n01)|{\cal F}_{\rm EM}(^3P_0)|\chi_Q(n01)\rangle  
=
{1\over \langle {\bf P}=0| {\bf P}=0 \rangle }\;
2\int d^3 {r} \int d^3 {r}'\int d^3{R}\int d^3{R}'\,
\langle {\bf P}=0| {\bf R}\rangle 
\nn
\\
&&\qquad
\times
\langle n011|{\bf r}\rangle \, 
\left[
\int d^3\xi \, 
\langle \underbar{0}; {\bf x}_1 {\bf x}_2 |  
{ \psi^\dagger \bfsigma\cdot g {\bf E} \chi 
|{\rm vac} \rangle \langle {\rm vac}| 
\chi^\dagger {\bfsigma}\cdot {\bf D} \psi \over 6} ({\bfxi}) \;
| \underbar{0}; {\bf x}_1^\prime {\bf x}_2^\prime \rangle 
\right]
\nn
\\
&&\qquad
\times
\, \langle{\bf r}'
| n011 \rangle
\langle  {\bf R}'|{\bf P}=0\rangle, 
\label{matrixma1}
\eea
where $|n011\rangle$ is the Schr\"odinger wave function of the state $\chi_Q(n01)$.
Now we expand the state $\langle \underbar{0}; {\bf x}_1{\bf
  x}_2|$ according to Sec. \ref{sec3c}. The first non-vanishing 
contribution comes from the $1/m$ correction given in Eq. (\ref{state1}).
Inserting that expression into Eq. (\ref{matrixma1}) 
and keeping in mind that only the term with the derivative projects onto the 
$|n011\rangle$ state, we obtain
\bea
&&
\langle \chi_Q(n01)|{\cal F}_{\rm EM}(^3P_0)|\chi_Q(n01)\rangle 
= 
{1\over \langle {\bf P}=0| {\bf P}=0 \rangle }\;
2\int d^3 {r} \int d^3 {r}'\int d^3{R}\int d^3{R}'\,
\langle {\bf P}=0| {\bf R}\rangle 
\nn
\\
&&\qquad
\times {1 \over m}\, \langle n011| {\bf r} \rangle
\sum_{k\neq 0} 
{-\bfnabla_{{\bf x}_1}\cdot {}^{(0)}\langle  0|g{\bf E}_1|k\rangle^{(0)}
+\bfnabla_{{\bf x}_2}\cdot {}^{(0)}\langle  0|g{\bf E}_2^T|k\rangle^{(0)}
\over (E_0^{(0)}-E_k^{(0)})^2} 
\nn
\\
&&\qquad
\times 
{}^{(0)}\langle \underbar{k}; {\bf x}_1 {\bf x}_2|
\int d^3 \xi  
{ \psi^\dagger \bfsigma\cdot g {\bf E} \chi 
|{\rm vac} \rangle \langle {\rm vac}| 
\chi^\dagger {\bfsigma}\cdot {\bf D} \psi \over 6} (\bfxi) \; 
| \underbar{0}; {\bf x}_1^\prime {\bf x}_2^\prime \rangle^{(0)} 
\, \langle {\bf r}'| n011 \rangle
\langle  {\bf R}'|{\bf P}=0\rangle, 
\nn
\\
&&~~
= 
- {2\over 3 m}
\sum_{k\neq 0} 
{{}^{(0)}\langle  0|g{\bf E}^{\ell^\prime}|k\rangle^{(0)}
{}^{(0)}\langle  k|g{\bf E}^\ell|{\rm vac}\rangle \, \langle {\rm vac}|0\rangle^{(0)} 
\over (E_0^{(0)}-E_k^{(0)})^2}
\langle n011| \bfsigma^{\ell} \bfnabla^{\ell^\prime} \delta^{(3)}({\bf r}) 
\bfsigma\cdot\bfnabla |n011\rangle.
\eea
In the second equality we have made use of Eq. (\ref{basis0M1}) and of the
Wick theorem. Finally from the fact that 
$\delta^{(3)}({\bf r}) |0\rangle^{(0)} = 
\delta^{(3)}({\bf r})\onec/\sqrt{N_c}|{\rm vac}\rangle$ 
and from Eq. (\ref{corrEn}) we get 
\be
\langle \chi_Q(n01)|{\cal F}_{\rm EM}(^3P_0)|\chi_Q(n01)\rangle = 
- C_A {|R_{n1}^{(0)\,\prime}({0})|^2\over \pi} {{\cal E}_1 \over m}, 
\ee
or equivalently, using Eq. (\ref{chio1}),  
\be
{ \langle \chi_Q(n01)|{\cal F}_{\rm EM}(^3P_0)|\chi_Q(n01)\rangle \over 
m \langle \chi_Q(n01) | O_{\rm EM}(^3P_0 ) | \chi_Q(n01) \rangle}
= - {2\over 3} {{\cal E}_1 \over m^2} .
\ee
 
Similar considerations may in principle also be applied to the matrix elements 
needed at relative order $v^4$ for $S$-wave decays. For a complete set 
of these operators and for considerations concerning their relevance in 
phenomenological studies, see Ref. \cite{bope}.

\section{Running equations of the matching coefficients}
\label{ARG}
The running equations obtained in Appendix B.3 of Ref. \cite{nrqcd} for
the NRQCD 4-fermion operators give us information on the running of
their matching coefficients. The running equations read as follows

\bea
\nu {d\over d\nu}{\rm Im\,}g_1(^1S_0)&=&{8 \over 3}C_f{\als \over
\pi}{\rm Im\,}f_1(^1S_0) 
\,,
\\
\nu {d\over d\nu}{\rm Im\,}g_1(^3S_1)&=&{8 \over 3}C_f{\als \over
\pi}{\rm Im\,}f_1(^3S_1) 
\,,
\\
\nu {d\over d\nu}{\rm Im\,}g_{\rm EM}(^1S_0)&=&{8 \over 3}C_f{\als \over
\pi}{\rm Im\,}f_{\rm EM}(^1S_0) 
\,,
\\
\nu {d\over d\nu}{\rm Im\,}g_{\rm EM}(^3S_1)&=&{8 \over 3}C_f{\als \over
\pi}{\rm Im\,}f_{\rm EM}(^3S_1) 
\,,
\\
\nu {d\over d\nu}{\rm Im\,}g_8(^1S_0)&=&{4 \over 3}\left(2C_f-{C_A \over
2}\right) {\als \over \pi}{\rm Im\,}f_8(^1S_0) 
\,,
\\
\nu {d\over d\nu}{\rm Im\,}g_8(^3S_1)&=&{4 \over 3}\left(2C_f-{C_A \over
2}\right) {\als \over \pi}{\rm Im\,}f_8(^3S_1) 
\,,
\\
\nu {d\over d\nu}{\rm Im\,}f_1(^1P_1)&=&{8 \over 3}C_f\left(C_f-{C_A \over
2}\right) {\als \over \pi}{\rm Im\,}f_8(^1S_0)
\,,
\\
\nu {d\over d\nu}{\rm Im\,}f_1(^3P_2)&=&{8 \over 3}C_f\left(C_f-{C_A \over
2}\right) {\als \over \pi}{\rm Im\,}f_8(^3S_1)
\,,
\\
\nu {d\over d\nu}{\rm Im\,}f_1(^3P_1)&=&{8 \over 3}C_f\left(C_f-{C_A \over
2}\right) {\als \over \pi}{\rm Im\,}f_8(^3S_1)
\,,
\\
\nu {d\over d\nu}{\rm Im\,}f_1(^3P_0)&=&{8 \over 3}C_f\left(C_f-{C_A \over
2}\right) {\als \over \pi}{\rm Im\,}f_8(^3S_1)
\,,
\\
\nu {d\over d\nu}{\rm Im\,}f_8(^1P_1)&=&-{8 \over 3}{\als \over
\pi}{\rm Im\,}f_1(^1S_0)-{4 \over 3}\left(4C_f-3{C_A \over
2}\right) {\als \over \pi}{\rm Im\,}f_8(^1S_0)
\,,
\\
\nu {d\over d\nu}{\rm Im\,}f_8(^3P_2)&=&-{8 \over 3}{\als \over
\pi}{\rm Im\,}f_1(^3S_1)-{4 \over 3}\left(4C_f-3{C_A \over
2}\right) {\als \over \pi}{\rm Im\,}f_8(^3S_1)
\,,
\\
\nu {d\over d\nu}{\rm Im\,}f_8(^3P_1)&=&-{8 \over 3}{\als \over
\pi}{\rm Im\,}f_1(^3S_1)-{4 \over 3}\left(4C_f-3{C_A \over
2}\right) {\als \over \pi}{\rm Im\,}f_8(^3S_1)
\,,
\\
\nu {d\over d\nu}{\rm Im\,}f_8(^3P_0)&=&-{8 \over 3}{\als \over
\pi}{\rm Im\,}f_1(^3S_1)-{4 \over 3}\left(4C_f-3{C_A \over
2}\right) {\als \over \pi}{\rm Im\,}f_8(^3S_1),
\eea
and 0 otherwise.

The imaginary pieces of the dimension-6 operator matching coefficients 
($f(S)$) do not run at leading non-vanishing order:
\be
\nu{d\over d\nu}{\rm Im\,}f(S)=0
\,.
\ee
Therefore, the above equations can be easily solved in that case. We
obtain at leading non-vanishing order:
\bea
{\rm Im\,}g_1(^1S_0)(\nu)&=&{\rm Im\,}g_1(^1S_0)(m)-{16 \over
3\beta_0}C_f {\rm Im\,}f_1(^1S_0)(m)\ln\left[{\als(\nu)\over \als(m)}\right] 
\,,
\\
{\rm Im\,}g_1(^3S_1)(\nu)&=&{\rm Im\,}g_1(^3S_1)(m)-{16 \over
3\beta_0}C_f {\rm Im\,}f_1(^3S_1)(m)\ln\left[{\als(\nu)\over \als(m)}\right] 
\,,
\\
{\rm Im\,}g_{\rm EM}(^1S_0)(\nu)&=&{\rm Im\,}g_{\rm EM}(^1S_0)(m)-{16 \over
3\beta_0}C_f {\rm Im\,}f_{\rm EM}(^1S_0)(m)\ln\left[{\als(\nu)\over \als(m)}\right] 
\,,
\\
{\rm Im\,}g_{\rm EM}(^3S_1)(\nu)&=&{\rm Im\,}g_{\rm EM}(^3S_1)(m)-{16 \over
3\beta_0}C_f {\rm Im\,}f_{\rm EM}(^3S_1)(m)\ln\left[{\als(\nu)\over \als(m)}\right] 
\,,
\\
{\rm Im\,}g_8(^1S_0)(\nu)&=&{\rm Im\,}g_8(^1S_0)(m)-{8 \over 3\beta_0}\left(2C_f-{C_A \over
2}\right) {\rm Im\,}f_8(^1S_0) (m)\ln\left[{\als(\nu)\over \als(m)}\right]
\,,
\\
{\rm Im\,}g_8(^3S_1)(\nu)&=&{\rm Im\,}g_8(^3S_1)(m)-{8 \over 3\beta_0}\left(2C_f-{C_A \over
2}\right) {\rm Im\,}f_8(^3S_1)(m)\ln\left[{\als(\nu)\over \als(m)}\right] 
\,,
\\
{\rm Im\,}f_1(^1P_1)(\nu)&=&{\rm Im\,}f_1(^1P_1)(m)-{16 \over
  3\beta_0} C_f\left(C_f-{C_A \over
2}\right) {\rm Im\,}f_8(^1S_0)(m)\ln\left[{\als(\nu)\over \als(m)}\right] 
\,,
\\
{\rm Im\,}f_1(^3P_2)(\nu)&=&{\rm Im\,}f_1(^3P_2)(m)-{16 \over
  3\beta_0} C_f\left(C_f-{C_A \over
2}\right) {\rm Im\,}f_8(^3S_1)(m)\ln\left[{\als(\nu)\over \als(m)}\right] 
\,,
\\
{\rm Im\,}f_1(^3P_1)(\nu)&=&{\rm Im\,}f_1(^3P_1)(m)-{16 \over
  3\beta_0} C_f\left(C_f-{C_A \over
2}\right) {\rm Im\,}f_8(^3S_1)(m)\ln\left[{\als(\nu)\over \als(m)}\right] 
\,,
\\
{\rm Im\,}f_1(^3P_0)(\nu)&=&{\rm Im\,}f_1(^3P_0)(m)-{16 \over
  3\beta_0} C_f\left(C_f-{C_A \over
2}\right) {\rm Im\,}f_8(^3S_1)(m)\ln\left[{\als(\nu)\over \als(m)}\right]
\,,
\\
{\rm Im\,}f_8(^1P_1)(\nu)&=&{\rm Im\,}f_8(^1P_1)(m)
\nn
\\
&&
\hspace{-5mm}
+\left(2 {\rm Im\,}f_1(^1S_0)(m)+\left(4C_f-3{C_A \over
2}\right) {\rm Im\,}f_8(^1S_0)(m)\right){8 \over 3\beta_0}
\ln\left[{\als(\nu)\over \als(m)}\right]
\,,
\\
{\rm Im\,}f_8(^3P_2)(\nu)&=&{\rm Im\,}f_8(^3P_2)(m)
\nn
\\
&&
\hspace{-5mm}
+\left(2 {\rm Im\,}f_1(^3S_1)(m)+\left(4C_f-3{C_A \over
2}\right) {\rm Im\,}f_8(^3S_1)(m)\right){8 \over 3\beta_0}
\ln\left[{\als(\nu)\over \als(m)}\right]
\,,
\\
{\rm Im\,}f_8(^3P_1)(\nu)&=&{\rm Im\,}f_8(^3P_1)(m)
\nn
\\
&&
\hspace{-5mm}
+\left(2 {\rm Im\,}f_1(^3S_1)(m)+\left(4C_f-3{C_A \over
2}\right) {\rm Im\,}f_8(^3S_1)(m)\right){8 \over 3\beta_0}
\ln\left[{\als(\nu)\over \als(m)}\right]
\,,
\\
{\rm Im\,}f_8(^3P_0)(\nu)&=&{\rm Im\,}f_8(^3P_0)(m)
\nn
\\
&&
\hspace{-5mm}
+\left(2 {\rm Im\,}f_1(^3S_1)(m)+\left(4C_f-3{C_A \over
2}\right) {\rm Im\,}f_8(^3S_1)(m)\right)
{8 \over 3\beta_0}\ln\left[{\als(\nu)\over \als(m)}\right],
\eea
where we have chosen $m$ as the starting point of the evolution
equation; the matching conditions at this scale at ${\cal O}(\als^2)$ can 
be read from Ref. \cite{nrqcd}.

\section{Regularizing products of distributions}
\label{Appreg}
In the intermediate steps of the calculation we find ill-defined products of
distributions. We first show how dimensional regularization makes sense out of these
expressions by setting them to zero, and next how they amount to renormalizations
of local terms when a cut-off regularization is used instead.

Consider, first, the product of two delta functions:
\bea
\delta^{(3)} ({\bf r})\, \delta^{(3)} ({\bf r})&=& 
\int d^D {\bf p}\int d^D {\bf p'}\int d^D {\bf p''} \, \vert {\bf
  p}\rangle\langle{\bf p}\vert \delta^{(3)} 
({\bf r})\vert {\bf p'}\rangle\langle{\bf p'}\vert \delta^{(3)} ({\bf r})
\vert {\bf p''}\rangle\langle{\bf p''}\vert
\cr  
&=& \int d^D {\bf p}\int d^D {\bf p'}\int d^D {\bf p''} \, 
\vert {\bf p}\rangle\langle{\bf p''}\vert =0,
\eea
since the integral over ${\bf p'}$ has no scale.

Consider next:
\bea
\delta^{(3)} ({\bf r}){1 \over r^s}
&=&\int d^D {\bf p}\int d^D {\bf p'}\int d^D {\bf p''} \, 
\vert {\bf p}\rangle\langle{\bf p}\vert \delta^{(3)} ({\bf r})\vert {\bf p'}
\rangle\langle{\bf p'}\vert {1\over r^s} \vert {\bf p''}\rangle\langle{\bf
  p''}\vert 
\cr 
&=& \int d^D {\bf p}\int d^D {\bf p'}\int d^D {\bf p''}\, \vert{\bf
  p}\rangle{{\rm const.}\over \vert
{\bf p'}-{\bf p''}\vert^{3-s}}\langle{\bf p''}\vert=0,
\eea
since, upon the translation ${\bf p'}\rightarrow {\bf p'}+{\bf p''}$, the integral over 
${\bf p'}$ has no scales.

Alternatively, if we use a cut-off regularization, for instance by smoothing 
the delta in momentum space, like:
\be
\langle{\bf p}\vert \delta^{(3)} ({\bf r})\vert {\bf p'}\rangle
=1\longrightarrow e^{-{({\bf p}-{\bf p'})^2\over \Lambda^2}} \,,
\ee
we obtain:
\be
\delta^{(3)} ({\bf r})\, \delta^{(3)} ({\bf r})\sim {\pi \sqrt{2\pi}\over 4} \Lambda^3
\delta^{(3)} ({\bf r}) - {\pi\over 4} \sqrt{{\pi \over 2}} \Lambda 
\left( \{\bfnabla^2 ,\delta^{(3)} ({\bf r})\} + 2\bfnabla^i\delta^{(3)} ({\bf r})\bfnabla^i
\right) + {\mathcal O} \left ( {1\over \Lambda } \right ) \, , 
\ee
which can be removed by local counterterms. 
Hence DR implements nothing but a suitable subtraction prescription.
Analogously, it is easy to see that 
$\delta^{(3)} ({\bf r})/r^s$ for $s=0,1,2,...$ reduces to local terms.

\section{Unitary transformations}
\label{Appunit}
It is well known that quantum-mechanical Hamiltonians, which are related by
unitary transformations, lead to the same physics.
This fact is particularly relevant to quantum-mechanical Hamiltonians, 
which are derived from a field theory, which is our case.
It is also well known that the quantum-mechanical potentials 
that are obtained from QED depend on the gauge one uses in the calculation 
(this is also so for QCD in perturbation theory), but physical 
observables computed with either potential turn out to be the same.
It is perhaps not so well known that the potentials obtained in one gauge 
can be related to the ones obtained with a different 
gauge by means of a unitary transformation. In fact the arbitrariness in the 
form of the potentials is not only due to gauge dependence.
It depends in general on the way one carries out the matching calculation. 
Any correct result is related to any other one by 
means of a unitary transformation. 

We shall use this fact here to prove that the result obtained in
Sec. \ref{sec5d} is equivalent to the one obtained in Sec. \ref{sec3g}.

Consider the following unitary transformation:
\be
U=e^{i\{ {\bf r},{\bf p}\} {\Lambda \over m}}\, .\label{unitary}
\ee
Consider next a delta function in the Hamiltonian: 
\be U^{\dagger} {\delta^{(3)} ({\bf r}) \over m^2} U \sim{\delta^{(3)} ({\bf r}) \over m^2}-{i \Lambda\over m^3} 
\left[ \{ {\bf r},{\bf p}\},\delta^{(3)} ({\bf r}) \right] \, , \label{hamu}
\ee
which on $S$-wave states reduces to:
\be
U^{\dagger} {\delta^{(3)} ({\bf r}) \over m^2} U \sim {\delta^{(3)} ({\bf r})
  \over m^2}+ {6\Lambda \over m^3}\delta^{(3)} ({\bf r}) \, . \label{hamu2}
\ee
This shows that a suitable unitary transformation may 
induce terms at ${\mathcal O } (1/m^3)$ proportional to $\vert \phi (0)\vert^2$. 
Of course, physics should not change. If $\phi ({\bf r})$ is an eigenfunction of $h$, then 
${\tilde \phi} ({\bf r})= U^{\dagger}\phi ({\bf r})$ is an eigenfunction of
$\tilde h = U^{\dagger} h U$. Then: 
\be
{\tilde \phi} ({\bf 0})=e^{-i\left( 2{\bf r}{\bf p} -3i\right){\Lambda \over
    m}}   \phi ({\bf 0}) \left(1 + {\cal O}(\vert {\bf r}\vert )\right) \vert_{{\bf r}=0}
=e^{-3{\Lambda \over m}} \phi ({\bf 0}) \, . \label{wfu}
\ee
Clearly:
\be
\left( {1 \over m^2} + {6\Lambda \over m^3}\right) \vert {\tilde \phi}
(0)\vert^2 \sim  {1 \over m^2}\vert\phi (0)\vert^2  \, . 
\label{wfu2}
\ee

We will illustrate this issue further with an example. Recall the two
different results, namely (\ref{m2e2sym}) and (\ref{m2e2asym}), 
we obtained from the first diagram
of Fig. 2 at second order in the expansion $mv^2/\lQ$. 
More explicitly, from (\ref{m2e2sym}) we get a real result:
\be
-{1 \over E-h_s} {4{\bfnabla}^2 \over m^2} 
\int_0^{\infty} dt\, t^2 \, \langle g{\bf E}(t) \cdot g {\bf E} (0) \rangle{1 \over E-h_s} \, ,
\ee
and from (\ref{m2e2asym}) a result containing an imaginary part:
\be
{1 \over E-h_s} 2 \left ( {\bf r}\cdot {( {\bfnabla}V_s) \over m}  
+3i f_1(^{2S+1} S_S) { \delta^{(3)}({\bf r}) \over m^3} \right ) 
\int_0^{\infty} dt \,  t^2 \, \langle g{\bf E}(t) \cdot g {\bf E} (0) \rangle {1 \over E-h_s} \, .
\ee
Both results are correct. They are indeed related by the following unitary transformation:
\bea
&& U = e^{i \{ {\bf r},{\bf p} \} {N_c{\mathcal E}_{2} \over m}} \, , 
\nonumber \\
&& {\mathcal E}_{2} 
= {1 \over N_c} \int_0^{\infty} dt \,  t^2 \, \langle g{\bf E}(t) \cdot g {\bf E} (0) \rangle  \, ,
\eea
and hence lead to the same decay width. This can even be further confirmed 
by an explicit calculation in the case of the Coulomb potential, since  
the induced terms then retain the same form as the original ones.

The unitary transformation, which brings the result 
of Sec. \ref{sec3g} to the one of Sec. \ref{sec5d} reads:
\bea
&& U = e^{-i \{ {\bf r},{\bf p} \} {\vartheta^2 \over m^2}} \, , 
\nonumber \\
&&\vartheta^2 = {1 \over 9} ({\cal E}_3^{(2,t)}+{\cal E}_3^{(2,{\rm EM})}
 - {\bar{\cal E}}_3^{(2,t)} - {\bar {\cal E}}_3^{(2,{\rm EM})}) \, .  
\eea

Clearly this transformation also reshuffles $1/m$ real
potentials into $1/m^3$ real potentials. This means that, in the 
more conservative counting considered in Sec. \ref{sec3h}, the 
whole set of potentials up to ${\cal O}(1/m^3)$, which are formally 
given in Sec. \ref{sec3b}, are expected to be relevant 
to calculate the wave function at the origin
with an accuracy that matches the NNLO terms in (\ref{hadrV})--(\ref{electrchi}).

\vfill\eject

\vfill\eject

\begin{figure} 
\centerline{\epsfxsize = 15cm \epsfbox{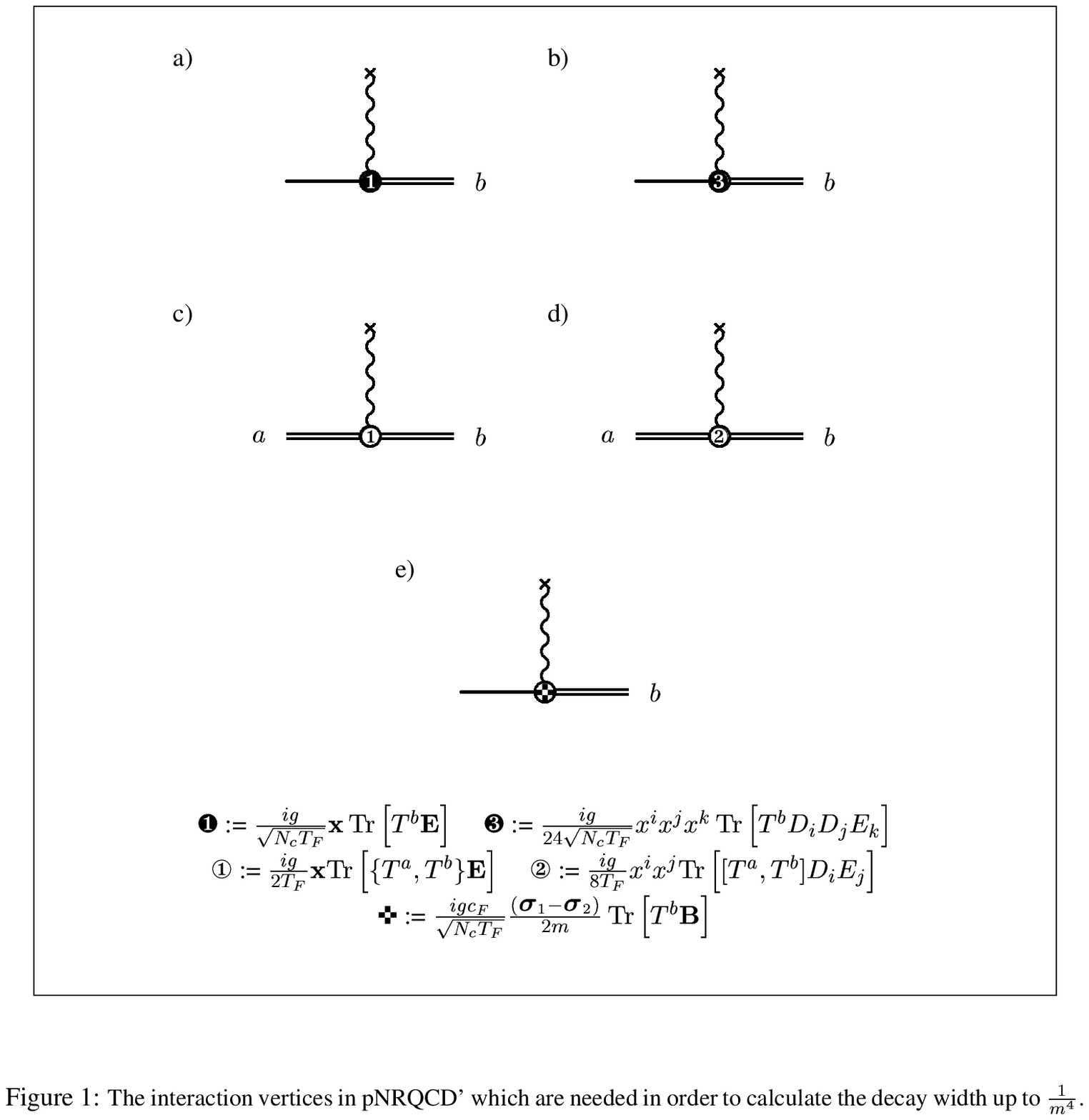}  
}  
\end{figure}    

\newpage
\begin{figure} 
\vspace{1cm}
\centerline{\epsfxsize = 17cm \epsfbox{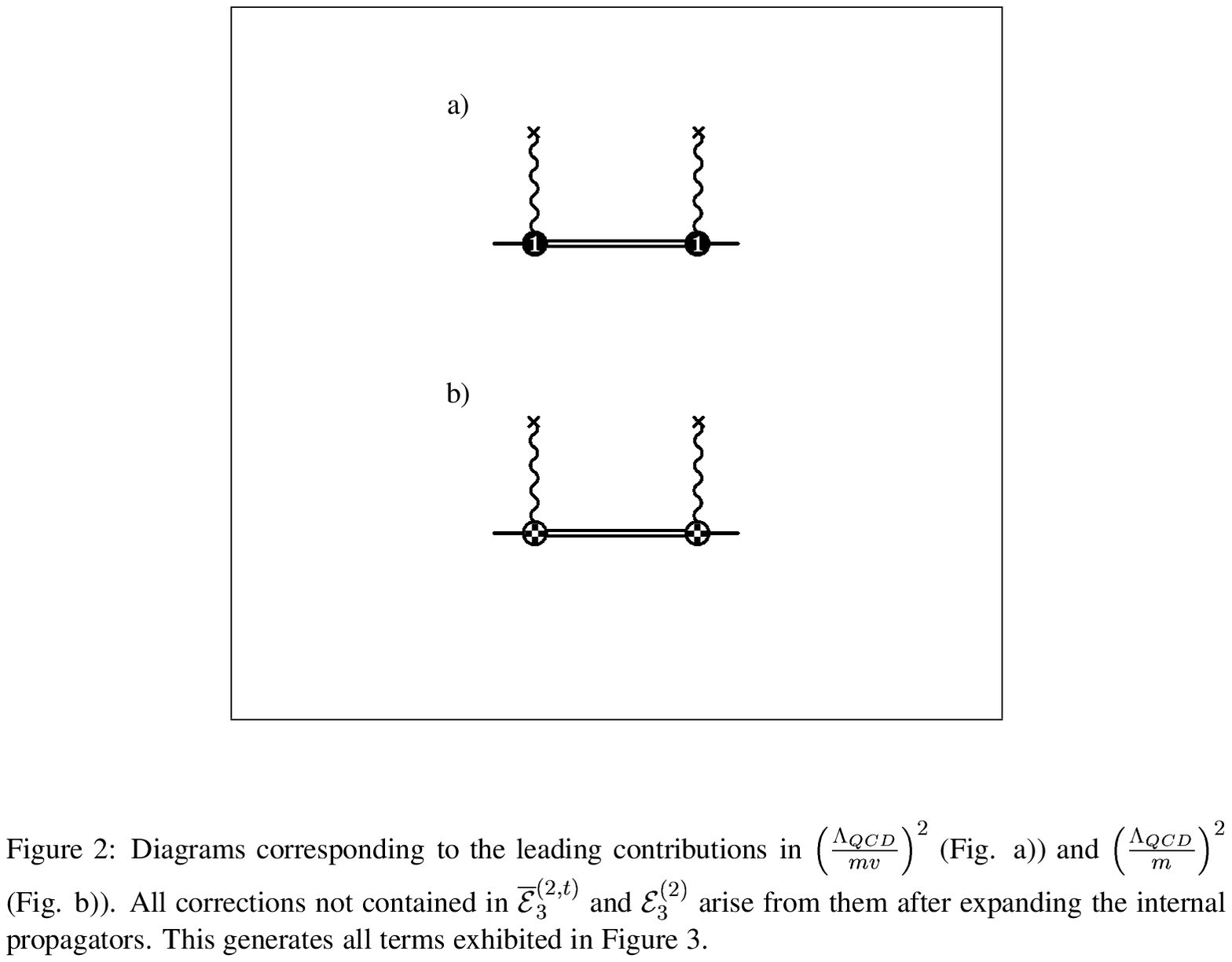}  
}  
\vspace{.5cm}
\end{figure}    

\newpage

\begin{figure} 
\vspace{1cm}
\centerline{\epsfxsize = 17cm \epsfbox{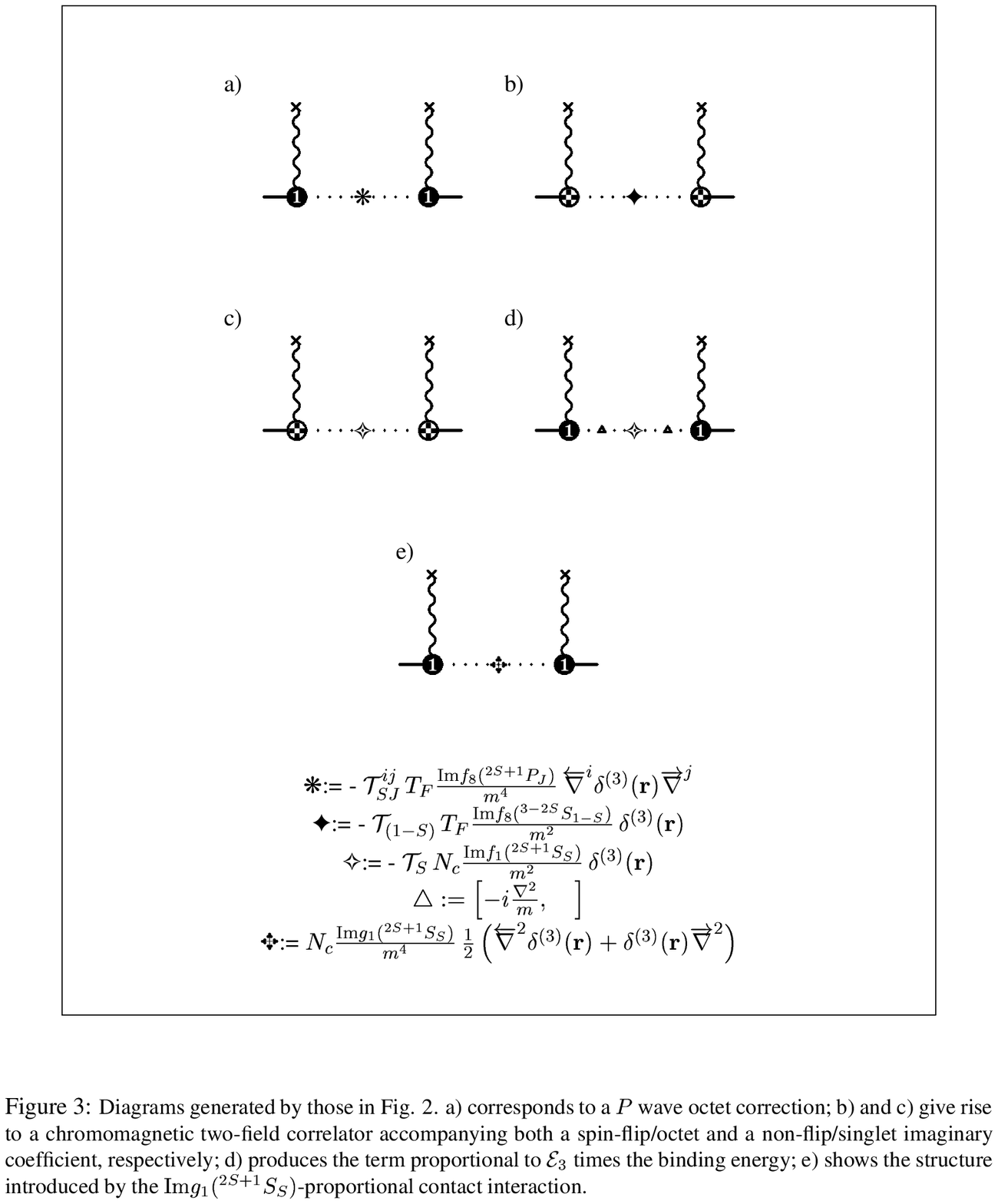}        
}  
\vspace{.5cm}
\end{figure}    

\newpage

\begin{figure} 
\vspace{1cm}
\centerline{\epsfxsize = 17cm \epsfbox{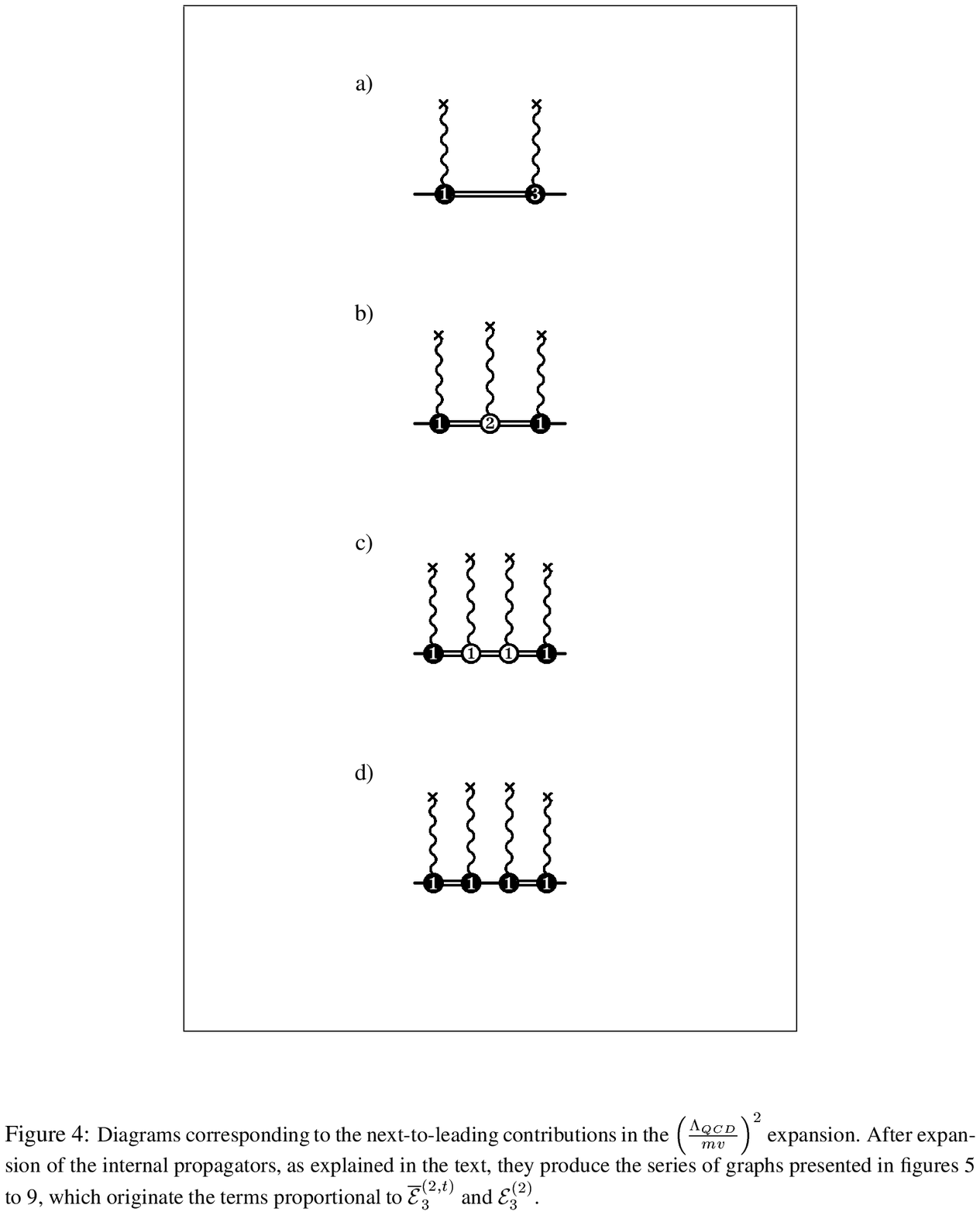} 
}  
\vspace{.5cm}
\end{figure}    

\newpage
\begin{figure} 
\vspace{1cm}
\centerline{\epsfxsize = 19cm \epsfbox{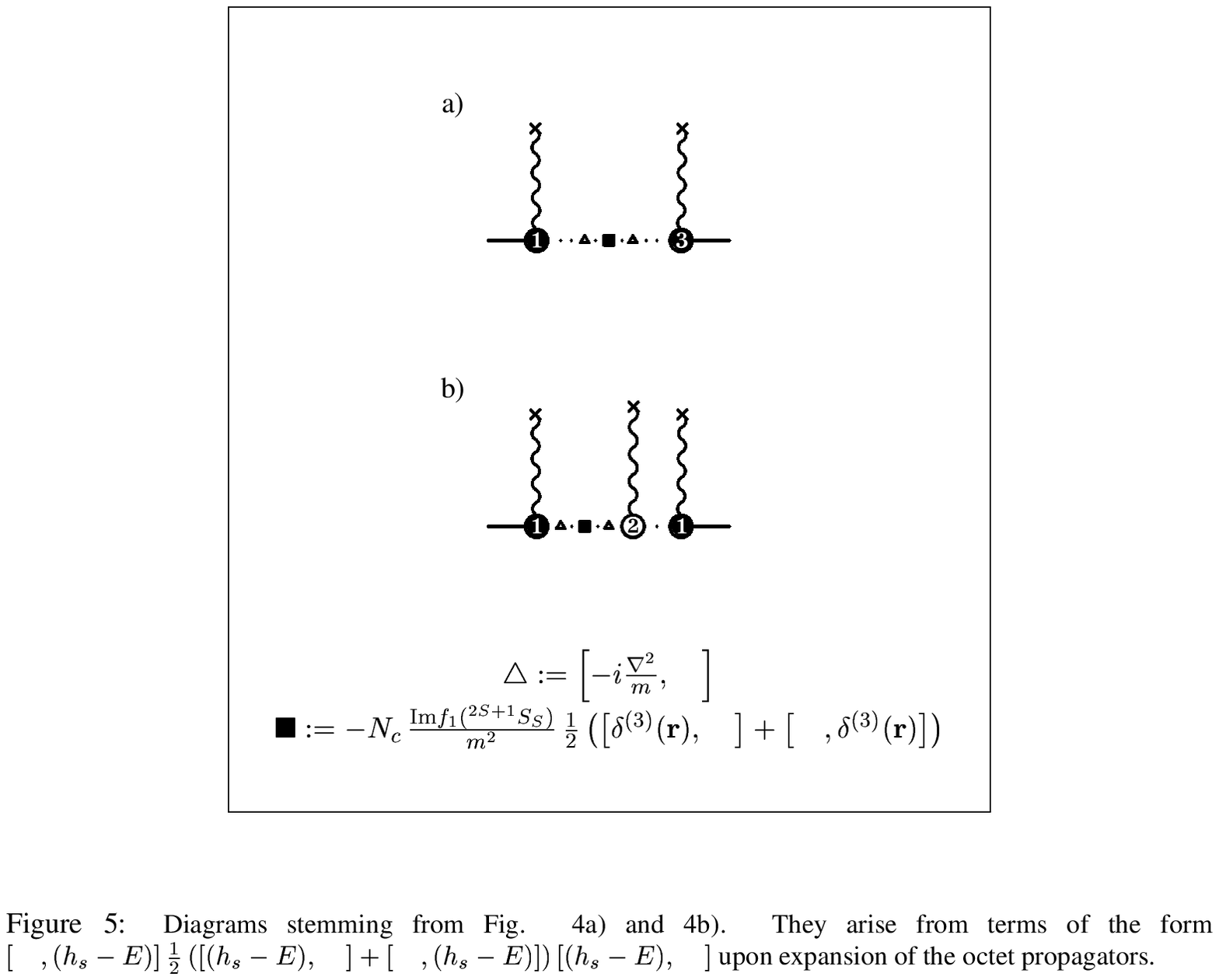} 
}  
\vspace{.5cm}
\end{figure}    
\newpage
\begin{figure} 
\vspace{1cm}
\centerline{\epsfxsize = 17cm \epsfbox{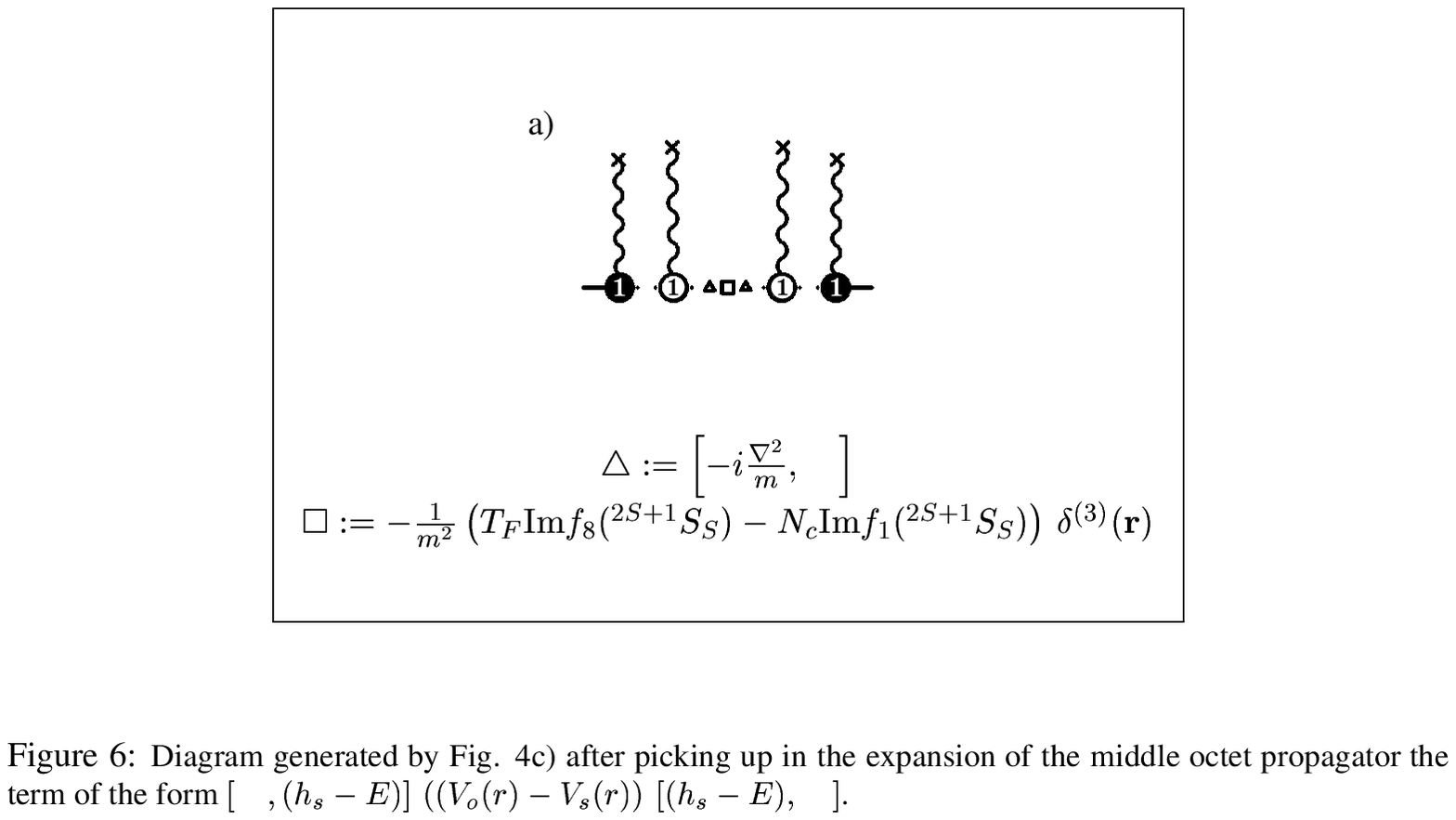}  
}  
\vspace{.5cm}
\end{figure}    

\newpage

\begin{figure} 
\vspace{1cm}
\centerline{\epsfxsize = 17cm \epsfbox{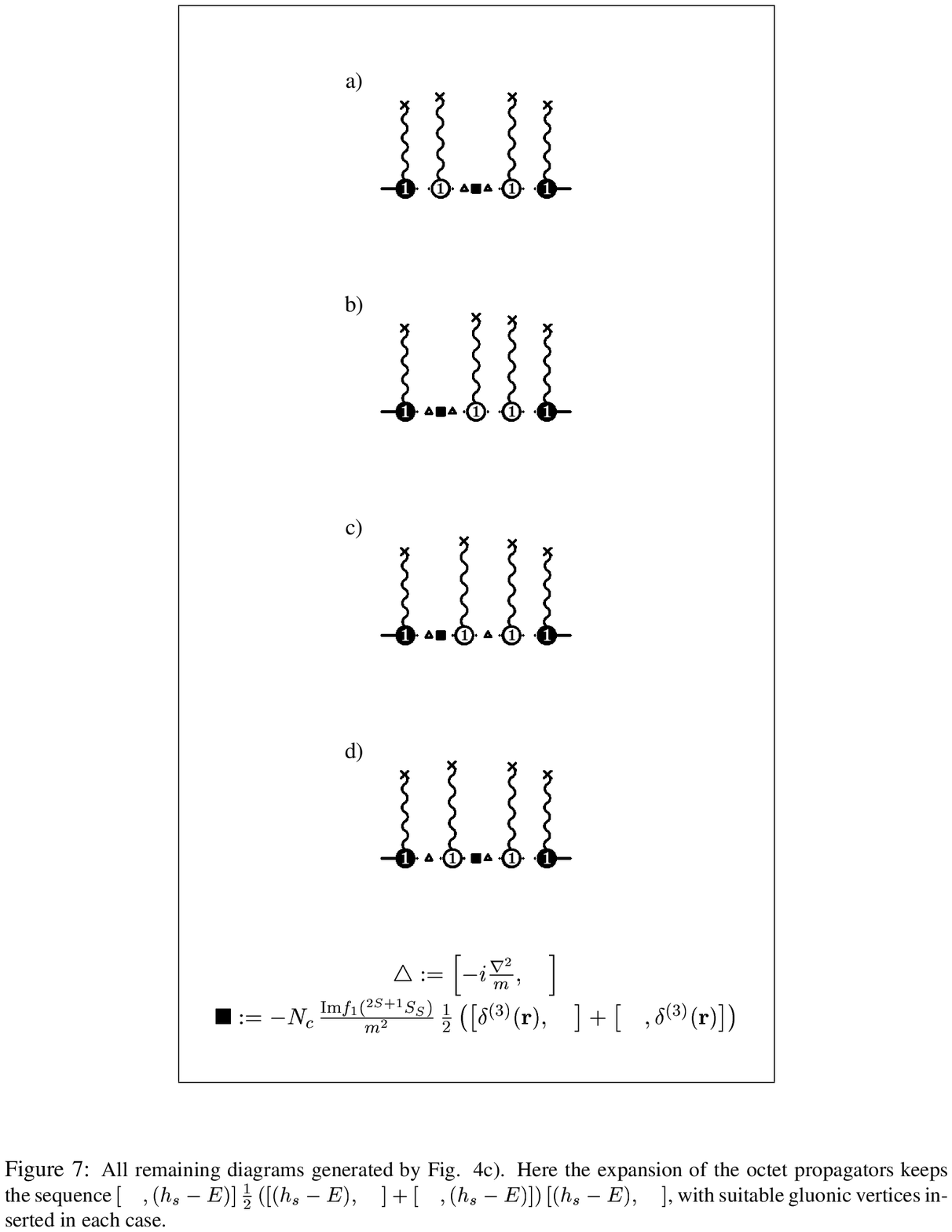}        
}  
\vspace{.5cm}
\end{figure}    

\newpage

\begin{figure} 
\vspace{1cm}
\centerline{\epsfxsize = 17cm \epsfbox{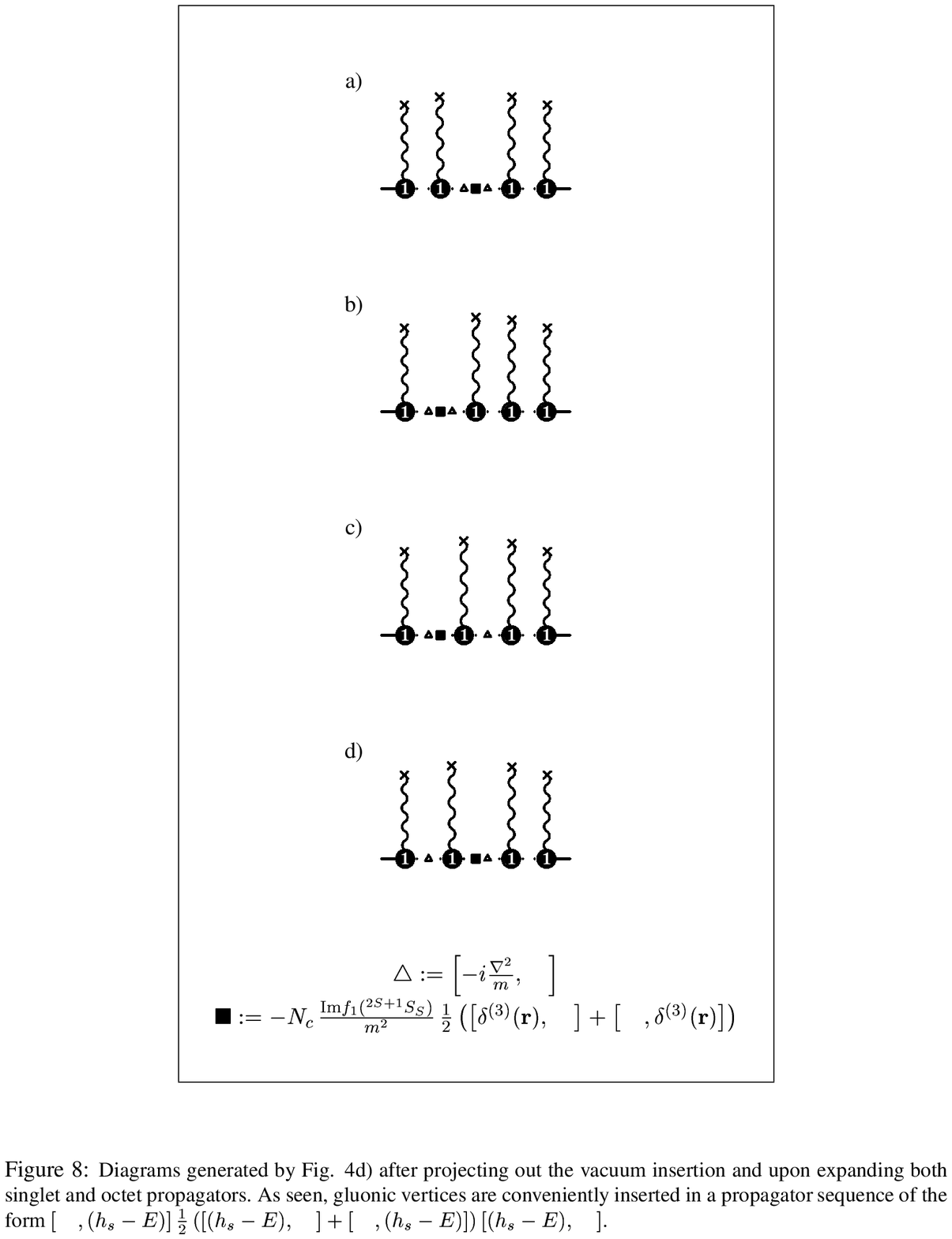}        
}  
\vspace{.5cm}
\end{figure}    

\newpage

\begin{figure} 
\vspace{1cm}
\centerline{\epsfxsize = 17cm \epsfbox{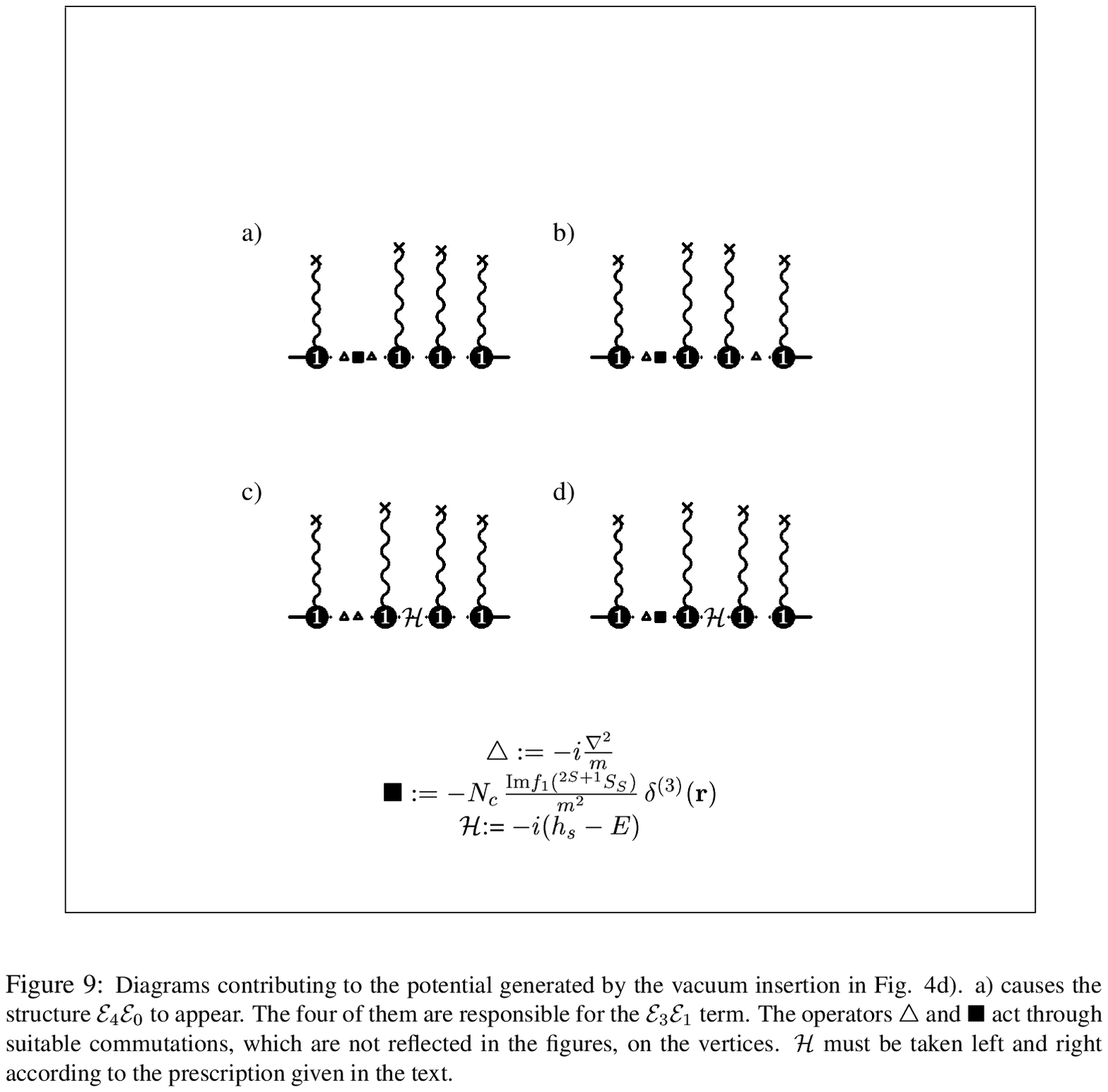} 
}  
\vspace{.5cm}
\end{figure}


\begin{thebibliography}{999}

\bibitem{nrqcd} G.T. Bodwin, E. Braaten and G.P. Lepage,
 Phys. Rev. {\bf D51}, 1125  (1995); Erratum, {\it ibid.} {\bf D55}, 5853 (1997).

\bibitem{potentials} K.G. Wilson, Phys. Rev. {\bf D10}, 2445 (1974); 
 L. Susskind, in Les Houches, Session XXIX, 
 ed. R. Balian and C.H. Llewellyn Smith (North-Holland Publishing Company,
 Amsterdam, 1977); 
 E. Eichten and F.L. Feinberg, Phys. Rev. {\bf D23}, 2724 (1981); 
 M.E. Peskin, in Proc. 11th SLAC Institute, SLAC Report No. 207,
 151, ed. P. Mc Donough (1983); 
 D. Gromes, Z. Phys. {\bf C26}, 401 (1984); 
 A. Barchielli, E. Montaldi and G.M. Prosperi, Nucl. Phys. {\bf B296}, 625 (1988); 
 Erratum, {\it ibid.} {\bf 303}, 752 (1988); 
 A. Barchielli, N. Brambilla and G. Prosperi, Nuovo Cimento {\bf 103A}, 59 (1990); 
 Y. Chen, Y. Kuang and R. J. Oakes, Phys. Rev. {\bf D52}, 264 (1995);
 A. P. Szczepaniak and E. S. Swanson, Phys. Rev. {\bf D55}, 3987 (1997).

\bibitem{potpert} W. Fischler, Nucl. Phys. {\bf B129}, 157 (1977);
 S.N. Gupta, S.F. Radford and W.W. Repko, Phys. Rev. {\bf D26}, 3305 (1982);
 S. Titard and F.J. Yndur{\'a}in, Phys. Rev. {\bf D49}, 6007 (1994); Y. Schr\"oder, Phys. 
 Lett. {\bf B447}, 321 (1999); M. Peter, Phys. Rev. Lett. {\bf 78}, 602  (1997).
 
\bibitem{Mont} A. Pineda and J. Soto, Nucl. Phys. {\bf B} (Proc. Suppl.) {\bf 64}, 428 (1998).

\bibitem{nucl} G.P. Lepage, nucl-th/9706029.

\bibitem{long} N. Brambilla, A. Pineda, J. Soto and A. Vairo, Nucl. Phys. {\bf B566}, 275 (2000).

\bibitem{VLP}  M. B. Voloshin, Nucl. Phys. {\bf B154}, 365 (1979); 
 H. Leutwyler, Phys. Lett. {\bf B98}, 447 (1981);
 A. Pineda, Nucl. Phys. {\bf B494}, 213 (1997).

\bibitem{short} N. Brambilla, A. Pineda, J. Soto and A. Vairo, Phys. Rev. {\bf D60}, 091502 (1999). 

\bibitem{logs} N. Brambilla, A. Pineda, J. Soto and A. Vairo, Phys. Lett. {\bf B470}, 215 (1999). 

\bibitem{KP} B.A. Kniehl and A.A. Penin, Nucl. Phys. {\bf B563}, 200 (1999). 

\bibitem{RGstatic} A. Pineda and J. Soto, Phys. Lett. {\bf B495}, 323 (2000).

\bibitem{RG} A. Pineda, Phys. Rev. {\bf D65}, 074007 (2002).

\bibitem{KPSS} B.A. Kniehl, A.A. Penin, V.A. Smirnov and 
 M. Steinhauser, Phys. Rev. {\bf D65}, 091503 (2002); 
 Nucl. Phys. {\bf B635}, 357 (2002); 
 A.A. Penin, M. Steinhauser, Phys. Lett. {\bf B538}, 335 (2002).

\bibitem{M1} N. Brambilla, A. Pineda, J. Soto and A. Vairo, Phys. Rev. {\bf D63}, 014023 (2001).
        
\bibitem{M2} A. Pineda and A. Vairo, Phys. Rev. {\bf D63}, 054007 (2001); 
 Erratum, {\it ibid.} {\bf D64}, 039902 (2001).

\bibitem{pw} N. Brambilla, D. Eiras, A. Pineda, J. Soto and A. Vairo, 
 Phys. Rev. Lett. {\bf 88}, 012003 (2002). 

\bibitem{ABN} G. Amoros, M. Beneke and M. Neubert, Phys. Lett. {\bf B401}, 81 (1997).

\bibitem{ManBauer} C. Bauer and A.V. Manohar, Phys. Rev. {\bf D57}, 337 (1998).

\bibitem{Hatfield} A.A Slavnov and L.D. Faddeev, {\it Gauge Fields:
 Introduction to quantum theory} (Benjamin-Cummings, Menlo Park, 1980);
 B. Hatfield, {\it Quantum Field Theory of Point Particles and Strings}
 (Addison-Wesley, Boston, 1992); R. Jackiw, in {\it Current Algebra and
 Anomalies}, eds. S.B. Treiman, R. Jackiw, B. Zumino and E. Witten (World Scientific,
 Singapore, 1985).

\bibitem{4f} A. Pineda and J. Soto, Phys. Rev. {\bf D58}, 114011 (1998).
 
\bibitem{lattice} TXL Collaboration and T(X)L Collaboration (G.S. Bali et
  al.), Phys. Rev. {\bf D62}, 054503 (2000).

\bibitem{Neubert} M. Neubert, Phys. Rep. {\bf 245}, 259 (1994).

\bibitem{CHARMONIUM} E. Eichten, K. Gottfried, T. Kinoshita, K.D. Lane and
 Tung-Mow Yan, Phys. Rev. {\bf D17}, 3090 (1978) and  Erratum, {\it ibid.} {\bf D21}, 313 (1980).

\bibitem{CMY} A. Czarnecki, K. Melnikov and A. Yelkhovsky, Phys. Rev. {\bf
    A59}, 4316 (1999).

\bibitem{pcnrqcd} G.P. Lepage, L. Magnea, C. Nakhleh, U. Magnea and
 K. Hornbostel, Phys. Rev. {\bf D46}, 4052 (1992).

\bibitem{GP} A. Galindo and P. Pascual, {\it Mec\'anica Cu\'antica} (Eudema Universidad, Madrid, 1989).

\bibitem{Beneke} M. Beneke, in Proc. 24th SLAC Institute, SLAC Report No. 508,
  549, eds. J. Chan, L. DePorcel and L. Dixon (1997); hep-ph/9703429.

\bibitem{FRL} S. Fleming, I.Z. Rothstein and A.K. Leibovich, Phys. Rev. {\bf D64}, 036002 (2001). 

\bibitem{LukeManohar} M.E. Luke and A.V. Manohar, Phys. Lett. {\bf B286}, 348 (1992). 

\bibitem{BGV} N. Brambilla, D. Gromes and A. Vairo, Phys. Rev. {\bf D64}, 076010 (2001). 

\bibitem{BV} M. Beneke and V.A. Smirnov, Nucl. Phys. {\bf B522}, 321 (1998).  

\bibitem{GK} M. Gremm and  A. Kapustin, Phys. Lett. {\bf B407}, 323 (1997). 

\bibitem{latbo} G.T. Bodwin, D.K. Sinclair and S. Kim,
  Int. J. Mod. Phys. {\bf A12}, 4019 (1997);
  Phys. Rev. {\bf D65}, 054504 (2002).

\bibitem{Maltoni} F. Maltoni, hep-ph/0007003.

\bibitem{pdg} K. Hagiwara et al., Phys. Rev. {\bf D66}, 010001 (2002).

\bibitem{Simonov} A. Di Giacomo, H.G. Dosch, V.I. Shevchenko and
 Yu.A. Simonov, Phys. Rept. {\bf 372}, 319 (2002); 
 M. Baker, N. Brambilla, H. G. Dosch and
 A. Vairo, Phys. Rev. {\bf D58}, 034010 (1998).

\bibitem{EichtenQuigg} E. J. Eichten and C. Quigg, Phys. Rev. {\bf D52}, 1726 (1995).

\bibitem{lattice2} G.S. Bali, K. Schilling and A. Wachter, Phys. Rev. {\bf
    D56}, 2566 (1997); see also G.S. Bali, Phys. Rep. {\bf 343}, 1 (2001). 

\bibitem{mawang} J.P. Ma and Q. Wang, Phys. Lett. {\bf B537}, 233 (2002).

\bibitem{bope} G.T. Bodwin and A. Petrelli, Phys. Rev. {\bf D66}, 094011 (2002).

\end{thebibliography}
\end{document}